\documentclass[10pt]{article}
\usepackage{fancyhdr}
\usepackage{extramarks}
\usepackage{amsmath}
\usepackage{amsthm}
\usepackage{amsfonts}
\usepackage{siunitx}
\usepackage{tikz}
\usepackage[plain]{algorithm}
\usepackage{algpseudocode}
\usepackage{multirow}
\usepackage{booktabs}
\usepackage{graphicx, subfig}
\usepackage[colorlinks,linkcolor=black,anchorcolor=black,citecolor=black,urlcolor=blue]{hyperref}
\usepackage{amsmath,bm}
\usepackage{booktabs}
\usepackage{mathtools}
\usepackage{amssymb}
\usepackage{caption}
\usepackage{capt-of}
\usepackage{mciteplus}
\usepackage{cite}
\usepackage{mathrsfs}
\usepackage[title,titletoc,toc]{appendix}
\usepackage{xr}
\usepackage{parskip}
\usepackage{soul}
\usepackage{textcomp}
\usepackage[colaction]{multicol}
\usepackage[switch]{lineno}
\usepackage{lipsum}
\usepackage{etoolbox}
\usepackage{longtable}
\usepackage{array}
\usepackage{tablefootnote}
\usepackage{ragged2e}
\usepackage{soul}
\usepackage{xr}

\usepackage{pdflscape}
\usepackage{afterpage}

\newcolumntype{C}[1]{>{\centering\arraybackslash}p{#1}}
\usetikzlibrary{automata,positioning}
\usetikzlibrary{automata,positioning}

\usepackage{xr}

\begin{document}
	\title{Revealing the Shape of Genome Space via $K$-mer Topology}
	
	\author{Yuta Hozumi$^1$\footnote{Current address: School of Mathematics, Georgia Institute of Technology, Atlanta, GA 30332} ~and  Guo-Wei Wei$^{1,2,3}$\footnote{
			Corresponding author.		Email: weig@msu.edu} \\
		\\
		$^1$ Department of Mathematics, \\
		Michigan State University, East Lansing, MI 48824, USA.\\
		$^2$ Department of Electrical and Computer Engineering,\\
		Michigan State University, East Lansing, MI 48824, USA.\\
		$^3$ Department of Biochemistry and Molecular Biology,\\
		Michigan State University, East Lansing, MI 48824, USA. \\
	}
	\date{\today} 
	
	\maketitle
	
	\begin{abstract}
		Despite decades of effort, understanding the shape of genome space in biology remains a challenge due to the similarity, variability, diversity, and plasticity of evolutionary relationships among species, genes, or other biological entities. We present a $k$-mer topology method, the first of its kind, to delineate the shape of the genome space. $K$-mer topology examines the topological persistence and the evolution of the homotopic shape of the sequences of $k$ nucleotides in species, organisms, and genes using persistent Laplacians, a new multiscale combinatorial approach. We also propose a topological genetic distance between species by their topological invariants and non-harmonic spectra over scales. This new metric defines the topological phylogenetic trees of genomes, facilitating species classification and clustering. $K$-mer topology substantially outperforms state-of-the-art methods on a variety of benchmark datasets, including mammalian mitochondrial genomes,  Rhinovirus, SARS-CoV-2 variants, Ebola virus,  Hepatitis E virus, Influenza hemagglutinin genes, and whole bacterial genomes.  $K$-mer topology reveals the intrinsic shapes of the genome space and can be directly applied to the rational design of viral vaccines.

	\end{abstract}
	keywords: {Shape of genome space, Phylogenetic analysis, Topological genetic distance,   Persistent Laplacian}
	
	\newpage
	
	{\setcounter{tocdepth}{4} \tableofcontents}
	\setcounter{page}{1}
	\newpage
	
	\section{Introduction}

	Phylogenetic analysis of genetic sequences is vital for understanding evolutionary relationships among species and within species \cite{nei1996phylogenetic, bellgard1999dynamic}. Genetic sequences can be regarded as sequences of letters, that is, four letters for nucleotides and 20 letters for amino acids. Comparison of sequences can be challenging due to nonuniform lengths, mutations, and sequencing and/or assembly errors. In particular, in full-genome comparisons, different species often have dramatically varying lengths, making it challenging to define genetic metrics between sequences. Therefore, having a robust and reliable method to extract meaningful features that encode complex patterns and capture the shape of the genome space is essential and challenging.
	
	Traditional phylogenetic analysis relies on `sequence alignment' using multiple sequence alignment (MSA) tools such as Clustal Omega \cite{sievers2011fast}, MAFFT \cite{katoh2013mafft}, and MUSCLE \cite{edgar2004muscle}. Alignment-based methods are effective for identifying mutations in sequences \cite{hozumi2021umap, chen2022omicron, chen2022persistent, bleher2021topological, patino2021recombination}. However, these methods may fail when conserved segments are not properly aligned or have different lengths, a common challenge with real-world data. Additionally, alignment-based methods are time-intensive and memory-demanding, making large-scale sequence comparisons difficult.  
	Alternatively, `alignment-free methods' transform variable-length sequences into uniform objects, such as vectors, for sequence analysis \cite{vinga2014alignment}. This approach enables the effective comparison of sequences regardless of sequence lengths \cite{zielezinski2017alignment}. Moreover, because alignment-free methods extract features from individual sequences, their computational complexity scales only with the sequence length and the number of sequences \cite{bonham2014alignment}. This makes whole-genome comparisons across species more efficient and easily parallelizable \cite{bernard2016alignment, zielezinski2019benchmarking, jun2010whole, sims2009whole}.

	Numerous alignment-free methods have been proposed, primarily categorized as word frequency- and information theory-based methods. The former includes $k$-mer methods \cite{blaisdell1986measure}, where the frequency of words or motifs is counted and concatenated into a vector.  Subsequently, the similarity or difference between sequences can be computed using a vector-based distance metric. Several extensions have been proposed to enhance efficiency and distance calculation \cite{wu1997measure, wu2001statistical, wu2005optimal, korf2009applying, jun2010whole}. In information theory-based methods, the distance between sequences is estimated by the amount of shared information between the two sequences using entropy and/ or complexity measures such as Kolmogorov complexity \cite{li2008introduction}, Lempel-Ziv complexity \cite{otu2003new}, Shannon entropy \cite{tribus1971energy}, etc. Comprehensive overviews of these methods can be found in \cite{zielezinski2017alignment, vinga2003alignment, bonham2014alignment}. However, these methods do not incorporate positional information from nucleotides.
	
	Several advanced alignment-free methods have also been proposed. Among them, the natural vector method (NVM) analyzes the moments of the $k$-mer positions \cite{yu2013real, deng2011novel}. The chaos game representation (CGR) \cite{jeffrey1990chaos} represents the sequence as an iterative function, which can be visualized as an image \cite{randic2013milestones}. Various generations of CGR have been proposed \cite{burma1992genome,almeida2001analysis,deschavanne1999genomic, hao2000fractals}. In addition, other methods such as the discrete Fourier power spectrum method\cite{hoang2015new, yin2014measure}, fuzzy integral similarity method \cite{saw2019alignment}, and others \cite{yu2010novel}  have been developed for phylogenetic analysis.  
	
	Persistent homology is an algebraic topology tool for analyzing the shape of complex data\cite{lum2013extracting}. Chan et al. \cite{chan2013topology} utilized persistent homology to compute sequence dissimilarity 
	in viruses. 
	The 0th-order homology and the 1st-order homology corresponded to vertical and horizontal evolution, respectively. This approach was used in analyzing the evolution of SARS-CoV-2 through the topological recurrence index (tRI)  
	\cite{bleher2021topological}. Nguyen et al. \cite{nguyen2022topological}  applied persistent homology to the CGR of viral sequences. 	 
	However, persistent homology neglects the intrinsic biological characteristics of genetic sequences.  
	To overcome the limitation of persistent homology,  persistent topological Laplacians (PTLs), including persistent combinatorial Laplacian or persistent spectral graph for point cloud data\cite{wang2020persistent} and persistent Hodge Laplacian for data on manifolds \cite{chen2021evolutionary}  were introduced.    By tracking the changes in the harmonic and non-harmonic spectra of a series of PTLs induced by filtration, one can uncover the topological and geometrical shape of the data. Specifically, the multiplicity of the harmonic spectra, i.e., the zero-eigenvalues, gives the Betti number of persistent homology at appropriate topological dimensions, while the non-harmonic spectra reveal the homotopic shape evolution.  
	
	In this work, we introduce the $k$-mer topology, a novel topological approach for sequence analysis. $K$-mer topology extracts the topological persistence of $k$-mer segments in a genome through persistent homology and/ or persistent Laplacian, the latter further captures the evolution of the homotopic shape of $k$-mers during filtration. We also define a novel topological genetic distance to define phylogenetic trees and position individual species, organisms, or genes in the genome space. 
	To benchmark the proposed method, we consider both classification tasks and phylogenetic analysis. Our method significantly outperforms other competing methods in viral classification and whole-genome phylogenetic analysis. The proposed topological genetic distance not only sheds light on genome evolution, but also provides a reliable antigenetic distance for the rational design of viral vaccines.  
	
	\section{Results}
	\subsection{Overview of \textit{K}-mer topology}
	
	\begin{figure}[H]
		\centering
		\includegraphics[width = \textwidth]{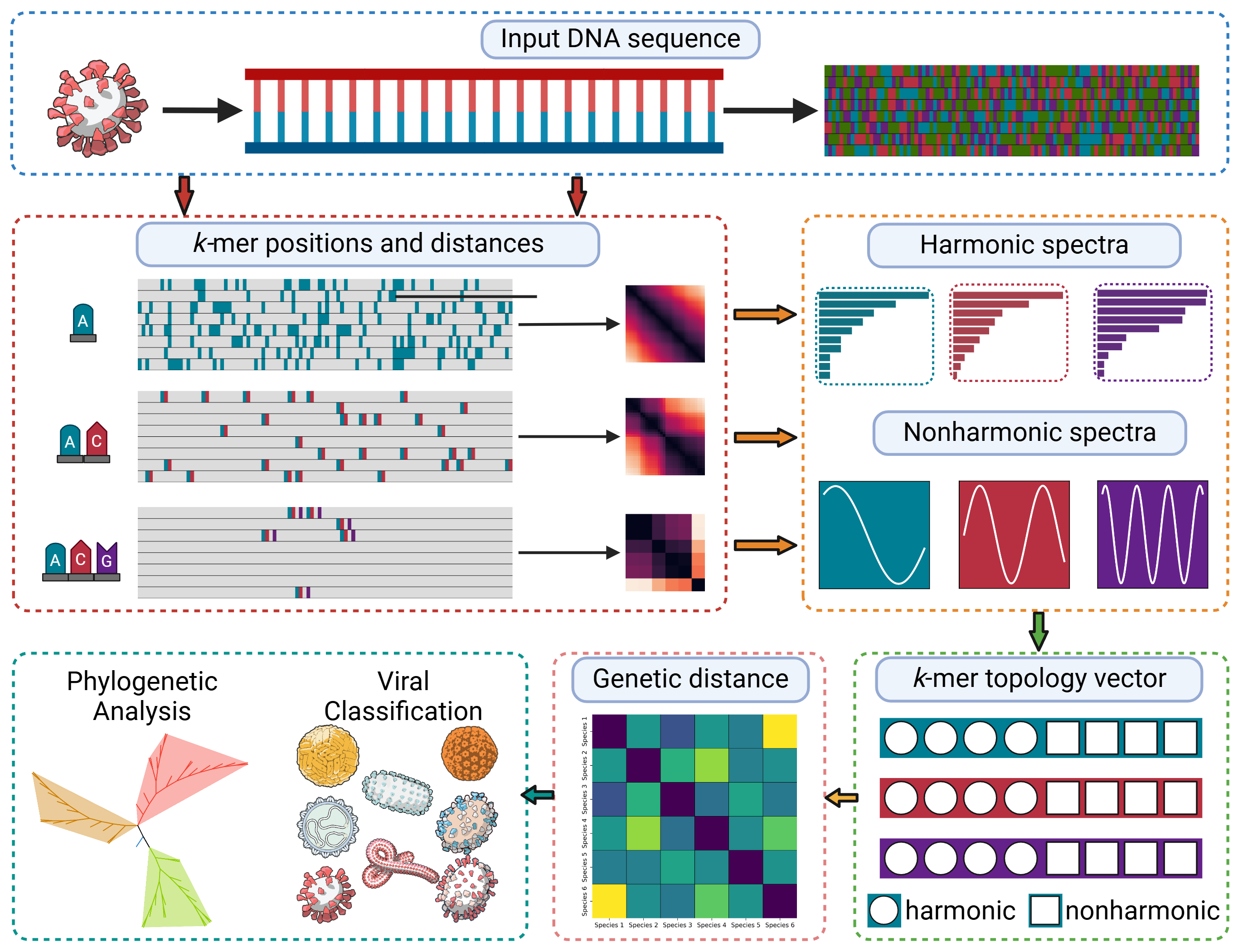}
		\caption{Workflow of $k$-mer topology. From a query sequence, k-mers are extracted. The positions of individual k-mers are used to compute the k-mer specific distances. Using these distances,  persistent Laplacian computes the persistent spectra of $k$-mers
			The harmonic persistent spectrum of the persistent Laplacian provides the persistent Betti numbers, which represent the number of connected $k$-mer components across multiple scales, while the non-harmonic persistent spectrum yields the homotopic shape evolution of the $k$-mers over scales. These features are converted into a  genome topological vector. Topological genetic distances are defined between genome topological vectors for genome classification and/or phylogenetic analysis.}
		\label{fig:flowchart}
	\end{figure}
	The workflow of the $k$-mer topology is depicted in \autoref{fig:flowchart}\footnote[1]{\label{note1}Illustration from NIAID NIH BIOART Source \#64, 144, 155, 156, 187, 391, 464, 545.  Accessed by \url{https://bioart.niaid.nih.gov/bioart/***}, where *** is the \#.}. Starting with a given sequence, the positions of the $k$-mers are extracted. For each set of $k$-mers, the pairwise distances between its positions are calculated, resulting in $k$-mer-specific distances. Using these $k$-mer-specific distances, a family of persistent Laplacian spectra is computed according to filtration. The kernels of persistent Laplacians give rise to harmonic persistent spectra which contain the same topological invariants as persistent homology does. The non-harmonic persistent spectra, namely the non-zero eigenvalues, capture additional shape information of the $k$-mers across multiple scales. These features are used to construct a $k$-mer-specific topological vector for each sequence. Topological genetic distances between the  sequence vectors are defined and used for genome classification and phylogenetic analysis. More details of the proposed $k$-mer topology are provided in \autoref{sec:method}.

	\subsection{Genome classification}\label{sec:classification}
	
	We first validate $k$-mer topology on a viral classification problem using reference sequences collected from the NCBI Virus Database. The viral families served as classification labels. Additionally, the labels in the NCBI Virus Database evolve because the International Committee on Taxonomy of Viruses (ICTV) continuously updates them based on new research findings, indicating the challenge of the problem. 
	A summary of recent changes is given in Table \ref{tab:supporting_classification_data}. Therefore, it is crucial for viral classification methods to remain robust to these updates.
	
	For the classification task, we followed the methodology described by Sun et al. \cite{sun2021geometric}. A 1-nearest neighbor (1-NN) classification was performed on the distance matrix generated from the $k$-mer topology. Specifically, for a given sequence, if the closest sequence according to our metric belongs to the same viral family, the classification is considered correct. This approach simulates real-world applications, where for a query sequence, the most similar sequence is identified and can be further analyzed in detail.
	
	Additionally, we performed 5-nearest neighbor (5-NN) classification on the same dataset to verify that our method can also detect similar sequences. Since some families contained a small number of samples, we excluded all families with fewer than 15 sequences. We then applied 5-fold cross-validation with 30 random seeds to obtain the 5-NN classification results.
	
	\begin{figure}[H]
		\centering
		\includegraphics[width = \textwidth]{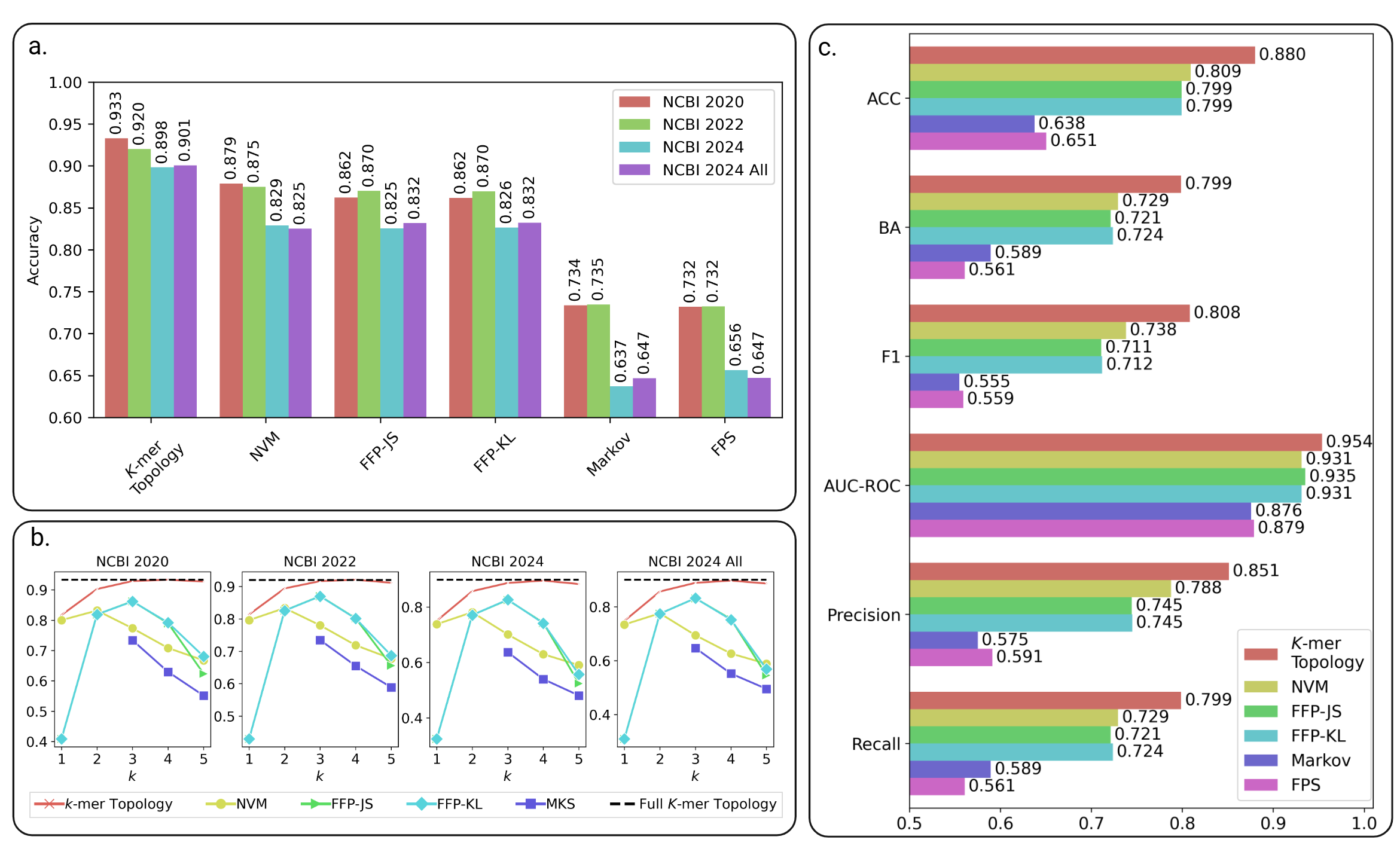}
		\caption{Benchmark of $k$-mer topology on NCBI viral reference sequences from 2020, 2022, and 2024. The NCBI 2024 All dataset includes sequences with invalid nucleotides as well as sequences assigned to unranked families.
			(a) Accuracy comparison of our method against the Natural Vector Method (NVM) \cite{sun2021geometric}, Jensen-Shannon divergence (FFP-JS) \cite{jun2010whole, sims2009alignment}, Kullback–Leibler divergence (FFP-KL) \cite{vinga2003alignment},  Markov K-String (Markov)\cite{qi2004whole}, and Fourier Power Spectrum (FPS) methods\cite{hoang2015new}. 
			(b) Classification accuracy for each $k$ from $k = 1$ to $k = 5$. The black dotted line indicates the accuracy of the weighted sum of $k$-mer topologies. The Markov K-String method starts from $k = 3$ because it requires $k-1$-mers and $k-2$-mers to compute features for a given $k$.
			(c) 5-NN classification performance on the NCBI 2024 dataset, including all sequences. Stratified 5-fold cross-validation was performed using 30 random seeds. The bars represent the average scores for each method. Metrics including accuracy (ACC), balanced accuracy (BA), macro-F1 (F1), area under the receiver operating characteristic curve (AUC-ROC), recall, and precision were computed for each method.}
		\label{fig:classification_accuracy}
	\end{figure}
	
	\autoref{fig:classification_accuracy} compares $k$-mer topology with the Natural Vector Method (NVM) \cite{sun2021geometric}, Jensen-Shannon Divergence (FFP-JS) \cite{jun2010whole, sims2009alignment}, Kullback-Leibler Divergence (FFP-KL) \cite{vinga2003alignment}, Markov K-String (Markov) \cite{qi2004whole}, and Fourier Power Spectrum (FPS) \cite{hoang2015new}.
	FFP-JS and FFP-KL are information theory-based methods that compute sequence dissimilarity by analyzing feature frequency profiles (FFPs), which are normalized $k$-mer counts. The Markov K-String method predicts the expected frequency of $k$-mers using a Markov model and compares it with the observed $k$-mer frequencies. FPS converts sequences into signals, applies a Fourier transform, and uses statistics derived from the Fourier coefficients as features.
	
	The upper row of \autoref{fig:classification_accuracy} shows the classification accuracy for each dataset using the parameters specified in the original work. For $k$-mer topology, $k=5$ was used, while for FFP-JS and FFP-KL, $k=3$ was selected. The bottom row displays the classification accuracy as a function of varying $k$-mer sizes, with the dashed black line representing the accuracy achieved by the weighted sum of different $k$-mer sizes.
	
	\autoref{fig:classification_accuracy} (a) presents the 1-NN classification accuracy in all four datasets. Our method consistently outperforms all other methods in all datasets. In particular, the accuracy of the $k$-mer topology in the NCBI 2024 All and NCBI 2024 datasets is comparable. Although NCBI 2024 excludes sequences with invalid nucleotides and unranked families, NCBI 2024 All includes sequences with these anomalies, highlighting the robustness of our method in handling real-world scenarios where sequence errors are common.
	
	A general decrease in accuracy from 2020 to 2024 is observed in all methods. This decline can be attributed to changes in viral classification, such as the abolition of large families like Reoviridae in 2020, resulting in its members being redistributed into smaller families. As a result, the NCBI 2024 dataset poses greater classification challenges compared to the NCBI 2020 dataset.
	
	The superior performance of the $k$-mer topology in the NCBI 2024 dataset demonstrates its ability not only to accurately classify viral families but also to effectively distinguish between closely related sequences within each family.
	
	Furthermore, we evaluated the impact of varying the $k$ values, as shown in \autoref{fig:classification_accuracy} (b). In individual k-mer analysis, the $k$-mer topology consistently outperformed all other methods by a significant margin. Interestingly, our method exhibited improved accuracy as $k$ increased, while other methods showed a decrease in accuracy after $k=3$. This trend aligns with the biological encoding of amino acids through nucleotide triplets (codons), suggesting that $k=3$ may provide the most informative features for viral classification.
	
	\autoref{fig:classification_accuracy} (c) presents the 5-NN classification results on the NCBI 2020 All dataset, with comparisons for the other three datasets detailed in Section \ref{sec:supporting_classification} of the supplementary materials. Our method achieved superior performance across all metrics, including accuracy (ACC), balanced accuracy (BA), macro F1 score (F1), macro area under the receiver operating characteristic curve (AUC-ROC), recall, and precision for all four datasets. Macro scores were used to ensure equal contribution from all viral families. For AUC-ROC, we used the one-versus-rest approach, where each family was compared against all the others. The outstanding performance of $k$-mer topology in both 1-NN and 5-NN classification tasks demonstrates its robustness in identifying sequences similar to the queried sequence.
	
	\subsection{Phylogenetic analysis }
	In order to further validate our method, we consider phylogenetic analysis of 7 benchmark problems.  
	Figures \ref{fig:supporting_hrv_ebola} and \ref{fig:supporting_influenza_mammalian_bacteria_hev} summarize the phylogenetic analysis of the $k$-mer topology. Further details of the comparative analysis can be found in the subsequent sections and in section \ref{sec:supporting_phylo} of the supporting materials. Essentially, the $k$-mer topology provides a completely correct phylogenetic analysis of all problems.
	Comparisons with state-of-the-art methods, such as NVM \cite{sun2021geometric},  FFP-JS  \cite{jun2010whole, sims2009alignment}, FFP-KL \cite{vinga2003alignment}, Markov \cite{wu2001statistical}, and FPS  \cite{hoang2015new}, are described below.

	\subsubsection{Mammalian mitochondrial genomes}
	
\begin{figure}[H]
		\centering
		\includegraphics[width = \textwidth]{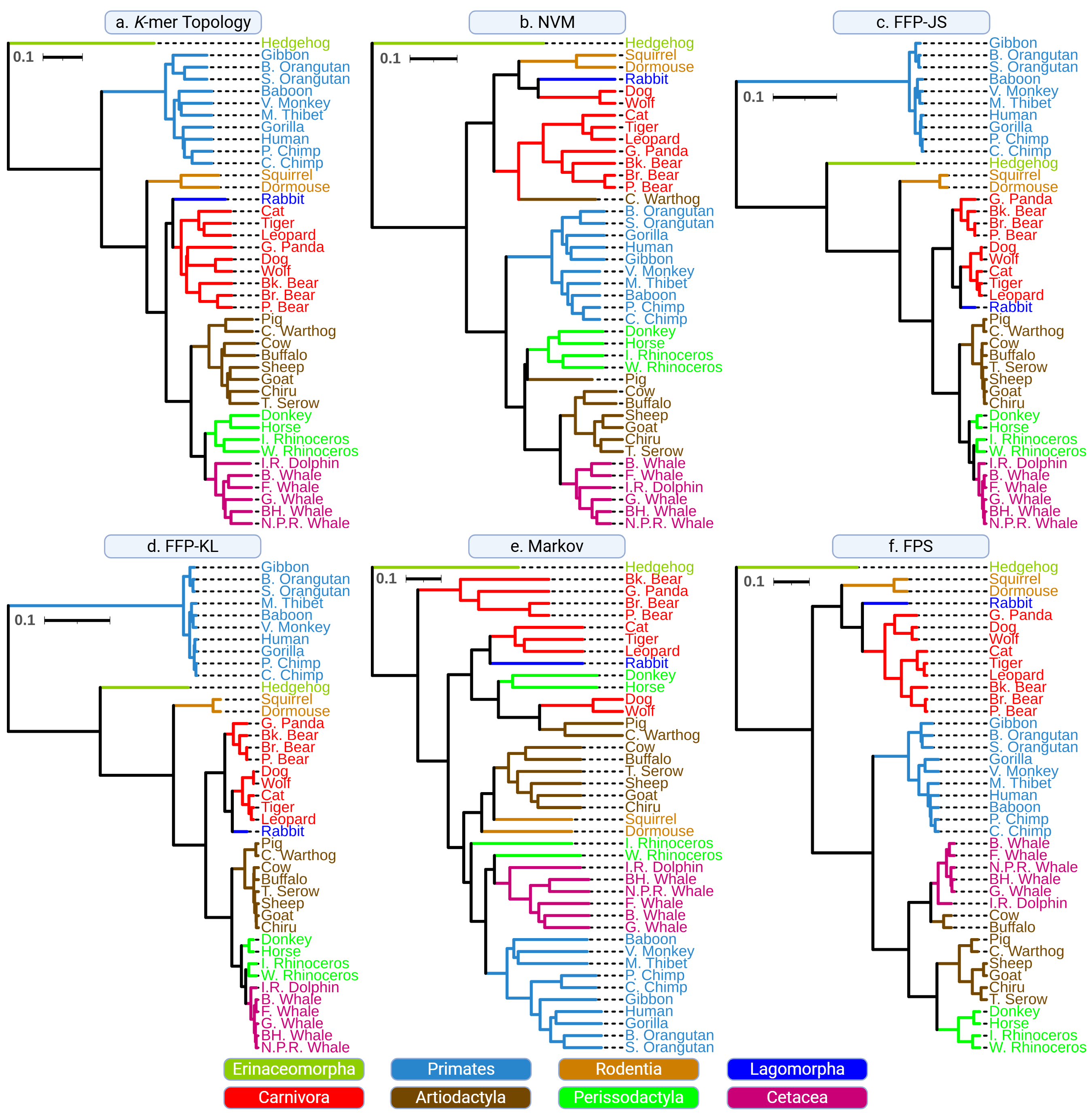}
		\caption{Phylogenetic tree analysis of 42 complete mammalian mitochondrial genomes. The mitochondrial genomes are categorized according to their host species classifications, including Artiodactyla, Carnivora, Cetacea, Erinaceomorpha, Lagomorpha, Perissodactyla, Primates, and Rodentia. The branches and labels are colored based on the host species classification.
			(a) $K$-mer topology with $K = 5$: All mitochondrial genomes are correctly classified into clades.
			(b) NVM using $k = 5$: The warthog and pig mitochondrial genomes are not clustered with the rest of the Artiodactyla mitochondrial genomes, and two separate clades of Carnivora mitochondrial genomes exist.
			(c) FFP-JS using $k = 3$: Two separate clades of Carnivora exist. Additionally, both rhinoceros mitochondrial genomes form a clade that is separated from the donkey and horse mitochondrial genomes.
			(d) FFP-KL using $k = 3$: Similar to FFP-JS, two clades of Carnivora and Perissodactyla mitochondrial genomes exist.
			(e) Markov K-String using $k = 3$: Three clades of Carnivora and two clades of Artiodactyla exist. Both Rodentia and Perissodactyla mitochondrial genomes do not form clades.
			(f) Fourier power spectrum method: Artiodactyla mitochondrial genomes form two clades.}
		\label{fig:mammalian_miochondria}
	\end{figure}

	\autoref{fig:mammalian_miochondria} shows the comparison of six methods for the phylogenetic tree analysis of 42 complete mitochondrial genomes. The mitochondrial genomes are classified according to the host species' classification, which includes Artiodactyla, Carnivora, Cetacea, Erinaceomorpha, Lagomorpha, Perissodactyla, Primates, and Rodentia. Our $k$-mer topology successfully clusters all genomes into their respective clades. NVM, FPS-JS, FPS-KL, and Markov all separate Carnivora mitochondrial genomes into multiple clades. NVM, Markov, and FPS have misclassified Artiodactyla mitochondrial genomes.
	
	\subsubsection{Rhinovirus}
	\begin{figure}[H]
		\centering
		\includegraphics[width = \textwidth]{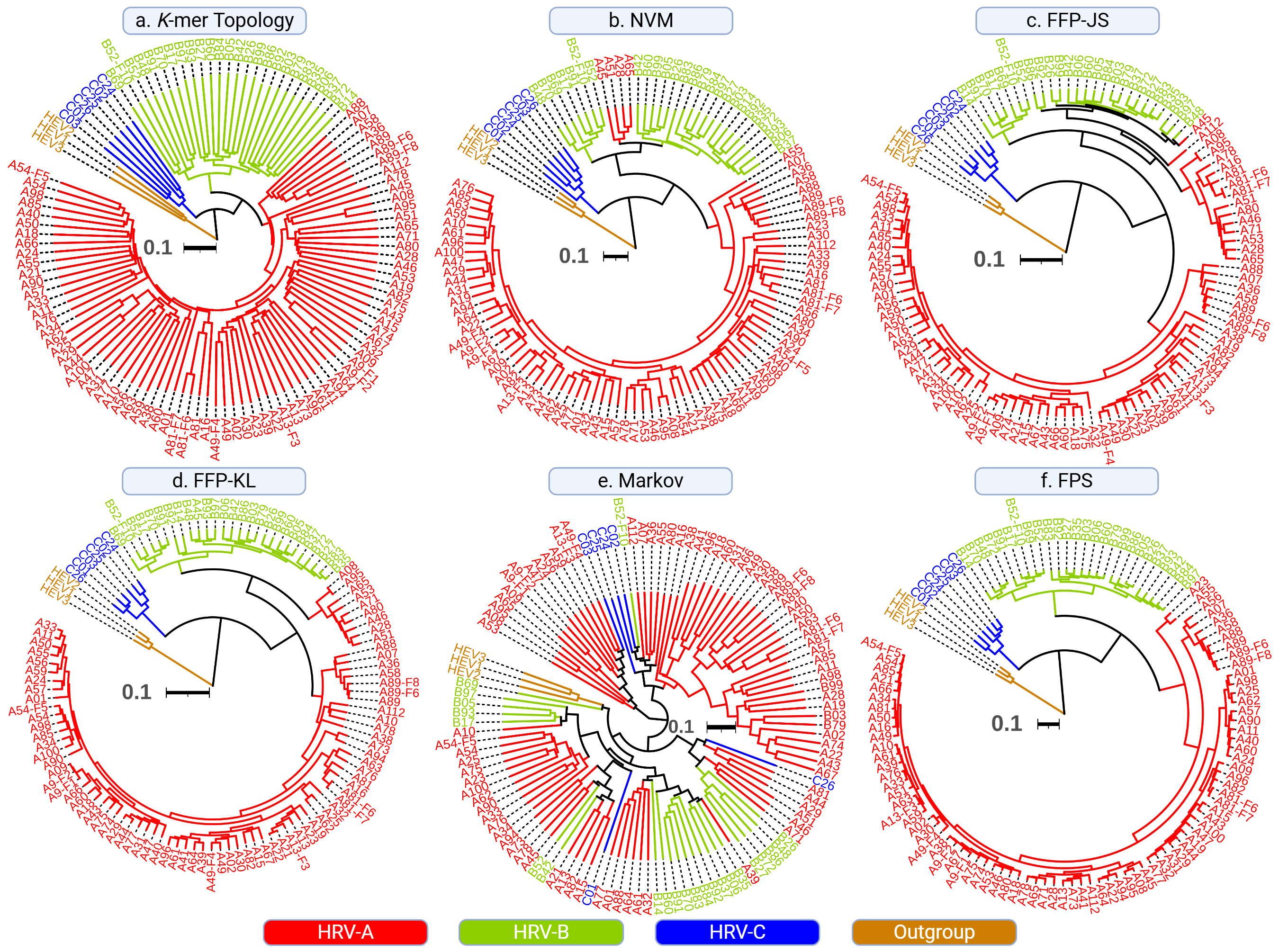}
		\caption{Phylogenetic tree analysis of 113 complete rhinovirus (HRV) genomes and 3 outgroup genomes. The HRV genomes are categorized into groups A, B, and C. The branches and labels are colored based on the HRV group, and the outgroup genomes are colored yellow.
			(a) $K$-mer topology using $K = 5$: All HRV genomes are correctly classified into clades, and the outgroup sequences form a separate clade from the HRV genomes.
			(b) NVM using $k = 5$: HRV genomes form a separate clade from the outgroup genomes. One HRV-A genome is placed within the HRV-B clade.
			(c) FFP-JS using $k = 3$: HRV genomes form a separate clade from the outgroup genomes. One HRV-A genome is placed within the HRV-B clade, and another HRV-A genome is within the HRV-B clade.
			(d) FFP-KL using $k = 3$: HRV genomes form a separate clade from the outgroup genomes. One HRV-A genome is placed within the HRV-B clade.
			(e) Markov K-String using $k = 3$: The outgroup genomes form a clade, but they do not form a separate clade from the HRV sequences. Although some groups form clusters, no uniform clade exists for each HRV group.
			(f) Fourier power spectrum method: All HRV genomes are correctly classified into clades, and the outgroup sequences form a separate clade from the HRV genomes.}
		\label{fig:rhinovirus}
	\end{figure}
	
	\autoref{fig:rhinovirus} shows the comparison of six methods for the phylogenetic tree analysis of 113 complete genomes of  rhinovirus (HRV) and 3 outgroup genomes. The HRV genomes are classified into 3 groups: A, B, and C. The $k$-mer topology and FPS successfully cluster all genomes into their respective clades. NVM, FPS-JS, and FPS-KL show similar results, where a small HRV-A clade is formed within the HRV-B clade. Markov does not form a meaningful clade for any of the HRV groups.

	\subsubsection{SARS-CoV-2 variants}
\begin{figure}[H]
		\centering
		\includegraphics[width = \textwidth]{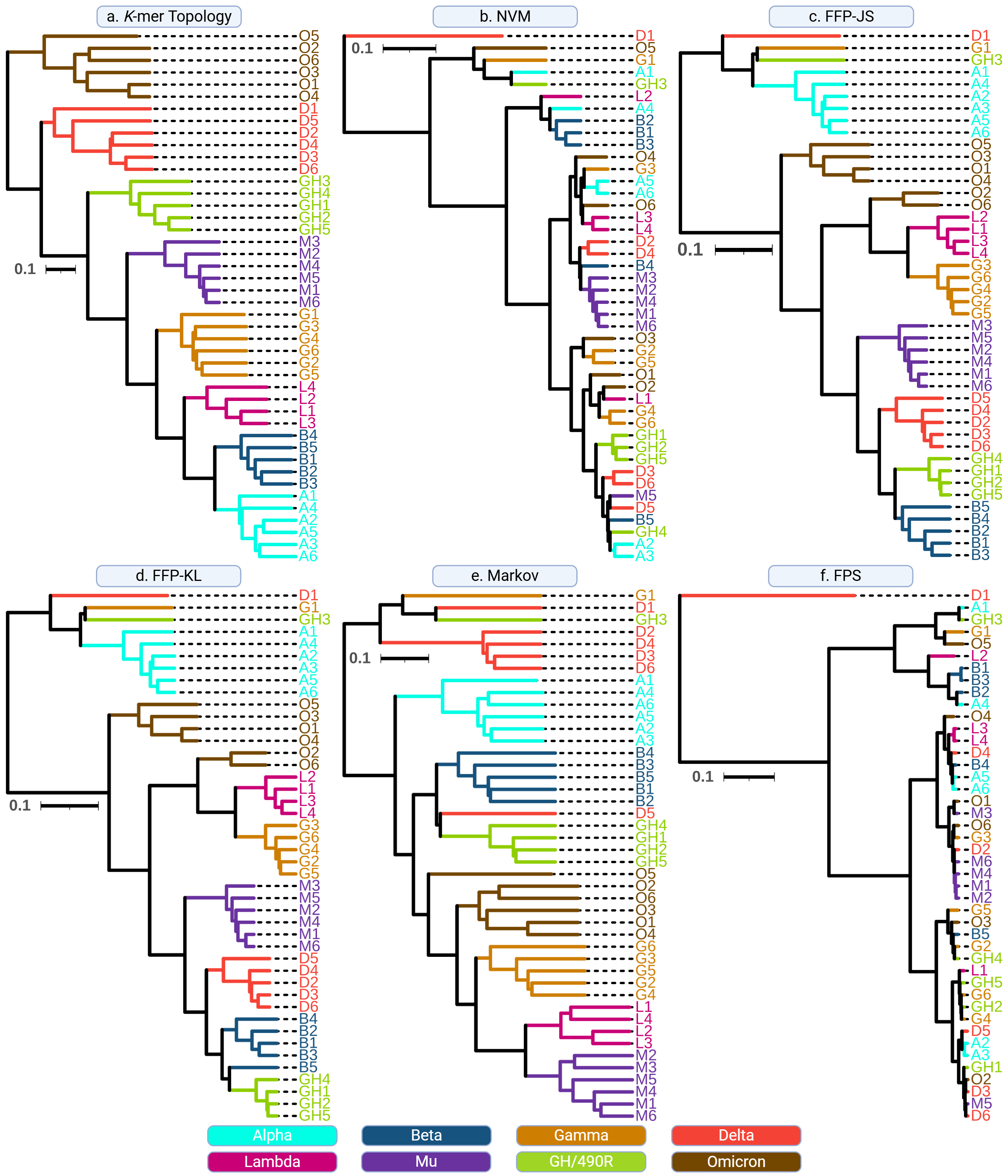}
		\caption{Phylogenetic tree analysis of 44 complete severe acute respiratory syndrome coronavirus 2 (SARS-CoV-2) genomes. The viral genomes are categorized according to their variants, including Alpha, Beta, Gamma, Delta, Lambda, Mu, GH/490R, and Omicron. Branches and labels are colored based on variant types.
			(a) $K$-mer topology using $k = 4$: All sequences are corrected clustered. 
			(b) NVM using $k = 5$: Only the Mu and GH/490R variants show signs of clustering.
			(c) FFP-JS using $k = 3$: Three sequences—D1, G1, and GH3—are misclassified. Omicron sequences form two clades.
			(d) FFP-KL using $k = 3$: Three sequences—D1, G1, and GH3—are misclassified. Omicron sequences form two clades.
			(e) Markov K-String using $k = 3$: Five sequences—G1, D1, GH3, D5, and O5—are misclassified. All other sequences are clustered within their respective clades.
			(f) Fourier power spectrum method: No meaningful clusters are formed.}
		\label{fig:sarscov2_variants}
	\end{figure}

	\autoref{fig:sarscov2_variants} shows the comparison of six methods for the phylogenetic tree analysis of 44 complete Severe Acute Respiratory Syndrome Coronavirus 2 (SARS-CoV-2) genomes. The sequences were obtained from GISAID and are labeled according to their variants, including Alpha, Beta, Gamma, Delta, Lambda, Mu, GH/490R, and Omicron. The branches and labels are colored according to their variants.
	
	Analyzing variants within a species is often difficult for alignment-free methods due to the key difference between the variants being mutations in just a few nucleotides. However, these key mutations have been shown to increase the infectivity of SARS-CoV-2 \cite{chen2022persistent}. Interestingly, all methods, except for our $k$-mer topology, fail to fully separate the variants into their respective clades. FFP-KL and FFP-JS have 4 misclassified sequences and 2 clades of Omicron variants. Markov has 5 misclassified sequences. NVM and FPS fail to show meaningful clustering of most variants. In fact, persistent homology-based $k$-mer topology has 4 misclassified sequences as shown in \autoref{fig:supporting_PhyloLaplacian}. This motivates the use of persistent Laplacian to enhance the clustering result, and more details can be found in \autoref{sec:PL_enhancement}.
	
	\subsubsection{Hepatitis E virus}
	Figure \ref{fig:supporting_HEV_benchmark} shows the comparison of six methods for the analysis of phylogenetic trees of 48 complete genomes of the hepatitis E virus (HEV). HEV viruses are categorized into  4  groups, and  the branches and labels are colored according to their groups. $K$-mer topoology, FFP-JS, and FFP-KL successfully cluster all sequences into their respective clade. NVM has 2 separate clades of group 3. Markov has 1 misclassified group 3 sequence (G3-17). FPS is unable to distinguish groups 3 and 4 HEV sequences.
	
	\subsubsection{Influenza A hemagglutinin genes}
	Figure \ref{fig:supporting_influenzaHAgene_benchmark} compares six methods for the phylogenetic tree analysis of 30 Influenza A hemagglutinin (HA) genes. The genes are categorized into six groups: H1N1, H2N2, H3N2, H5N1, H7N9, and H7N3. The branches and labels are color-coded according to the influenza A classifications.
	
	$K$-mer topology, FFP-JS, and FFP-KL successfully cluster the genes into their respective clades. A notable difference is that, in FFP-JS and FFP-KL analyses, H3N3 and H2N2 share a node, while, in $k$-mer topology analysis, H3N3 and H2N2 are placed in different clades. NVM fails to group all H1N1 HA genes into a single clade and misclassifies one H2N2 sequence. Similarly, the Markov method misclassifies one H1N1 sequence. Lastly, FPS does not demonstrate clear clustering for most groups.
	
	\subsubsection{Ebola virus}
	Figure \ref{fig:supporting_ebolavirus_benchmark} compares six methods for the phylogenetic tree analysis of 59 ebola virus genomes. Ebola viruses are categorized into five types: Bundibugyo virus (BDBV), Reston virus (RESTV), Ebola virus (EBOV), Sudan virus (SUDV), and Tai Forest virus (TAFV). In addition, EBOV is further categorized according to the epidemic location and year. The branches are color-coded according to types of ebola viruses, and the labels are highlighted to represent specific epidemics.
	
	All six methods successfully separate the types of ebola viruses into their respective clades. Furthermore, $k$-mer topology, NVM, FFP-JS, FFP-KL, and Markov separate the EBOV epidemics into individual clades. However, FFP-JS and FFP-KL show minimal differences between the various EBOV epidemics, as indicated by the short branch lengths. In contrast, $k$-mer topology and NVM display much longer branch lengths between EBOV epidemics, suggesting that these two methods better capture the finer differences among the epidemics. Markov produces a slightly different tree structure, where EBOV and RESTV share a node. FPS, while capable of separating the different types of ebola viruses, does not fully distinguish between EBOV epidemics.
	
	\subsubsection{Bacteria}
	To test the computational limits of $k$-mer topology, we performed a phylogenetic analysis on whole bacterial genomes. Figure \ref{fig:supporting_bacteria_benchmark} shows the comparison of the phylogenetic tree of 30 bacterial genomes. The genomes are categorized into nine families: \textit{Bacillaceae, Borreliaceae, Burkholderiaceae, Clostridiaceae, Desulfovibrionaceae, Enterobacteriaceae, Rhodobacteraceae, Staphylococcaceae,} and \textit{Yersiniaceae}.
	
	Due to sequence lengths exceeding 1 million nucleotides, we utilized $k = 3, 4,$ and $5$ for $k$-mer topology. All six methods successfully separated the bacteria into their appropriate clades.

	\section{Discussion}
	
	We propose $k$-mer topology as a novel phylogenetic analysis method that utilizes persistent homology and/or persistent Laplacian for $k$-mer topological studies. Our method significantly outperforms other methods in viral classification tasks on different versions of viral data, indicating the robustness to changes in taxonomy. We have also validated our method on standard phylogenetic analysis, including the whole bacteria sequence, indicating the robustness, reliability, and scalability of our method. 
	
	\subsection{The shape of genome space from $k$-mer topology} 		
	It is interesting to understand why the proposed $k$-mer topology works so well for genome sequence analysis and prediction. In this approach, multiscale topological tools that can delineate the shape of the data \cite{lum2013extracting} are used to characterize the shape of genome space. Topology is one of the most abstract subjects in mathematics that offers insights that cannot be obtained from any other mathematical, physical, and statistical approaches. Additionally, its abstraction dramatically simplifies data complexity. Moreover,  persistent topology captures the multiscale relationship or intermittence patterns of $k$-mers in a genome sequence. Finally, the whole sequence topology does not contain enough information to depict the multifacet shape of a genome.  The success of the $k$-mer topology is crucially attributed to the $k$-mer specific persistent topology. 
	
	\subsection{Genome-specific $k$-mer distributions} 	
	We analyzed the $k$-mer distributions of 8 virus families: Adenoviridae, Circoviridae, Geminiviridae, Sedoreoviridae, Autographiviridae, Fiersviridae, Phenuiviridae, and Spinaeroviridae. To compare the distributions, we counted the average number of $k$-mers within a given radius of a nucleotide. Specifically, for each $k$-mer at $l_p$, we identified $k$-mer segments that were also within the radius $r$ of $l_p$, and tracked the changes in the number of neighbors as the radius increased. We then normalized the counts to obtain the average number of $k$-mers in a given radius. Figure \ref{fig:supporting_kmer_distribution} visualizes the 16 2-mers of these viruses. Each plot corresponds to one of the 16 possible 2-mers, with color indicating the virus family. The $x$-axis represents the radius, while the $y$-axis shows the average 2-mer count.	
	In particular, the distributions differ significantly between families. For example, Autographiviridae has the largest average AC count across all radii, but the lowest average TT count. These genome-specific distributions are characterized by the proposed $k$-mer topology. 		
	
	\subsection{Persistent Laplacian enhancement}\label{sec:PL_enhancement}
	While persistent homology provides detailed distributions of $k$-mers through persistent Betti numbers, it does not capture the complex geometric and topological information related to the evolution of homotopic shapes induced by filtration. To address this limitation, we incorporate persistent Laplacian \cite{wang2020persistent} to obtain topological Laplacian spectra at each stage of filtration. The harmonic spectra (zero eigenvalues) return the persistent Betti numbers from persistent homology, while the non-harmonic spectra (non-zero eigenvalues) capture the evolution of the homotopic shapes of the genome data. This additional information is utilized to improve our $k$-mer topology analysis of genome sequences. 
	A concise explanation of persistent Laplacian can be found in \autoref{sec:method}. 
	Figure \ref{fig:supporting_PhyloLaplacian} compares the characters derived from persistent homology (a) and persistent Laplacian (b) in the entire SARS-CoV-2 genome. SARS-CoV-2 variants were collected from GISAID (\url{https://gisaid.org/}), including Alpha, Beta, Gamma, Lambda, Mu, GH/490R, and Omicron variants. The clades and labels are color-coded according to the variants, with accession IDs for each label available in the supporting materials. In panel (a), Betti numbers for $k = 1, 2, 3,$ and 4 were calculated, and the genetic distance was calculated using \autoref{eq:genetic_distance} with $b_k = 0$ and $a_k = \frac{1}{\omega_k2^{4-k}}$, where $\omega_k$ is the average $k$-mer distance between species. In panel (b), we include the spectra (the smallest nonzero eigenvalue) for each filtration value and used a multiscale distance on the Betti features to obtain a more detailed genetic distance. Details of the distance calculation can be found in Section \ref{sec:supporing_genetic_distance} of the supporting materials. It is clear that the multiscale distance and the spectra feature significantly improve the clustering performance, enabling the $k$-mer topology to handle sequences with high similarities. However, it is important to note that calculating the Laplacian spectra is computationally expensive for larger sequences.
	
	\subsection{High-dimensional topological features}
	
	The inclusion of high-dimensional topological features has been examined, as detailed in the Methods section (see \autoref{fig:method}(c)). High-dimensional topological features can certainly enhance the accuracy of the proposed $k$-mer topology. However, most genomes do not naturally possess high-order simplices. In fact, we can artificially create high-order simplices by connecting both ends of a sequence, as demonstrated in the Methods section. This approach generates non-trivial higher-order topological invariants that can further improve the accuracy of $k$-mer topology. However, the computation of higher-order simplicial complexes is typically time-consuming, and a balance must be struck between accuracy and computational efficiency.
	
	\subsection{Topological antigenetic distance}
	Antigenetic distance serves as a basis for the rational design of viral vaccines \cite{hu2023rational}. The superior performance of  the $k$-mer topology in the phylogenetic tree analysis of various viruses suggests that the proposed topological genetic distance can be directly used as a reliable antigenetic distance. Our topological antigenetic distance can be applied to the rational design of viral vaccines to improve vaccine effectiveness against emerging viral variants.   
	
	\subsection{Generalizations and applications}
	The proposed $k$-mer topology utilized only persistent homology and persistent Laplacian.  
	It is important to explore further improvements of genome sequence analysis with other topological objects, such as path complexes, sheaf complexes, digraphs, hypergraphs, etc., and other topological formulations, such as quantum topology via Dirac, interaction topology, and Mayer topology. These improvements will be studied in future work. 
	
	Only the first non-harmonic eigenvalue of the persistent Laplacian was used in the present work. The inclusion of more non-harmonic eigenvalues may improve the present approach. This improvement is needed for the analysis of sequences with high similarity, such as SARS-CoV-2 variants.    
	
	Additionally, the computation of the proposed $k$-mer topology can be accelerated with parallel and GPU architectures because the calculations of each specific $k$-mer topology in a given sequence can be carried out independently of other $k$-mers and other sequences. Moreover, the computation of topological invariants can be improved. Fast algorithms for solving Laplacians can be implemented because only the first few eigenvalues are used. These accelerations are crucial for the whole-genome analysis of advanced biological species.
	
	The proposed $k$-mer topology has potential applications in protein sequence alignment-free analysis, coding region identification, enhancer classification, etc.

	\section{Methods}\label{sec:method}
	In this section, we provide an overview of persistent homology of persistent topological Laplacian. Then, we provide the construction of the $k$-mer topology.

	\subsection{Persistent homology}
	Persistent homology is a new tool to analyze the shape of data \cite{carlsson2009topology,lum2013extracting}. It utilizes concepts from algebraic topology, such as independent components, holes, and voids,  to extract topological invariants from the data by representing point cloud data as simplicial complex, constructed from vertices, edges, triangles, and/or higher-order simplexes\cite{edelsbrunner2008persistent}. 
	Filtration is applied to capture the persistence of these topological invariants across scales. Persistent homology was used in topological deep learning (TDL), coined in 2017 \cite{cang2017topologynet}, in predictions of  protein-ligand binding affinities and changes in protein stability upon mutation.  TDL outperforms other methods in the D3R Grand Challenges, a worldwide competition series in computer-aided drug design \cite{nguyen2019mathematical, nguyen2020mathdl}.  
	
	\subsubsection{Simplicial complex}
	We begin by providing a brief background on the simplicial complex.
	Let $\sigma_q = [v_0, ..., v_q]$ be a $q$-simplex, where $v_i$ is a vertex, and $\sigma_q$ consists of $q+1$ vertices. For example, $\sigma_0$ is a node, $\sigma_1$ is an edge, $\sigma_2$ is a triangle, $\sigma_3$ is a tetrahedron, and so on. A simplicial complex $K$ is a union of simplicies such that the following 2 properties hold.
	\begin{enumerate}
		\item If $\sigma_q \in K$, and $\sigma_p$ is a face of $\sigma_q$, then $\sigma_p \in K$
		\item The nonempy intersection of any 2 simplicies in $K$ is a face of both simplicies.
	\end{enumerate}
	In essence, one can view $K$ as gluing together low-order simplicies.
	
	\subsubsection{Chain complex and homology groups}
	A $q$-chain is a formal sum of $q$-simplicies in $K$ with coefficients $\mathbb{Z}_2$. The set of all $q$-chains contain the basis for the set of $q$-simplifies in $K$, and this set form a finitely generated Abelian group $C_q(K)$ called the chain group. These chain groups are related by a boundary operator $\partial_q$, which is a group homomorphism $\partial_q:C_q(K) \to C_{q-1}(K)$. This boundary operator is defined as the following
	\begin{align}
		\partial_q\sigma_q = \sum_{i=0}^q(-1)^i\sigma_{q-1}^i
	\end{align}
	where $\sigma_{q-1}^i = [v_0,...,v_i^*, ..., v_q]$ is a $(q-1)$-simplex with the vertex $v_i$ removed from $\sigma_q$. Then, one can define the chain complex as a sequence of chain groups connected by these boundary operators.
	\begin{align}
		...\xrightarrow{\partial_{q+2}} C_{q+1}(K) \xrightarrow{\partial_{q+1}} C_{q}(K)  \xrightarrow{\partial_{q}} ...
	\end{align}
	
	An important property of the boundary operator is that applying the boundary operator twice to any $q$-chain results in a mapping to a zero element. That is, $\partial_{q-1}\partial_{q} = \emptyset$. Additionally, the boundary operator for the $0$-chain maps to 0, i.e., $\partial_0 = \emptyset$.
	
	Using the boundary operators, one can define the $q$th cycle group $Z_q$ and the $q$th boundary group $B_q$. Both $Z_q$ and $B_q$ are subgroups of the $q$th chain group $C_q$, and are defined as 
	\begin{align}
		& Z_q = \text{Ker}\partial_q = \{ c \in C_q | \partial_q c = \emptyset\}\\
		& B_q = \text{Im}\partial_{k+1} = \{c \in C_q | \exists  d \in C_{q+1}: c = \partial _{q+1}d\}
	\end{align}
	Additionally, $\partial_{q-1}\partial_{q} = \emptyset$ implies that $B_q \subseteq Z_q \subseteq C_q$. Moreover, the $q$th cycle is the $q$-dimensional hole.
	
	Using the groups one has defined, one can now define the homology group. The $q$th homology group, denoted $H_q$, is the quotient group generated by $Z_q$ and $B_q$, i.e., $H_q = Z_q / B_q$. The rank of $H_q$ is $\beta_q$, or the $q$th Betti number. When $H_q$ is torsion free, the Betti number can be defined as 
	\begin{align}
		\beta_q = \text{rank}H_q = \text{rank} Z_q - \text{rank} B_q
	\end{align}
	The $q$th Betti number describes the $q$th dimensional hole. For example, $\beta_0$ is the number of connected components, $\beta_1$ is the number of loops, $\beta_2$ is the number of cavities, etc. Furthermore, the Betti numbers describe the topological property of the system.
	
	\subsubsection{Filtration and persistence} 
	One downside of utilizing the simplicial complex is that it does not provide sufficient information to describe the geometry of the data because one can only capture the data in a single scale. To this end, one uses simplicial complex induced by filtration,
	\begin{align}
		\emptyset = K_0 \subseteq ...\subseteq K_p \subseteq ...\subseteq K_P = K
	\end{align}
	where $P$ is the number of filtration. Using such filtration is the foundation of persistent homology, where persistence is observed through long-lasting topological features. For each filtration $p$, one constructs the simplicial complex, the chain group, the subgroup, and the homology group. In particular, the $p$-persistent $q$-th homology group $K^i$ is 
	\begin{align}
		H_q^{i,p} = Z_q^i / \left(B_q^{i+p} \bigcap Z_q^i\right)
	\end{align}
	Computing the Betti numbers gives the persistence of $q$-dimensional holes.
	
	\subsection{Persistent topological Laplacian}
	
	Although persistent homology can capture the persistence of topological invariants, it does not capture the homotopic shape evolution of the data. To this end, persistent spectral graph, also called persistent combinatorial Laplacian or persistent Laplacian, was proposed to obtain the spectra of topological Laplacians induced by filtration \cite{wang2020persistent}. These spectra not only return all the topological invariant, namely persistent Betti numbers, from persistent homology, but also capture the homotopic shape evolution across scales. Persistent Laplacians have stimulated rapid theoretical developments \cite{memoli2022persistent, liu2023algebraic,wang2021hermes, wei2024persistent, wang2023persistent, chen2023persistent} and successful applications \cite{qiu2023persistent,  meng2021persistent, hozumi2024analyzing, cottrell2023plpca}, 
including forecasting of the emerging dominant SARS-CoV-2 variant BA.4 and BA.5 about two months in advance \cite{chen2022persistent}.

\subsubsection{Topological Laplacian and its spectrum}
The topological Laplacian provides insight into the structure of the simplicial complex. 
One can view topological Laplacian as an extension to the traditional graph Laplacian by analyzing higher-order topological structures. Note that standard graph Laplacian analysis can be viewed as analyzing the 1-simplices. For example, the multiplicity of 0-eigenvalue is the number of connected components, the first nonzero eigenvalue is the spectral gap, which is related to the Cheeger constant, and the second smallest non-zero eigenvalue relates to the algebraic connectivity. Furthermore, the collection of the eigenvalues of the graph Laplacian is called the Laplacian spectrum.

Traditionally, the graph Laplacian is computed by observing the adjacency matrix $A$ and the degree matrix $D$ of the graph, and computing $\mathcal{L} = D- A$. If we instead consider the graph as $1$-simplex, we can compute the Laplacian matrix as $\mathcal{L} = B_1B_1^T$, where $B_1$ is the $1$-dimensional boundary operator matrix. Using this idea, we can extend this concept to higher-order simplicies.

In order to define the topological Laplacian, we first define the dual chain complex through the adjoint operator of $\partial_q$ which is defined in the dual spaces $C^q(K)\equiv C_q^*(K)$. The coboundary operator $\partial_q^*:C^{q-1}(K) \to C^q(K)$ is defined as 
\begin{align}
	\partial^*\omega^{q-1}(c_q) \equiv \omega^{q-1}(\partial c_q)
\end{align}
where $\omega^{q-1} \in C^{q-1}(K)$ and $c_q \in C_q(K)$. Here, $\omega^{q-1}$ is the $(q-1)$ cochain, which is a homomorphism mapping of a chain to the coefficient group. The homology of the dual chain complex is called cohomology.

We can now define the $q$-combinatorial Laplcian operator $\triangle_q:C^q(K) \to C^q(K)$ as
\begin{align}
	\triangle_q := \partial_{q+1}\partial{q+1}^* + \partial_q^*\partial_q.
\end{align}
Denoting $\mathcal{B}_q$ as the standard basis of the $q$-boundary operator from $C_q(K)$ and $C_{q-1}(K)$ and $\mathcal{B}_q^T$ as the basis for the $q$-boundary operator, we can obtain the matrix representation of the $q$-th order Laplacian operator $\mathcal{L}_q$, which is defined as 
\begin{align}
	\mathcal{L}_q = \mathcal{B}_{q+1}\mathcal{B}_{q+1}^T + \mathcal{B}_{q}^T \mathcal{B}_q.
\end{align}
The multiplicity of the zero eigenvalues of $\mathcal{L}_q$ is the $q$th Betti number, and the nonzero eigenvalues, or the non-harmonic spectra, contain other topological and geometrical features.

\subsubsection{Filtration and persistence}
Similarly to persistent homology, we can introduce filtration to obtain the persistence of topological invariants and the evolution of homotopic shapes. For each $K_p$ $0\le p \le P$, denote $C_q(K_p)$ as the chain group induced by $K_p$, and the corresponding boundary operator $\partial_q^p:C_q(K_p) \to C_{q-1}(K_p)$ defined as
\begin{align}
	\partial_q^p\sigma_q = \sum_{i=0}^q(-1)^i\sigma_{q-1}^i
\end{align}
for $\sigma_q \in K_p$. Similarly, the adjoint operator is defined as $\partial_q^{p*}:C^{q-1}(K_p) \to C^q(K_p)$. 

Now, we can define the spectra of the persistent Laplacian. Denote $\mathbb{C}_q^{p+t}$ as the $C_q^{p+t}$ whose boundary is in $C_{q-1}^p$, assuming an inclusion mapping $C_{q-1}^p \to C_{q-1}^{p+t}$. In such a set, we can define the $t$-persistent $q$-boundary operator $\hat{\partial}^{p,t}:\mathbf{C}_q]{p,t}\to C_{q-1}^p$, and its corresponding adjoint operator $(\hat{\partial}^{p,t})^*:C_{q-1}^p \to \mathbf{C}_q^{p,t}$. Then, the $q$-order $t$-persistent Laplacian operator can obtained via
\begin{align}
	\triangle_q^{p,t} = \hat{\partial}_{q+1}^{p,t}(\hat{\partial}_{q+1}^{p,t})^* + (\hat{\partial}_{q}^{p})^*\hat{\partial}_{q}^{p}.
\end{align}

Similarly, using the standard basis for the matrix representation of the operators, we obtain the $q$-order $t$-persistent Laplacian operator $\mathcal{L}_q^{p,t}$
\begin{align}
	\mathcal{L}_q^{p,t} = \mathcal{B}_{q+1}^{p,t}(\mathcal{B}_{q+1}^{p,t})^T + (\mathcal{B}_{q}^{p})^T(\mathcal{B}_{q}^{p}).
\end{align}

\subsection{$K$-mer topology for nucleotide sequences}
\begin{figure}[H]
	\includegraphics[width = \textwidth]{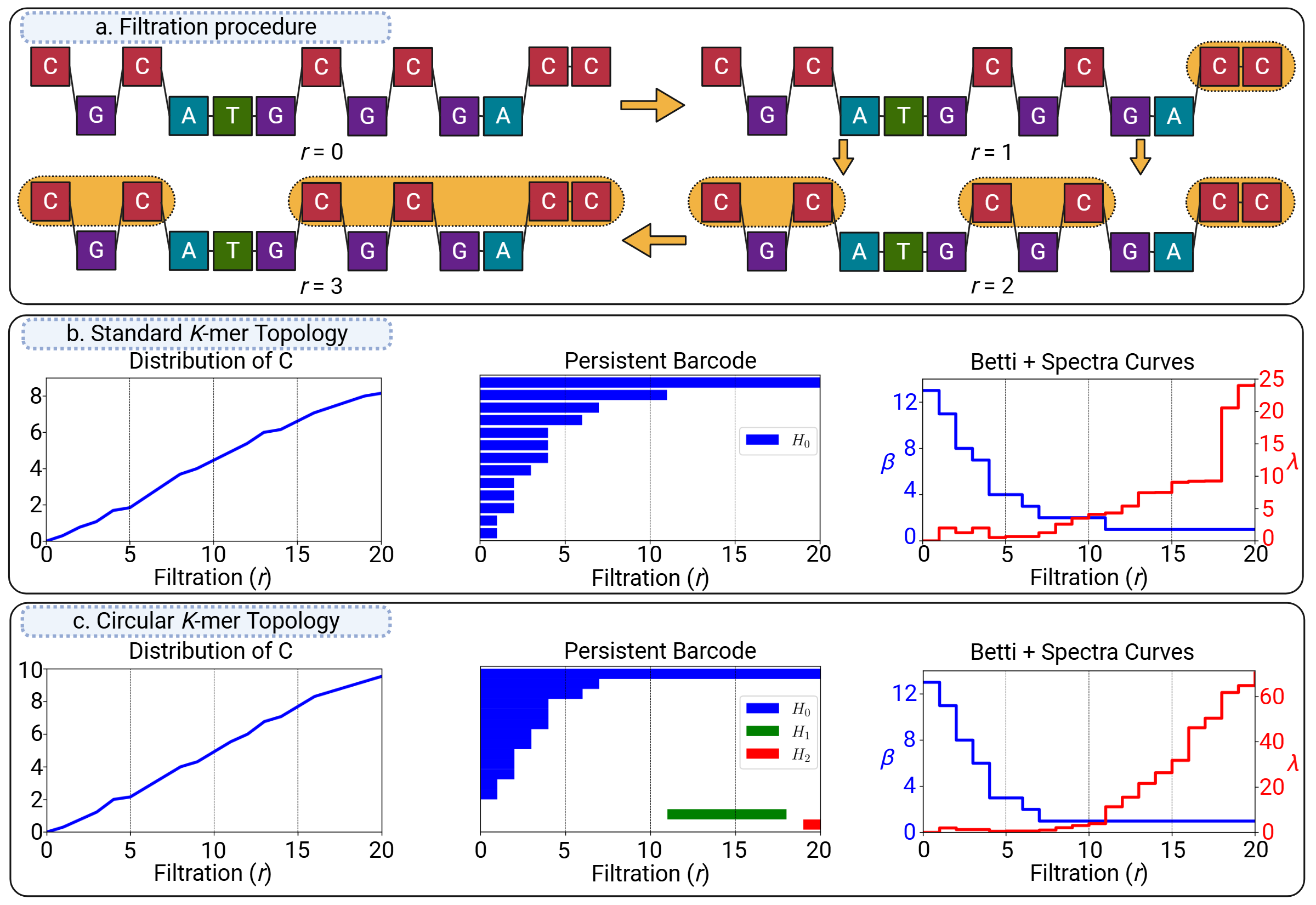}
	\caption{Feature extraction procedure using nucleotide $C$ with a randomly generated sequence of length 50. (a) The filtration procedure using nucleotide C. C's connected by the orange bar indicates a connected component. (b) The distribution of $C$, persistent barcode and the topological curves for the standard $K$-mer topology, where the sequence is considered a line. (c) The distribution of $C$, persistent barcode, and the topological curves for the circular $K$-mer topology, where the ends of the sequence is connected into a circle. For the distribution, the average number of nucleotide C was counted at different filtration radius. In the persistent barcode, the blue, green and red barcodes correspond to persistent Betti-0, Betti-1, and Betti-2, respectively. The blue and red line in the topological curves correspond to the Betti and smallest nonzero eigenvalue, respectively, at different filtration radius.}
	\label{fig:method}
\end{figure}

In this section, we describe the persistent topology for nucleotide sequences, and the extraction of topological and algebraic features. Then, we define a distance metric on the features for phylogenetic analysis. \autoref{fig:method} shows the feature extraction procedure using a randomly generated sequence.

\subsubsection{Position based distance}
Let $S = s_1s_2...s_N$ be a DNA sequence of length $N$, where $s_i \in$ \{A, C, G, T\}  is a nucleotide. Define the nucleotide-specific indicator function $\delta_l(s_i)$ as 
\begin{align}
	\delta_l(s_i) = \begin{cases}
		1, & s_i = l\\
		0, & {\rm otherwise},
	\end{cases}
\end{align}
where $l \in $\{A, C, G, T\}. Then, we can define a set of nucleotide specific positions $S^l$ as
\begin{align}
	S^l = \{i | \delta_l(s_i) = 1, 1\le i \le N\}.
\end{align}
The set $S^l$ contains all the global positions of nucleotide type $l$. Then, we can compute the pairwise distance matrix of the nucleotide type $l$
\begin{align}
	D^{l} = \{d_{ij}^{l}\}, \quad  ~ d^l_{ij} = |S^l(i) - S^l(j)|,
\end{align}
where $1 \le i,j \le |S^l|$, and $|S^l|$ is the total number of nucleotide type $l$.

In general, we can define a set of $k$-mer-specific positions, where instead of looking only at a specific nucleotide, we look at a given string of nucleotides of length $k$, i.e., a $k$-mer type. Moreover, there are $4^k$ different combinations of $k$-mers for a given $k$, which we denote as $l_1,l_2,...,l_{4^k}$. Similarly, we can define the $k$-mer specific indicator function as 
\begin{align}
	\delta_{l_p}(s_is_{i+1}...s_{i+k}) = \begin{cases}
		1, & s_is_{i+1}...s_{i+k} = l_p \\
		0, & \text{otherwise.}
	\end{cases}
\end{align}
Then, the set of  $k$-mer specific positions $S^{l_p}$ is given by
\begin{align}
	S^{l_p} = \{i | \delta_{l_p}(s_is_{i+1}...s_{i+k}) = 1, 1\le i \le N-k+1\}.
\end{align}

Then the $k$-mer specific distance matrix for $k$-mer type $l_p$ is given by 
\begin{align}
	D^{l_p} = \{d_{ij}^{l_p}\}, \quad ~ d^{l_p}_{ij}= |S^{l_p}(i) - S^{l_p}(j)|.
\end{align}
Distance matrices are used to compute both persistent homology and persistent Laplacians features.   

\subsubsection{Persistent Laplacian features}
For feature generation, we begin by constructing a series of graph Laplacian induced by filtration. For $0\le p \le P$, let $L_{0,r}^{l_p}$ be the $l_p$-specific $0$-order $r$-th graph Laplacian, which can be computed as the following
\begin{align}
	L_{0,r}^{l_p} = \{(l_{0,r}^{l_p})_{ij}\}, \quad (l_{0,r}^{l_p})_{ij} = \begin{cases}
		-1, & d_{ij}^{l_p} \le r, i \ne j \\
		\sum_{j}(l_{0,r}^{l_p})_{ij} & i =j.
	\end{cases}
\end{align}
We can order the eigenvalues of $L_{0,r}^{l_p}$ as $0 = \lambda_1 \le \lambda_2 \le ... \le \lambda_{|S^{l_p}|}$. The number of non-zero eigenvalues is the 0-Betti number, denoted $\beta_{0,r}^{l_p}$, and the smallest non-zero eigenvalue is denoted $\lambda_{r,0}^{l_p}$. 

For each $k$-mer type $l_p$, we can collect persistent Betti numbers and the smallest non-zero eigenvalues to obtain the $k$-mer specific $\mathbf{\beta}_0^{l_p}$ and $\mathbf{\lambda}_0^{l_p}$.
\begin{align}
	& \boldsymbol{\beta}_0^{l_p} = (\beta_{0,r_1}^{l_p}, \beta_{0,r_2}^{l_p}, ... ) \\
	& \boldsymbol{\lambda}_0^{l_p} = (\lambda_{0,r_1}^{l_p}, \lambda_{0,r_2}^{l_p}, ... )
\end{align}
where $r_1, r_2, ...$ is the increasing filtration radius.

Now, consider a sequence of length 13 shown in \autoref{fig:method}. The $k$-mer positions can be described in \autoref{tab:example_sequence}. 
\begin{table}[H]
	\centering 
	\caption{Sample sequence CGCATGCGCGACC and the positron of the $k$-mers.}
	\begin{tabular}{c|c c c c c c c c c  cccc}\hline
		Position & 1 & 2 & 3 & 4 & 5 & 6 & 7 & 8 & 9 & 10 & 11 & 12 & 13 \\ \hline
		Sequence  & C & G & C & A & T & G & C & G & C & G & A & C & C\\ \hline
		1-mer  & C & G & C & A & T & G & C & G & C & G & A & C & C \\
		2-mer & CG & GC & CA & AT & TG & GC & CG & GC & CG & GA & AC & CC & - \\
		3-mer & CGC & GCA & CAT & ATG & TGC & GCG & CGC & GCG & CGA & GAC & ACC & - & - \\ \hline
	\end{tabular}
	\label{tab:example_sequence}
\end{table}

We can compute $S^{\rm C}$ and $D^{\rm C}$ as 
\begin{align}
	S^{\rm C} = \{1, 3, 6, 8\}, \quad D^{\rm C} = \begin{pmatrix}
		0 & 2 & 5 & 7 \\
		2 & 0 & 3 & 5 \\
		5 & 3 & 0 & 2 \\
		7 & 5 & 2 & 0
	\end{pmatrix}.
\end{align}

Using the same sequence in \autoref{tab:example_sequence}, then 2-mer  CG  specific positions and distances can be defined as the following
\begin{align}
	S^{\rm CG} = \{1, 6, 8\}, \quad D^{\rm CG} = \begin{pmatrix}
		0 & 5 & 7\\ 5 & 0 & 2 \\ 7 & 5 & 0
	\end{pmatrix}.
\end{align}
Similarly, the 3-mer  CGC  specific positions and distances can be defined as the following
\begin{align}
	S^{\rm CGC} = \{1, 6\}, \quad D^{\rm CGC} = \begin{pmatrix}
		0 & 5 \\ 5 & 0
	\end{pmatrix}.
\end{align}

We illustrate an example of our method using the influenza haemagglutinin (HA) gene of H1N1. In Figure \ref{fig:supporting_ciruclar_vs_standard}, we generated 2 cases of the $K$-mer topology by treating the DNA sequence as a line and as a circle, and utilized the $k$-mer AC. The upper row shows the standard case, and the bottom row shows the circular case. In the circular case, the distance between the $k$-mers is modified, by taking 
\begin{align}
	d^l(i,j) = \min[ S^l(j)-S^l(i), N - S^l(j) + S^l(i)].
\end{align}
By doing so, the sequence is allowed to wrap around. The first column shows the average number of AC within the filtration radius $r$. Not surprisingly, the average number of AC is slightly higher for the circular case. The second column shows the Betti-0 number as we increase the filtration radius. Interestingly, the Betti-0 number remains similar for both cases; however, the circular case has lower values of Betti-0. This is because by allowing the sequence to wrap around, the ACs  at the 5' and 3' ends are allowed to form connected components. Nonetheless, the difference is very small. The third column shows the smallest non-zero eigenvalue as we increase the filtration radius. Interestingly, the behavior is similar until $r = 40$. However, because we allow the sequence to wrap around, the number of edges in our Laplacian graph is significantly higher for the circular case. This most likely led to a larger eigenvalue as the filtration radius increased.

\subsubsection{Topological genetic distance }
We now define a topological metric for perform phylogenetic analysis. For each $k$, we have $4^k$ different $k$-mers specific Betti-0  and minimal eigenvalue curves. We concatenate these curves to obtain the vector
\begin{align}
	\boldsymbol{\beta}^k = (\boldsymbol{\beta}_0^{l_1},..., \boldsymbol{\beta}_0^{l_{4^k}}) \\
	\boldsymbol{\lambda}^k = (\boldsymbol{\lambda}_0^{l_1},..., \boldsymbol{\lambda}_0^{l_{4^k}}).
\end{align}
Then, for sequences $i$ and $j$, we can define the topological metric between the $k$-mer specific curves as the following
\begin{align}
	\text{dist}_\beta^k(i, j) = \|\boldsymbol{\beta}_i^k - \boldsymbol{\beta}_j^k\|_2 \\
	\text{dist}_\lambda^k(i, j) = \|\boldsymbol{\lambda}_i^k - \boldsymbol{\lambda}_j^k\|_2,
\end{align}
where $\|\cdot\|_2$ denote the Euclidean distance.

Then, we can take a weighted sum over different $k$ to obtain a topological genetic distance between sequences $i$ and $j$
\begin{align}\label{eq:genetic_distance}
	\text{Dist}(i,j) = \sum_{k=1}^K a_k \text{dist}_\beta^k(i,j) + b_k\text{dist}_\lambda^k(i,j).
\end{align} 
If we only need Betti-0 features, we set $b_k = 0$, and in this work, we utilized $a_k = \frac{1}{\omega_k2^{K-k}}$ and  $b_k = 0$, where $K$ is the largest $k$, and $\omega_k$ is the average $\text{dist}_\beta^k$.
Alternative definitions of topological genetic/antigenetic distances are presented in the Supporting Information. 

\section*{Data  and code availability}
Datasets are available from the NCBI virus database, \href{ https://www.ncbi.nlm.nih.gov/labs/virus/vssi/}{ https://www.ncbi.nlm.nih.gov/labs/virus/vssi/}, which was accessed in March 2024. NCBI 2020 and 2022 can be accessed from \cite{sun2021geometric} and \cite{yu2024optimal}, respectively. All codes, including those of other methods compared in this work, are available on Github,
\linebreak \href{https://github.com/hozumiyu/KmerTopology}{https://github.com/hozumiyu/KmerTopology}, and are further described in the Supporting Information.

\section*{Supporting Information}		
Supporting Information is available.

\section*{Acknowledgment}	
This work was supported in part by NIH grants R01AI164266, and R35GM148196, NSF grants DMS-2052983 and IIS-1900473, MSU Research Foundation, and Bristol-Myers Squibb 65109.

\newpage
\setcounter{equation}{0}
\setcounter{figure}{0}
\setcounter{table}{0}
\setcounter{section}{0}
\renewcommand{\thesection}{S\arabic{section}}
\renewcommand{\thefigure}{S\arabic{figure}}
\renewcommand{\thetable}{S\arabic{table}}
\renewcommand{\theequation}{S\arabic{equation}}

\section{Topological genetic/antigenetic distances}\label{sec:supporing_genetic_distance}

Genetic distance and antigenetic distance have a wide variety of applications, including the rational design of vaccines against viruses, such as SARS-CoV-2 \cite{hu2023rational} and influenza \cite{skowronski2017serial}. Specifically, antigenetic distance can be used to predict vaccine effectiveness against viral variants. This approach becomes very powerful when combined with deep mutational scanning \cite{starr2020deep,chen2023topological} and the accurate forecasting of emerging viral variants \cite{chen2022omicron, chen2022persistent}, offering rapid responses to fast viral evolution \cite{wee2024rapid}. 

In the main text, we define a topological genetic distance for classification and clustering. That is, for sequences $i$ and $j$, the genetic distance between them is given by
\begin{equation}\label{TGD}
	\text{Dist}(i,j) = \sum_{k=1}^K a_k \text{dist}_\beta^k(i,j) + b_k\text{dist}_\lambda^k(i,j)
\end{equation}
for some hyperparameters $a_k$ and $b_k$. If we only consider Betti-0 features, we can set $b_k = 0$ for all $k$. In our benchmark, we set $a_k = \frac{1}{\omega_k2^{K-k}}$, where $K$ is the maximum $k$-mer we consider, and $\omega_k$ is the average $\text{dist}_\beta^k(i,j)$. One potential downside to this approach is that the distance can be dominated by some large Betti numbers (that is, those in the first few filtration steps) and large spectra. 

For a more detailed analysis, it may be more feasible to consider multiscale distance calculations, where the distance is calculated for each filtration radius. For each $k$, we have $l_{4^k}$ combinations of $k$-mers. Denote $\beta_{k,r}$ as the Betti number for a given $k$ and a given filtration radius.
\begin{align}
	\mathbf{\beta}_{k,r} = (\beta^{l_1}_r, \beta^{l_2}_r, ..., \beta^{l_{4^k}}_r)
\end{align}
Similarly, we can define $\lambda_{k,r}$ as the smallest nonzero eigenvalue for a given $k$ and a given filtration radius.
\begin{align}
	\mathbf{\lambda}_{k,r} = (\lambda^{l_1}_r, \lambda^{l_2}_r, ..., \lambda^{l_{4^k}}_r)
\end{align}

Then, we can consider the distance between sequences $i$ and $j$ for $k$-mer $l_{4^k}$ at radius $r$ to be
\begin{align}
	& 	\text{dist}_{\beta{k,r}}(i,j) = \|\mathbf{\beta}_{k,r}(i) - \mathbf{\beta}_{k,r}(j)\| \\
	&	\text{dist}_{\lambda_{k,r}}(i,j) = \|\mathbf{\lambda}_{k,r}(i) - \mathbf{\lambda}_{k,r}(j)\|.
\end{align}

Then, the genetic distance will be the weighted sum of these multiscale distances.

\begin{align}
	\text{Dist}(i,j) = \sum_{r}\sum_{k=1}^K a_{k,r} \text{dist}_{\beta_{k,r}}(i,j) + b_{k,r}\text{dist}_{\lambda_{k,r}}(i,j)
\end{align}
We validated this approach for SARS-CoV-2 classification in Figure \ref{fig:supporting_PhyloLaplacian}. We let $K=4, a_{k,r} = \frac{1}{\omega_{k,r}2^{4-k}}, b_{k,r} = \frac{1}{\gamma_{k}2^{4-k}}$, and $\omega_{k,r}$ is the average $\text{dist}_{\beta_{k,r}}$, and $\gamma_{k}$ is the average $\text{dist}_{\lambda_{k,r}}$. For this analysis, we do not use multiscale distances for the spectra curves. 

The classification of SARS-CoV-2 variants is extremely challenging for alignment-free methods because the defining mutations on the variants are only a few nucleotides. We can see in \autoref{fig:sarscov2_variants} that none of the 5 other alignment-free methods can classify the variants properly. However, by introducing the multiscale distance, $k$-mer topology can successfully capture the small mutational differences.

\section{Additional classification benchmark}\label{sec:supporting_classification}

\begin{figure}[H]
	\centering
	\includegraphics[width = \textwidth]{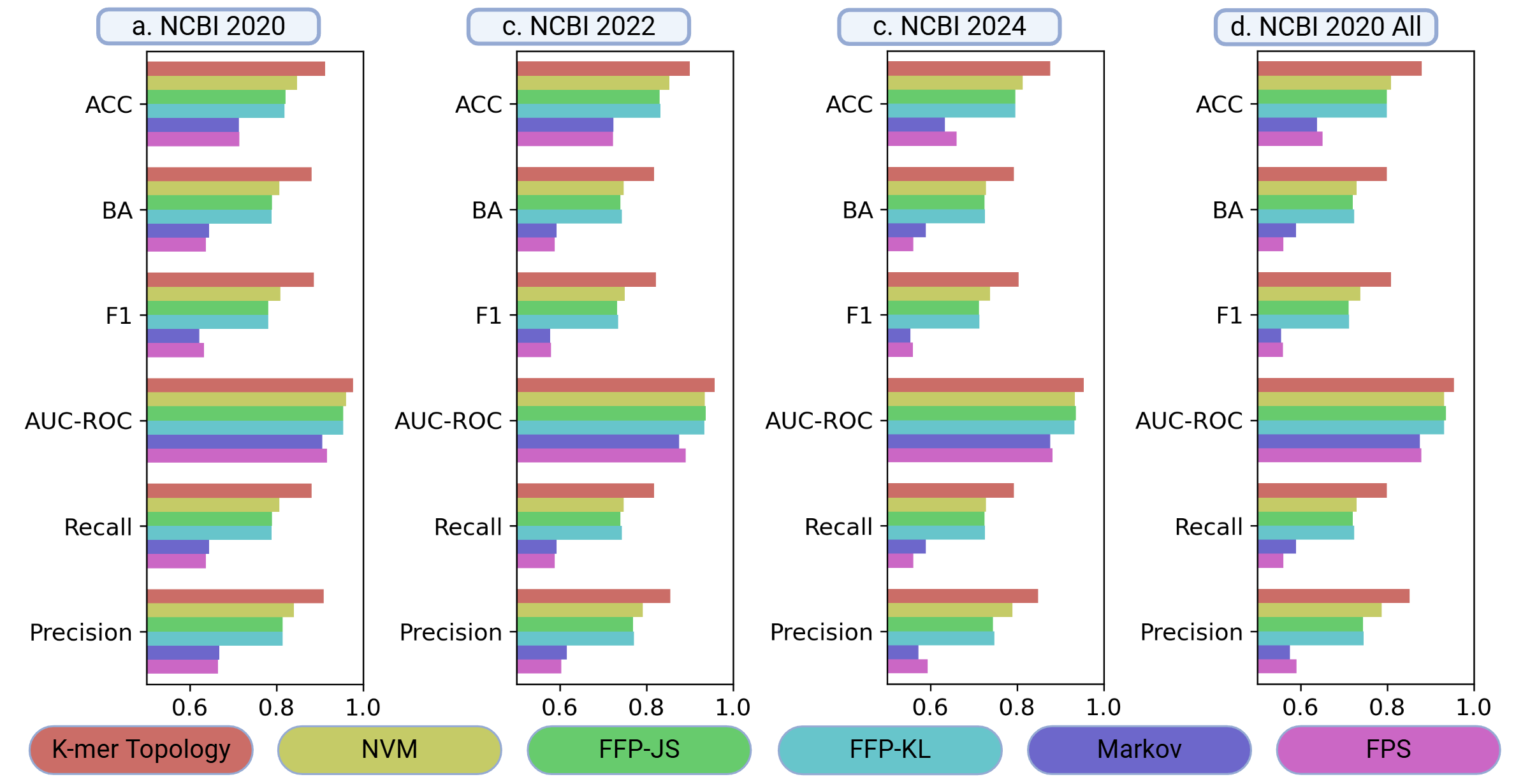}
	\caption{Comparison of six methods for 5-nearest neighbor  classification of the NCBI virus datasets. Stratified 5-fold cross validation with 30 random seeds was performed. (a)NCBI 2020. (b) NCBI 2022. (c) NCBI 2024. (d) NCBI 2024 All. Accuracy (ACC), balanced accuracy (BA), macri-F1 (F1), area under the receiving operator curve (AUC-ROC), recall, and precision were used to evaluate the classification performance.}
	\label{fig:supp_5NN_benchmark}
\end{figure}

In Tables 	\ref{tab:supporting_1NN}, 	\ref{fig:supp_5NN_benchmark}, and 	\ref{tab:supporting_5NN}, we present more details on the classification of four viral datasets. Six methods, including $k$-mer topology, NVM, FFP-JS, FFP-KL, Markov and FPS, were employed in the comparative analysis. 

\autoref{tab:supporting_1NN} shows the accuracy comparison of $k$-mer topology, NVM, FFP-JS, FFP-KL, Markov and FPS methods for the 1-nearest neighbor classification of 4 benchmark datasets. The bold values indicate the best accuracy for each data. Our method outperforms other methods by a large margin on all datasets.
\begin{table} [H]
	\centering
	\caption{1-NN classification accuracy of  6 methods. }
	\begin{tabular}{| c | c c c c c c |}\hline
		& \multicolumn{6}{c|}{Method}  \\ \hline
		Data & $k$-mer topology & NVM & FPS-JS & FFP-KL & Markov & FPS \\\hline
		NCBI 2020     & \textbf{0.933} & 0.879 & 0.862 & 0.862 & 0.734 & 0.732 \\
		NCBI 2022     & \textbf{0.920} & 0.875 & 0.870 & 0.870 & 0.735 & 0.732 \\
		NCBI 2024     & \textbf{0.898} & 0.829 & 0.825 & 0.826 & 0.637 & 0.656 \\
		NCBI 2024 All & \textbf{0.901} & 0.825 & 0.832 & 0.832 & 0.647 & 0.647 \\ \hline
	\end{tabular}
	\label{tab:supporting_1NN}
\end{table}

\autoref{fig:supp_5NN_benchmark} shows the benchmark comparison of our method with the 5 other methods on 4 datasets. For each dataset, stratified 5-fold cross validation with 30 random seeds was used to obtain the scores. Accuracy (ACC), balanced accuracy (BA), macri-F1 (F1), area under the receiving operator curve (AUC-ROC), recall, and precision were used to evaluate the classification performance. Macro scores were used for all tests because each viral family is equally important. Additionally, for AUC-ROC, one-versus-rest was used to obtain the scores. $K$-mer topology out performs all other methods in every metric across all datasets.

\autoref{tab:supporting_5NN} shows the ACC, BA, F1, AUC-ROC, recall, and precision of 5-NN classification. The bold values indicate the best scores for each dataset and scores.
\begin{table} [H]
	\centering
	\caption{Comparison of 5-NN classification scores of the 6 methods. }
	\begin{tabular}{| c | c | c c c c c c |}\hline
		& & \multicolumn{6}{c |}{ Scores} \\
		Data & Method & ACC & BA & F1 & AUC-ROC & Recall & Precision \\ \hline
		\multirow{6}{*}{NCBI 2020} & $k$-mer topology & \textbf{0.912} & \textbf{0.881} & \textbf{0.886} & \textbf{0.977} & \textbf{0.881} & \textbf{0.909}\\
		& NVM & 0.847 & 0.807 & 0.809 & 0.960 & 0.807 & 0.840\\
		& FFP-JS & 0.821 & 0.790 & 0.781 & 0.954 & 0.790 & 0.814\\
		& FFP-KL & 0.819 & 0.789 & 0.780 & 0.954 & 0.789 & 0.814\\
		& Markov & 0.713 & 0.644 & 0.622 & 0.905 & 0.644 & 0.668\\
		& FPS & 0.714 & 0.637 & 0.633 & 0.917 & 0.637 & 0.665\\  \hline
		\multirow{6}{*}{NCBI 2022} & $k$-mer topology & \textbf{0.900} & \textbf{0.818} & \textbf{0.821} & \textbf{0.957} & \textbf{0.818} & \textbf{0.855}\\
		& NVM & 0.852 & 0.747 & 0.750 & 0.935 & 0.747 & 0.791\\
		& FFP-JS & 0.830 & 0.740 & 0.733 & 0.937 & 0.740 & 0.769\\
		& FFP-KL & 0.832 & 0.743 & 0.735 & 0.934 & 0.743 & 0.771\\
		& Markov & 0.724 & 0.593 & 0.577 & 0.876 & 0.593 & 0.617\\
		& FPS &  0.723 & 0.588 & 0.580 & 0.890 & 0.588 & 0.603\\\hline
		\multirow{6}{*}{NCBI 2024} & $k$-mer topology & \textbf{0.877 }& \textbf{0.793} & \textbf{0.803} & \textbf{0.954} & \textbf{0.793} &\textbf{ 0.848}\\
		& NVM & 0.814 & 0.729 & 0.738 & 0.933 & 0.729 & 0.789\\
		& FFP-JS & 0.796 & 0.724 & 0.712 & 0.936 & 0.724 & 0.744\\
		& FFP-KL & 0.796 & 0.727 & 0.714 & 0.932 & 0.727 & 0.747\\
		& Markov & 0.633 & 0.589 & 0.554 & 0.876 & 0.589 & 0.573\\
		& FPS &  0.660 & 0.561 & 0.560 & 0.882 & 0.561 & 0.593\\\hline
		\multirow{6}{*}{NCBI 2024 All} & $k$-mer topology &\textbf{ 0.880 }& \textbf{0.799} & \textbf{0.808} & \textbf{0.954} & \textbf{0.799} & \textbf{0.851}\\
		& NVM & 0.809 & 0.729 & 0.738 & 0.931 & 0.729 & 0.788\\
		& FFP-JS & 0.799 & 0.721 & 0.711 & 0.935 & 0.721 & 0.745\\
		& FFP-KL & 0.799 & 0.724 & 0.712 & 0.931 & 0.724 & 0.745\\
		& Markov & 0.638 & 0.589 & 0.555 & 0.876 & 0.589 & 0.575\\
		& FPS &  0.651 & 0.561 & 0.559 & 0.879 & 0.561 & 0.591\\\hline
	\end{tabular}
	\label{tab:supporting_5NN}
\end{table}

\begin{enumerate}
	\item 
\end{enumerate}
\section{Additional materials for phylogenetic analysis using \textit{k}-mer topology}\label{sec:supporting_phylo}

In this section, we compare the phylogenetic trees generated from the $k$-mer topology with those from natural vector method (GNV) \cite{sun2021geometric}, Markov K-string \cite{wu2001statistical}, KL-and JS divergence of $k$-mers frequency profile, and Fourier power spectrum (FPS) method \cite{hoang2015new}. For Markov K-String, KL divergence and JS divergences, $k = 3$ was used to obtain the features. To build up phylogenetic trees, unweighted pair group method with arithmetic mean (UPGMA) was used, and interactive tree of life v6 \cite{letunic2024interactive} was used to generate and annotate the trees.

\begin{figure}[H]
	\centering
	\includegraphics[width=\textwidth]{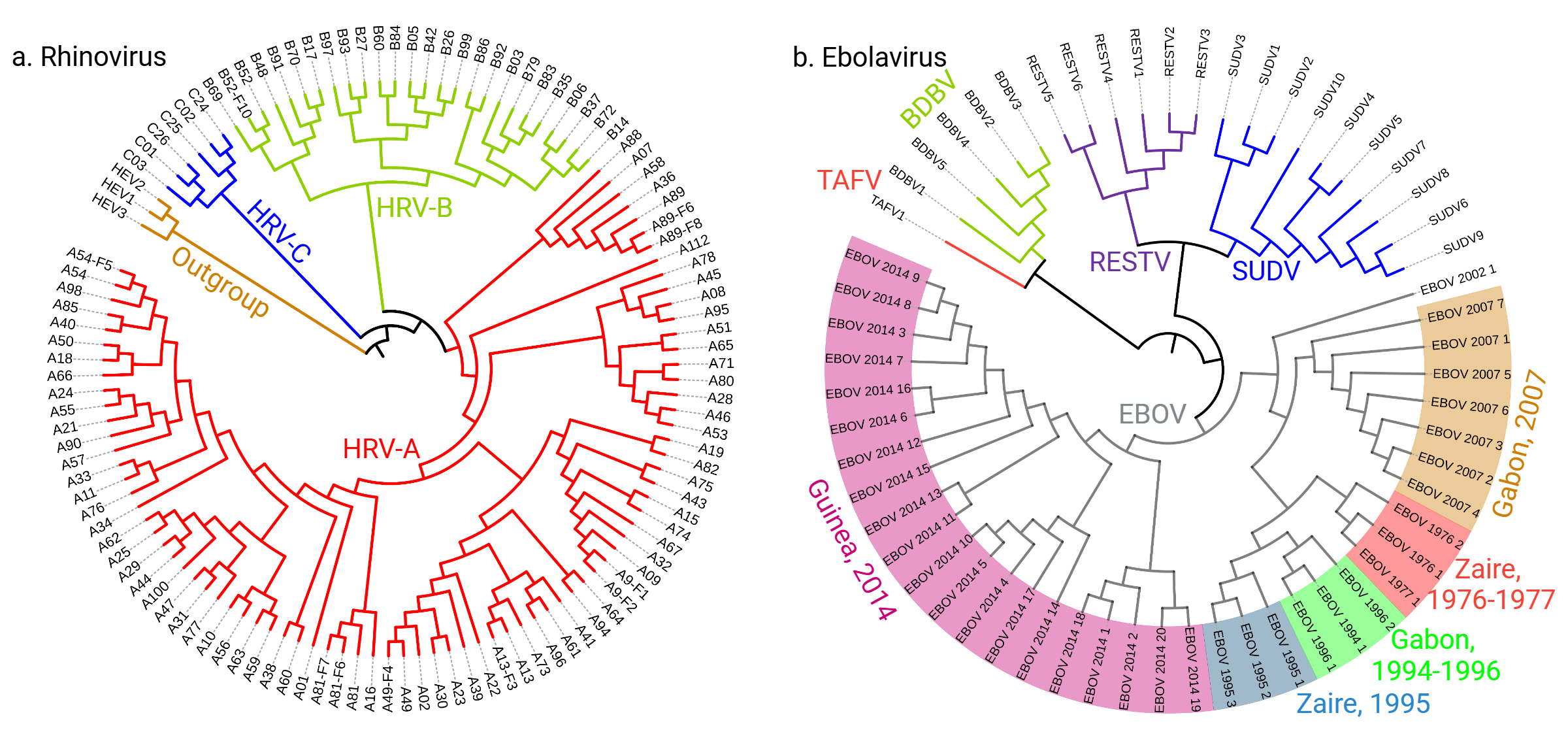}
	\caption{(a) Phylogenetic tree of rhinovirus (HRV) using $k = 5$.
		The data contains three HRV groups—A, B, and C—along with three outgroup sequences. The $K$-mer topology places the outgroup sequences in an entirely separate clade and distinguishes each HRV group into individual clades.
		(b) Phylogenetic tree of ebola virus using $k = 5$.
		The data contains five types: Tai Forest virus (TAVF), Sudan virus (SUDV), Bundibugyo virus (BDBV), Reston virus (RESTV), and ebola virus (EBOV). Additionally, EBOV is categorized by pandemic year and location, with labels highlighted according to their respective locations. The $K$-mer topology places each type into distinct clades and further separates the EBOV subtypes.}
	\label{fig:supporting_hrv_ebola}
\end{figure}

\begin{figure}[H]
	\centering
	\includegraphics[width=\textwidth]{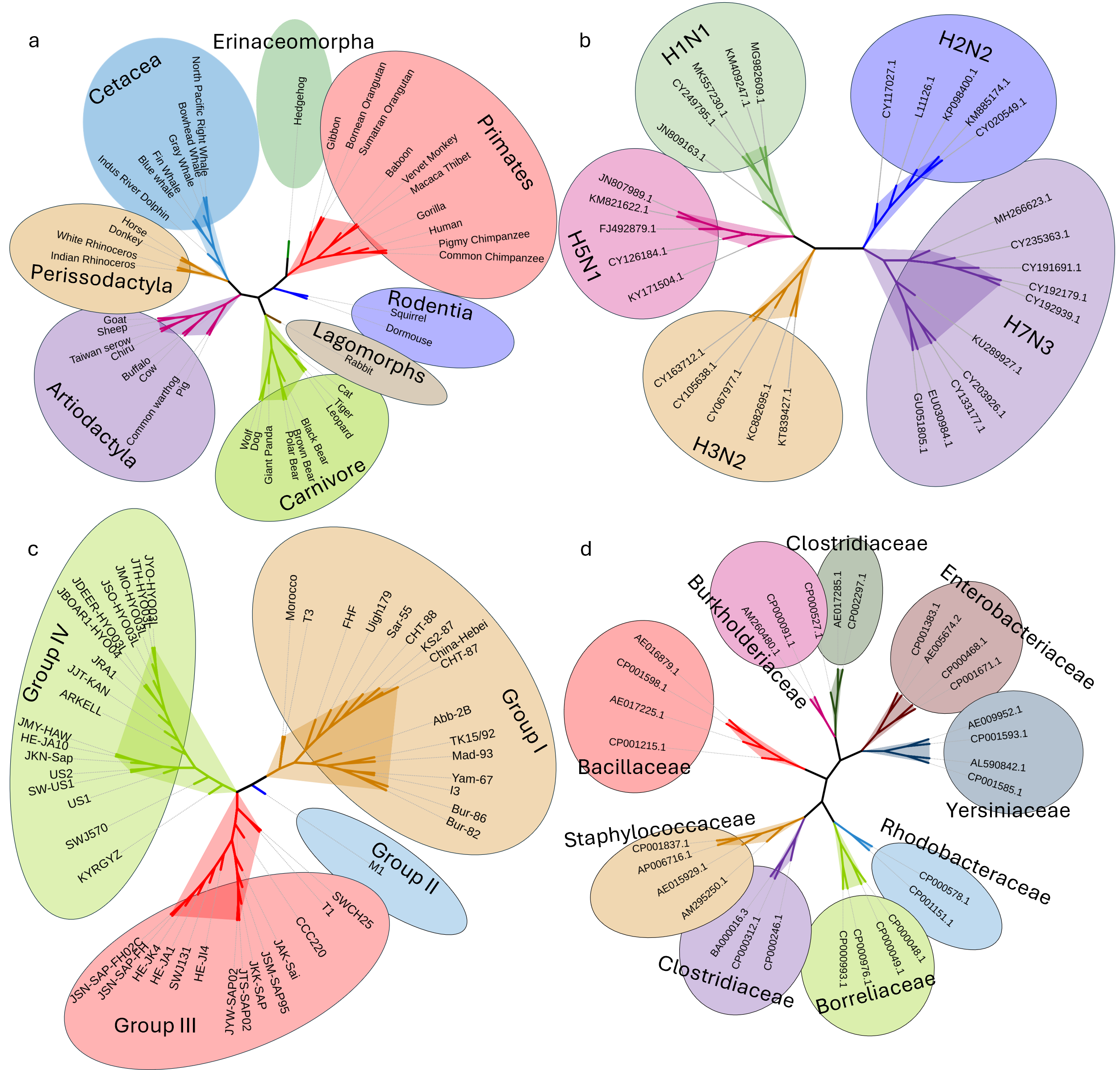}
	\caption{Phylogenetic analysis using $K$-mer topology.
		(a) Phylogenetic tree of the influenza A haemagglutinin genes using $k = 3$.
		The data contains six types: H1N1, H2N2, H7N9, H7N3, H3N3, and H5N1. The $K$-mer topology separates all six types into distinct clades.
		(b) Phylogenetic tree of complete mammalian mitochondrial genomes using $k = 5$.
		The data contains eight host species types, with mitochondrial genomes categorized according to their host species. The $K$-mer topology separates all eight types into distinct clades.
		(c) Phylogenetic tree of complete bacterial genomes using $k = 5$.
		Because the sequence length exceeds 1 million base pairs, 3-mers, 4-mers, and 5-mers were used. The data contains nine bacterial families, and the $K$-mer topology separates all nine families into distinct clades.
		(d) Phylogenetic tree of hepatitis E virus (HEV) using $k = 5$.
		The data contains four groups. The $K$-mer topology separates all four groups into distinct clades.}
	\label{fig:supporting_influenza_mammalian_bacteria_hev}
\end{figure}

\autoref{fig:supporting_hrv_ebola} shows the phylogenetic tree of rhinovirus and  ebola virus using our method  with $k  =  5$.  \autoref{fig:supporting_influenza_mammalian_bacteria_hev} shows  the phylogenetic  trees of  influenza A haemagglutinin genes,  mammalian mitochondria,  bacteria and hepatitis E virus generated using our method.  Our $K$-mer topology successfully clusters all genomes into their respective classifications. We provide a more thorough comparative analysis in the subsequent subsections.

\subsection{Hepatitis E virus genomes}

\begin{figure}[H]
	\centering
	\includegraphics[width = \textwidth]{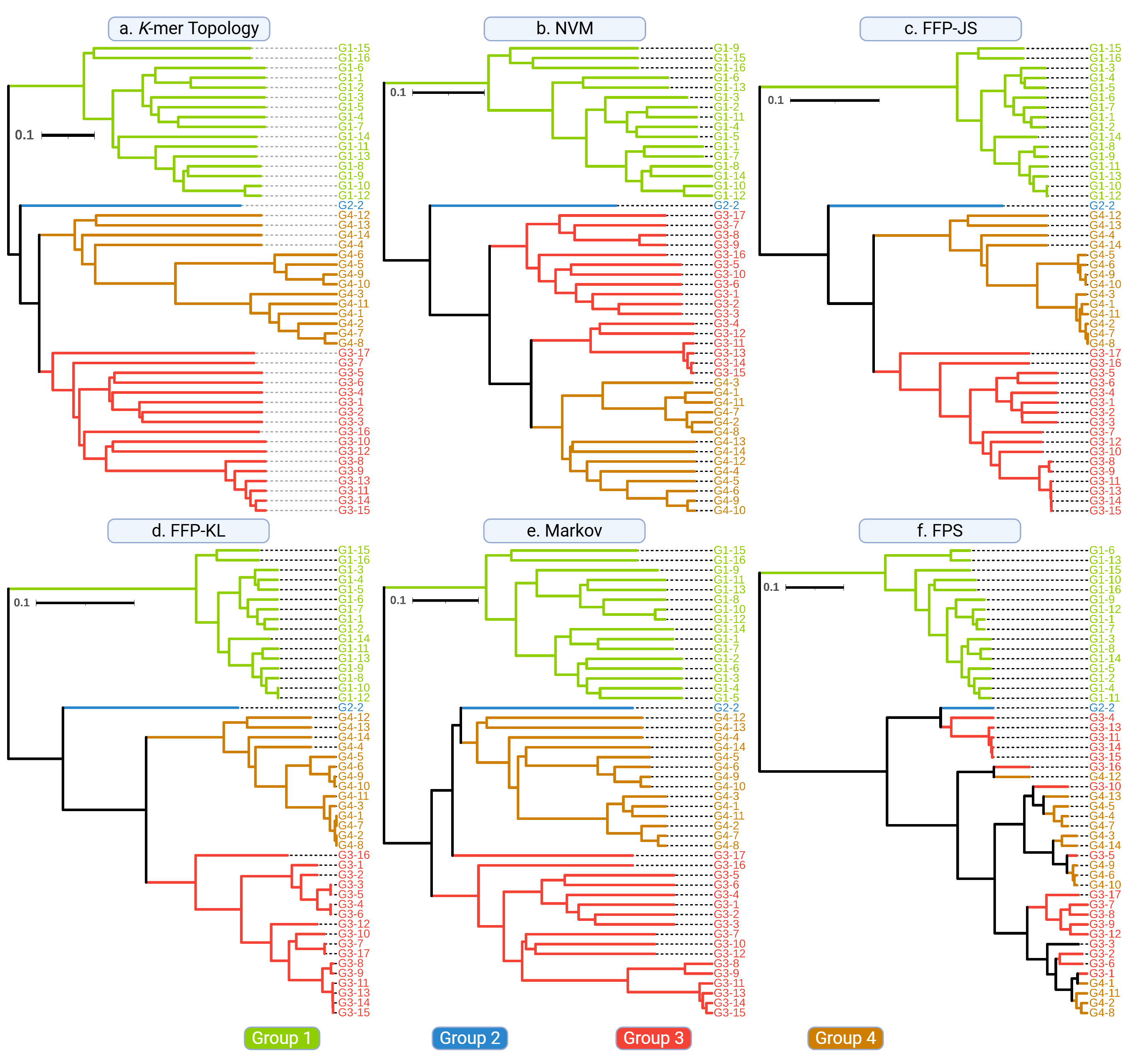}
	\caption{Phylogenetic trees of 48 complete hepatitis E virus  (HEV) genomes.
		HEV are categorized into 4 groups: 1, 2, 3 and 4. The branches and labels are colored based on the HEV group.
		(a) $K$-mer topology using $k = 5$: All HEV genomes are correctly classified into their respective clades.
		(b) NVM using $k = 5$: HEV groups 1, 2  and  4 are  correctly  classified into their respective clades. Group 3 genomes are in 2  distinct  clades.
		(c) FFP-JS using $k = 3$: All HEV genomes are correctly classified into their respective clades.
		(d) FFP-KL using $k = 3$: All HEV genomes are correctly classified into their respective clades.
		(e) Markov K-String using $k = 3$: One Group 3 genome (G3-17) is misclassified.
		(f) Fourier power spectrum (FPS): Only Group 1 HEV genomes are clustered together. The other groups do not form cohesive clades.}
	\label{fig:supporting_HEV_benchmark}
\end{figure}

\subsection{Influenza A hemagglutinin genes}
\begin{figure}[H]
	\includegraphics[width = \textwidth]{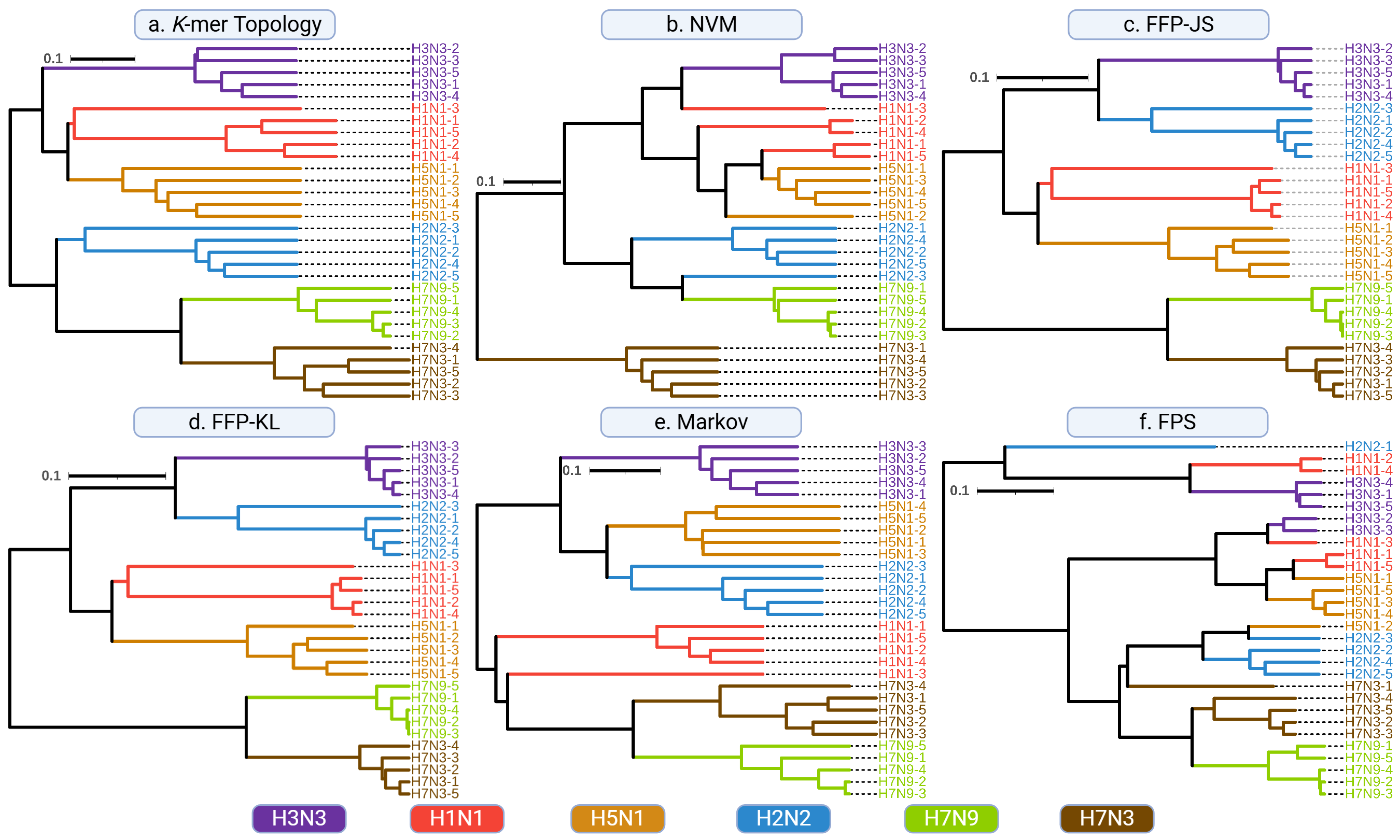}
	\caption{Phylogenetic trees of 30 influenza A  hemagglutinin (HA) genes.
		Influenza HA genes are categorized into 6 groups: H1N1, H2N2, H3N2, H5N1, H7N9, and H7N3. The branches and
		labels are color-coded according to the influenza A classifications.
		(a) $K$-mer topology using $k = 3$: All influenza A HA genes  are correctly classified into their respective clades. 
		(b) NVM using $k = 3$: H1N1 HA genes do not form aa clade. One H5N1 gene (H5N1-2) is misclassified.
		(c) FFP-JS using $k = 3$: All influenza A HA genes  are correctly classified into their respective clades. 
		(d) FFP-KL using $k = 3$: All influenza A HA genes  are correctly classified into their respective clades. 
		(e) Markov K-String using $k = 3$: One H1N1 gene (H1N1-3) is misclassified.
		(f) Fourier power spectrum (FPS): Only the H7N9 HA genes form a  clade.}
	\label{fig:supporting_influenzaHAgene_benchmark}
\end{figure}

\subsection{Ebola virus}
\begin{figure}[H]
	\includegraphics[width = \textwidth]{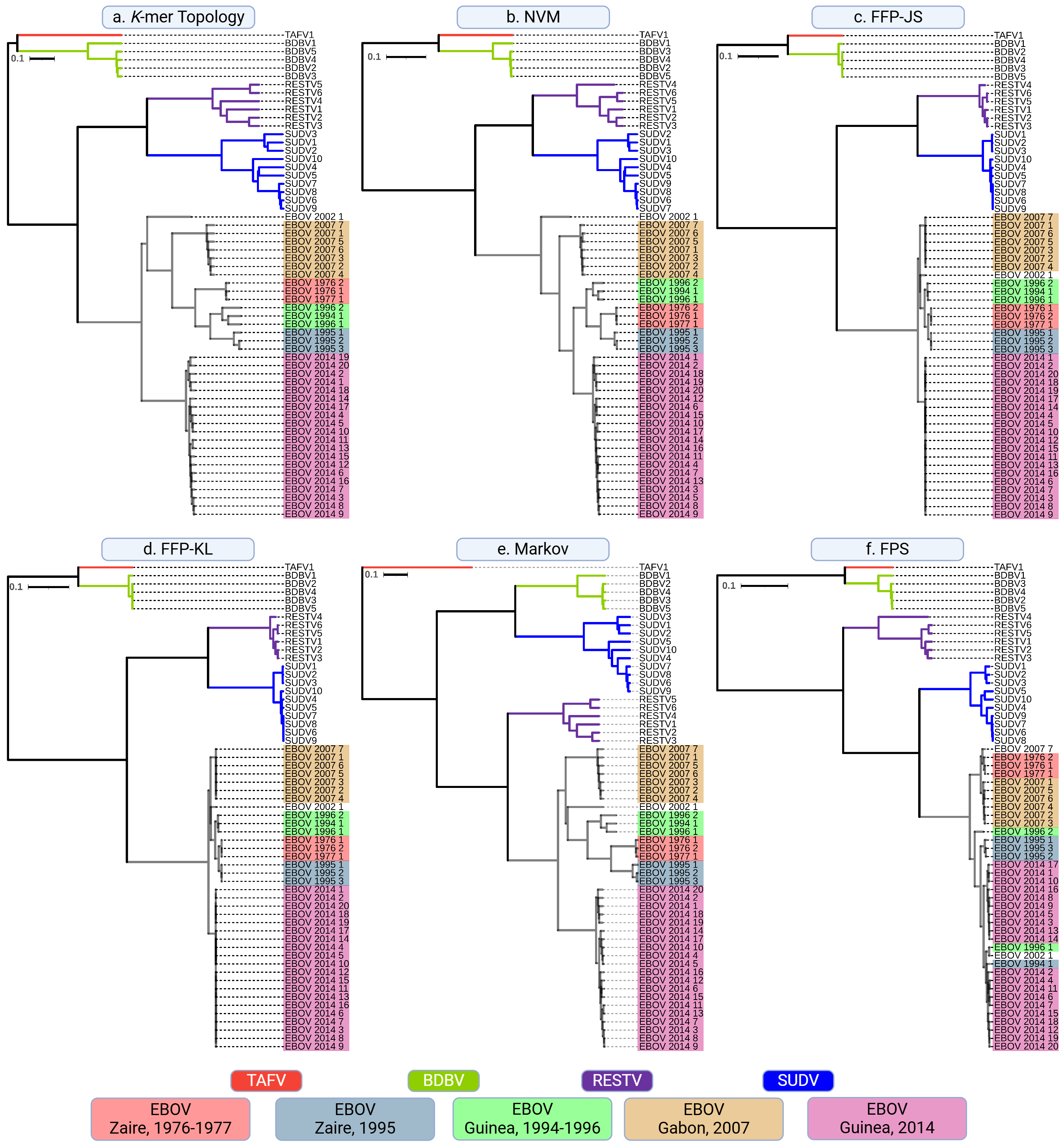}
	\caption{Phylogenetic trees of 59 complete ebola virus genomes.
		Ebolaviruses are categorized into five types: Bundibugyo virus (BDBV), Reston virus (RESTV), Ebola virus (EBOV), Sudan virus (SUDV), and Tai Forest virus (TAFV). EBOV is further subdivided based on epidemic location and year. The branches are colored based on the ebola virus types, and the labels are highlighted according to its EBOV epidemic.
		(a) $K$-mer topology using $k = 5$: All ebola virus genomes are correctly classified into their respective clades, with EBOV epidemics also clustered within the clades.
		(b) NVM using $k = 5$: All ebola virus genomes are correctly classified into their respective clades, with EBOV epidemics also clustered within the clades.
		(c) FFP-JS using $k = 3$: All ebola virus genomes are correctly classified into their respective clades, with EBOV epidemics also clustered within the clades.
		(d) FFP-KL using $k = 3$: All ebola virus genomes are correctly classified into their respective clades, with EBOV epidemics also clustered within the clades.
		(e) Markov K-String using $k = 3$:  All ebola virus genomes are correctly classified into their respective clades, with EBOV epidemics also clustered within the clades.
		(f) Fourier power spectrum (FPS): All ebola virus genomes are correctly classified into their respective clades; however, EBOV epidemics are not clustered together.}
	\label{fig:supporting_ebolavirus_benchmark}
\end{figure}

\subsection{Bacteria}

\begin{figure}[H]
	\includegraphics[width = \textwidth]{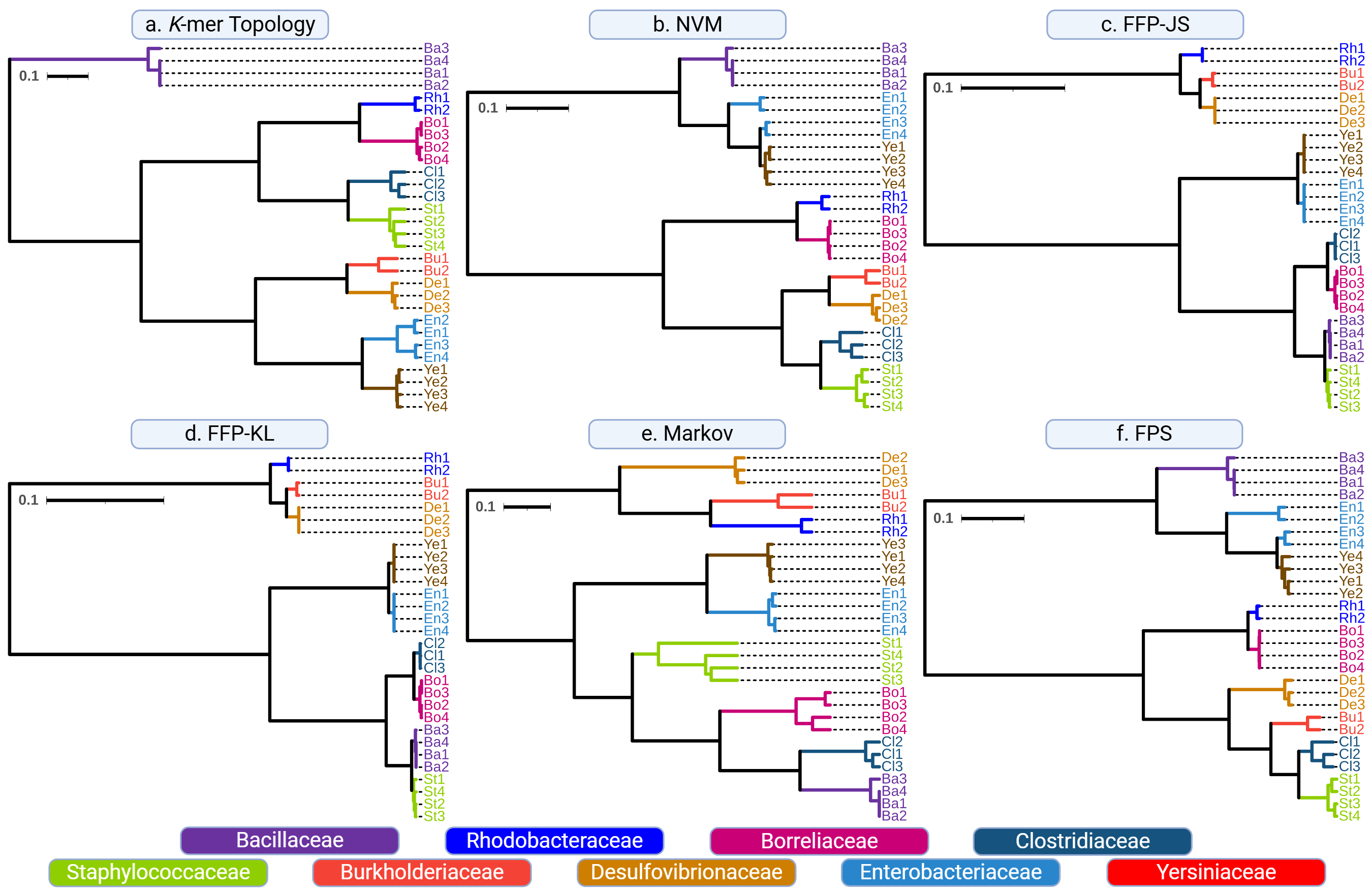}
	\caption{Phylogenetic trees of influenza A hemagglutinin (HA) genes generated using (a) $K$-mer Topology with $K = 5$, (b) NVM with $k = 7$, (c) FFP-JS, (d) FFP-KL, (e) Markov K-String, and (f) FPS. The labels and clades are colored according to their types. }
	\label{fig:supporting_bacteria_benchmark}
\end{figure}

\autoref{fig:supporting_bacteria_benchmark} compares $k$-mer topology with five other methods on the whole bacteria genome data. The dataset contains 30 sequences, categorized into nine families. The label colors correspond to each family.

All methods cluster each family into a separate clade; however, their divisions show significant differences.

\section{Distribution of \textit{k}-mers}

\textit{K}-mer based methods often calculate the statistics on the frequency of the individual $k$-mers but do not take into account the local distribution of the individual $k$-mer. For example, NVM takes positional information into account and calculates statistics on the distribution, such as the average position, the variance of the positions, and more. However, local structural information is not considered. 

\begin{figure}[H]
	\centering
	\includegraphics[width=\textwidth]{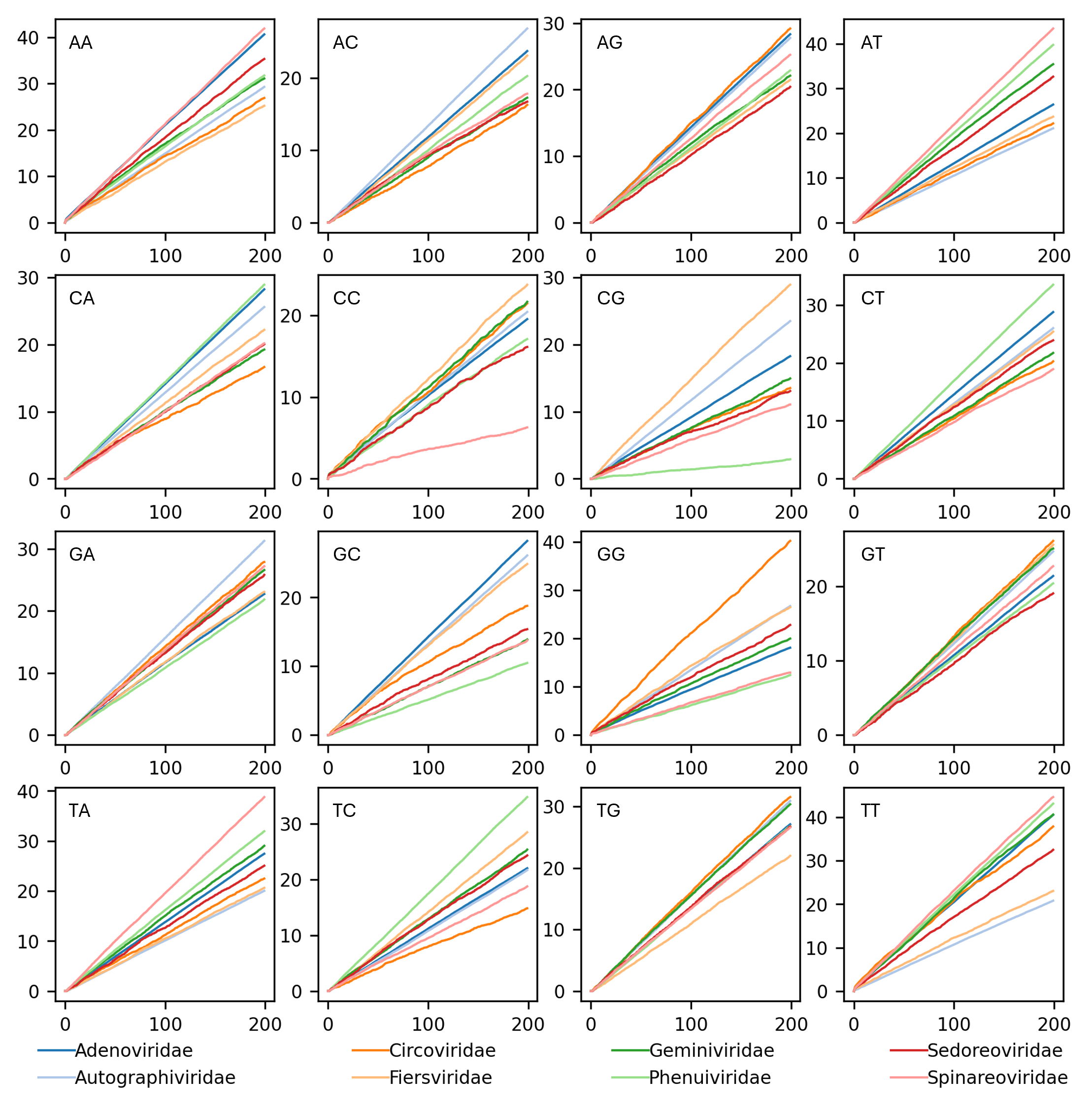}
	\caption{Distribution of 2-mers of 8 viral families: AC\_000001.1 (\textit{Adenoviridae}) NC\_001271.1 (\textit{Autographiviridae}) NC\_001792.2 (\textit{Circoviridae}) NC\_001417.2 (\textit{Fiersviridae}) NC\_000869.1 (\textit{Geminiviridae}) NC\_002323.1 (\textit{Phenuiviridae}) NC\_003760.1 (\textit{Sedoreoviridae}) NC\_002557.1 (\textit{Spinareoviridae}). Each plot corresponds to one of 16 2-mer types, and the lines are colored according to their families. The $x$-axis and $y$-axis correspond to the radius and averaged 2-mer counts. }
	\label{fig:supporting_kmer_distribution}
\end{figure}

\autoref{fig:supporting_kmer_distribution} shows the local distribution of each $k$-mer type. For a given $k$-mer type, we count how many $k$-mers of the same type exist within a distance of $r$ and then take the average. In \autoref{fig:supporting_kmer_distribution}, we calculated the distribution for the 16 possible 2-mer types in eight different viral families in the NCBI reference data. We can see that different sequences (or species) exhibit different local distributions. For example, \textit{Phenuviridae} exhibits a high local average frequency of TCs, whereas they show a very low local average frequency of CG. These local distributions are not taken into account in the other existing $k$-mer-based methods, which motivates the use of $k$-mer topology for a more detailed analysis of this local structural information.

\section{Persistent Laplacian improves persistent homology in phylogenetic analysis}

\begin{figure}[H]
	\centering
	\includegraphics[width = 0.95\textwidth]{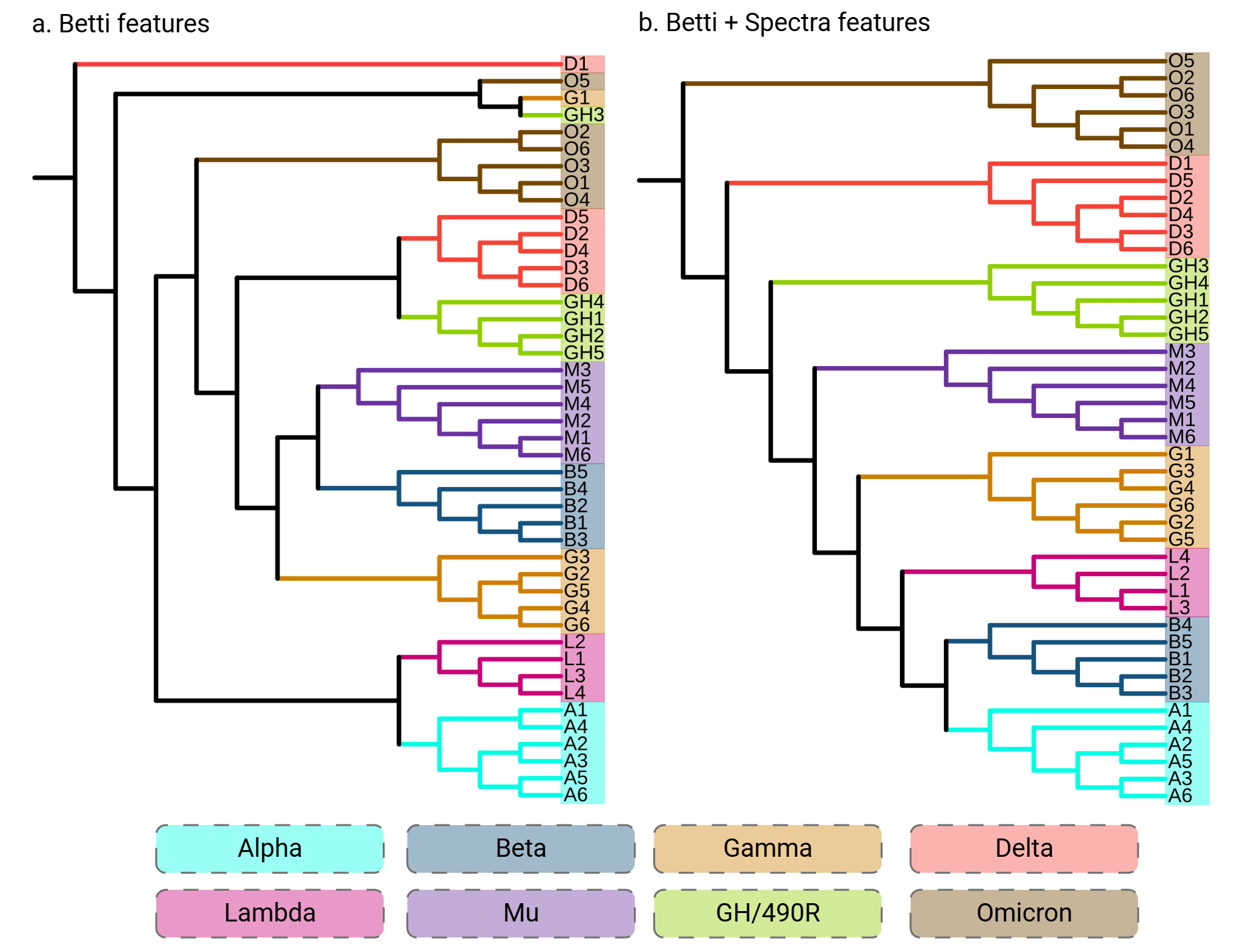}
	\caption{Comparison of persistent Laplacian-based and persistent homology-based phylogenetic trees of whole SARS-CoV-2 genomes. The clades are colored according to the SARS-CoV-2 variants, and the labels are the accession ID from GISAID. (a)  Phylogenetic tree generated from persistent homology. Four sequences—D1, O5, GH3, and G1—are misclassified. All other sequences are clustered within their respective clades. (b) Phylogenetic tree generated from persistent Laplacians, including the Betti numbers and the smallest non-zero eigenvalue. All sequences are corrected clustered.}
	\label{fig:supporting_PhyloLaplacian}
\end{figure}

Persistent homology provides persistent Betti numbers \cite{carlsson2009topology,lum2013extracting} that can facilitate the first term in our topological genetic distance, Eq. (\ref{TGD}). The $k$-mer topology utilizes only persistent Betti numbers is termed persistent homology-based $k$-mer topology. However, persistent homology has many limitations \cite{wang2020persistent,chen2023persistent}. Persistent Laplacians outperform persistent homology in protein engineering\cite{qiu2023persistent}. Therefore, the $k$-mer topology is proposed based on persistent Laplacians and is referred to the persistent Laplacian-based $k$-mer topology. In this case, both persistent Betti numbers and non-harmonic persistent eigenvalues are used.  
For most datasets studied in this work, the persistent homology-based $k$-mer topology works well. However, the phylogenetic analysis of SARS-CoV-2 variants is a very challenging case, for which all other existing methods do not work well. Figure \ref{fig:supporting_PhyloLaplacian} shows that the persistent homology-based $k$-mer topology encounters difficulty for the phylogenetic analysis of SARS-CoV-2 variants. However, the persistent Laplacian-based $k$-mer topology improves the phylogenetic analysis of the persistent homology-based $k$-mer topology and achieves fully correct phylogenetic genetic clustering.     

\section{Further analysis of persistent Laplacian on DNA sequences}

\begin{figure}[H]
	\centering
	\includegraphics[width = \textwidth]{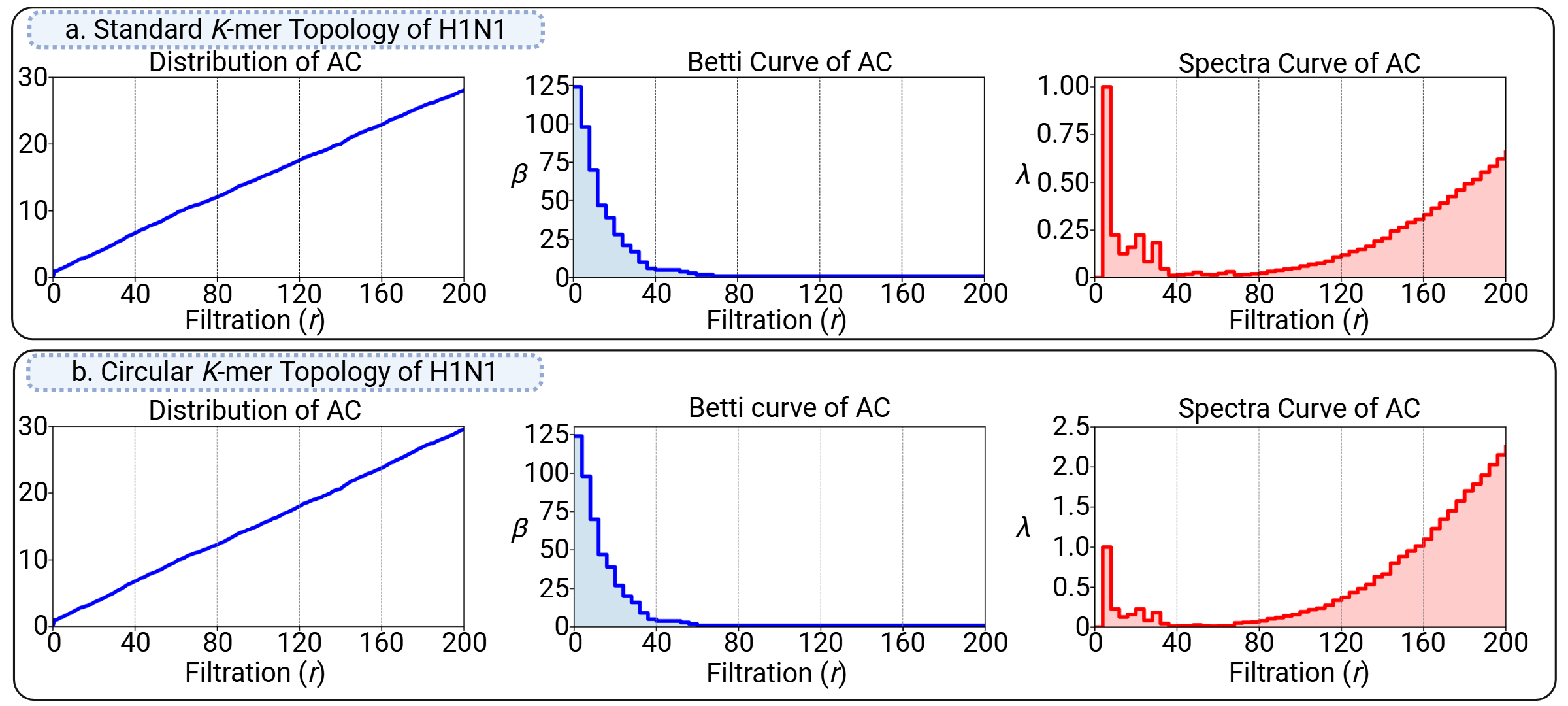}
	\caption{Comparison of the standard $k$-mer topology (a) and circular  $k$-mer topology (b) using the H1N1 HA gene sequence. The left figure shows the distribution of 2-mers AC across the filtration radius. Middle figure shows the Betti-0 curve of AC. The right figure shows the smallest nonzero eigenvalue across different filtration radius.}
	\label{fig:supporting_ciruclar_vs_standard}
\end{figure}

\autoref{fig:supporting_ciruclar_vs_standard} shows the comparison between H1N1 using the standard persistent Laplacian (\autoref{fig:supporting_ciruclar_vs_standard}(a)) and the circular persistent Laplacian (\autoref{fig:supporting_ciruclar_vs_standard}(b)). In the circular Laplacian, the 5' and 3' ends are glued together to form a loop. A 2-mer AC was used to compare the two methods.

Not surprisingly, when we consider the sequence as a circle, the local AC count increases. Additionally, Betti-0 decreases faster for the circular sequence. Most interestingly, we observe a significant difference between the spectral curves. For the standard case, the spectral curve gradually increases from $r = 40$, while in the circular case the growth is much faster, nearly tripling the value by $r = 200$. In the circular case, the connectivity of the sequence is much higher, as the 5' end is connected with the 3' end.

\section{Further details on the codes used in comparisons}\label{sec:supporting_code}
This work reproduced all the codes of the methods used in the comparisons. 
Specifically, the natural vector method (NVM) was reproduced on the basis of \cite{sun2021geometric}.
Jensen-Shannon (JS) and Kullback-Leibler (KL) divergences were calculated by first calculating the feature frequency profiles (FFPs), which are the normalized k-mer counts. The JS and KL divergences are calculated by treating the FFPs as a probability distribution. More details can be found in \cite{jun2010whole, vinga2003alignment,  sims2009alignment}.
Markov K-String method was reproduced on the basis of \cite{qi2004whole}.
The code for the Fourier power spectrum (FPS) was given in the original work\cite{hoang2015new} and is available on the Mathworks'  website \url{ http://www.mathworks.com/matlabcentral/fileexchange/49026m}. The codes we reproduced (i.e., NVM,  FFP-JS,  FFP-KL,  FPS) can be found at \url{https://github.com/hozumiyu/PhyloBenchmark}


\section{Classification data}

\autoref{tab:supporting_classification_data} shows the overview of the dataset used for our viral classification task. For NCBI 2020 and NCBI 2022, we utilized the dataset from Sun et al. \cite{sun2021geometric} and Yu et al. \cite{yu2024optimal}, respectively. Since some reference sequences from NCBI 2020 and NCBI 2022 were removed from the NCBI viral database, we removed these sequences. Additionally, some viral sequences have different taxonomy from the most recent classification, but we utilized the original taxonomy from the studies. For the NCBI 2024 dataset, we obtained the sequences from the NCBI viral database, which was accessed on January 20, 2024. For all the dataset, we removed any viral family which only had 1 reference sequence. 

\begin{table}[H]
	\centering
	\caption{Dataset, NCBI collection date, proprcessing procedure, number of families and number of sequences}
	\begin{tabular}{c| c |l c c } \hline
		Name & NCBI date & \multicolumn{1}{c}{Removed sequence} & \# families & \# sequence\\ \hline
		\multirow{3}{*}{NCBI 2020 \cite{sun2021geometric}} & \multirow{3}{*}{03/2020}  &  Unknown Baltimore class & \multirow{3}{*}{83} & \multirow{3}{*}{6,993} \\ 
		& & Unknown family & & \\
		& & families $<$2 sequence & & \\\hline
		\multirow{4}{*}{NCBI 2022 \cite{yu2024optimal}} & \multirow{4}{*}{03/2022} & Partial sequence & \multirow{4}{*}{123} & \multirow{4}{*}{11,428} \\
		& & Unknown family & & \\
		& & families $<$2 sequence & & \\
		& & Invalid nucleotides & & \\\hline
		\multirow{5}{*}{NCBI 2024} & \multirow{5}{*}{01/20/2024} & Partial sequence & \multirow{5}{*}{199} & \multirow{5}{*}{12,154} \\
		& & Unknown family & & \\
		& & Only '-viridae' & & \\
		& & families $<$2 sequence & & \\ 
		& & Invalid nucleotide & &\\ \hline
		
		\multirow{3}{*}{NCBI 2024 full} & \multirow{3}{*}{01/20/2024} & Partial sequence & \multirow{3}{*}{209} & \multirow{3}{*}{13,645} \\
		& & Unknown family & & \\
		& & families $<$2 sequence & & \\\hline
	\end{tabular}
	\label{tab:supporting_classification_data}
\end{table}

\afterpage{%
	\clearpage
	\begin{landscape}
		\begin{table}
			\centering
			\caption{Viral family name, number of sequence, minimum, maximum and average sequence length, and average $N_A$, $N_C$, $N_G$, $N_T$}
			\begin{tabular}{|l | c c c c c c c c |} \hline
				Family & \# sequence & Min length & Max length & Avg length & Avg $N_A$ & Avg $N_C$ & Avg $N_G$ & Avg $N_T$ \\\hline
				Ackermannviridae & 64 & 143249.00 & 167619.00 & 156691.41 & 42614.16 & 35865.44 & 36011.62 & 42200.19\\
				Adenoviridae & 124 & 415.00 & 48395.00 & 31944.69 & 8050.70 & 8074.22 & 8017.19 & 7802.57\\
				Adintoviridae & 2 & 12571.00 & 12754.00 & 12662.50 & 3797.50 & 2582.50 & 2572.50 & 3710.00\\
				Aliceevansviridae & 129 & 30664.00 & 48800.00 & 36318.58 & 12465.62 & 6405.73 & 7710.43 & 9736.81\\
				Aliusviridae & 8 & 13163.00 & 15277.00 & 14150.62 & 4418.38 & 2773.00 & 2861.50 & 4097.75\\
				Alloherpesviridae & 17 & 451.00 & 295146.00 & 135007.88 & 32656.82 & 34600.47 & 34880.06 & 32870.53\\
				Allomimiviridae & 2 & 186217.00 & 668031.00 & 427124.00 & 128975.00 & 84000.00 & 86135.00 & 128014.00\\
				Alphaflexiviridae & 67 & 762.00 & 8831.00 & 6742.87 & 1897.00 & 1938.70 & 1435.27 & 1471.90\\
				Alphasatellitidae & 122 & 619.00 & 1531.00 & 1270.06 & 385.78 & 239.97 & 291.65 & 352.66\\
				Alphatetraviridae & 8 & 2055.00 & 6625.00 & 4027.88 & 919.62 & 1195.38 & 1024.00 & 888.88\\
				Alternaviridae & 24 & 1420.00 & 3594.00 & 2710.62 & 506.08 & 660.92 & 843.17 & 700.46\\
				Amalgaviridae & 23 & 3156.00 & 3478.00 & 3389.30 & 865.65 & 721.48 & 918.78 & 883.39\\
				Amesuviridae & 3 & 2707.00 & 3442.00 & 2952.00 & 764.00 & 597.00 & 747.33 & 843.67\\
				Amnoonviridae & 10 & 465.00 & 1641.00 & 1032.00 & 260.60 & 263.10 & 217.10 & 291.20\\
				Ampullaviridae & 3 & 22609.00 & 28489.00 & 24970.00 & 8106.67 & 4104.00 & 4158.00 & 8601.33\\
				Anelloviridae & 174 & 1744.00 & 3907.00 & 2739.02 & 860.40 & 668.24 & 601.16 & 609.22\\
				Arenaviridae & 131 & 1924.00 & 7417.00 & 5027.60 & 1436.51 & 1088.17 & 952.91 & 1550.02\\
				Arteriviridae & 25 & 12704.00 & 15728.00 & 14978.60 & 3342.24 & 4098.16 & 3634.80 & 3903.40\\
				Artoviridae & 8 & 6367.00 & 12395.00 & 10072.50 & 2959.00 & 2227.88 & 2130.12 & 2755.50\\
				Ascoviridae & 8 & 741.00 & 199721.00 & 129876.00 & 35897.38 & 28953.75 & 29567.75 & 35457.12\\
				Asfarviridae & 20 & 170101.00 & 193886.00 & 184524.00 & 56550.50 & 35563.90 & 35537.35 & 56872.25\\
				Aspiviridae & 14 & 136.00 & 8182.00 & 3002.50 & 823.86 & 613.07 & 449.71 & 1115.86\\
				Assiduviridae & 3 & 53385.00 & 57447.00 & 54832.00 & 21130.00 & 7363.33 & 9868.00 & 16470.67\\
				Astroviridae & 59 & 2403.00 & 7722.00 & 5978.46 & 1621.12 & 1416.78 & 1453.34 & 1487.22\\
				Atkinsviridae & 91 & 3364.00 & 4535.00 & 3895.51 & 919.65 & 947.07 & 975.75 & 1053.04\\
				Autographiviridae & 385 & 10313.00 & 47868.00 & 41458.66 & 10671.61 & 10054.84 & 10850.82 & 9881.40\\
				Autolykiviridae & 5 & 10046.00 & 10636.00 & 10428.80 & 3035.00 & 2268.20 & 2251.60 & 2874.00\\
				Avsunviroidae & 5 & 247.00 & 434.00 & 350.00 & 79.40 & 89.20 & 90.40 & 91.00\\
				Bacilladnaviridae & 21 & 3492.00 & 6000.00 & 4955.76 & 1285.00 & 1164.71 & 1067.43 & 1438.62\\
				Baculoviridae & 103 & 1182.00 & 178733.00 & 127394.97 & 37115.49 & 26505.71 & 26608.93 & 37164.84\\
				Benyviridae & 15 & 1320.00 & 7038.00 & 4319.27 & 1076.07 & 782.27 & 1103.60 & 1357.33\\
				Betaflexiviridae & 151 & 276.00 & 9654.00 & 7604.56 & 2223.48 & 1457.15 & 1847.44 & 2076.48\\
				Bicaudaviridae & 8 & 48774.00 & 76107.00 & 62491.50 & 19691.50 & 12002.00 & 11801.88 & 18996.12\\
				Bidnaviridae & 4 & 6022.00 & 6543.00 & 6283.00 & 2263.25 & 915.75 & 968.50 & 2135.50\\
				Birnaviridae & 29 & 2605.00 & 3579.00 & 3076.34 & 911.93 & 838.45 & 751.72 & 574.24\\ \hline
			\end{tabular}
			\label{tab:supporting_family_stat1}
		\end{table}
	\end{landscape}
	\clearpage
}
\afterpage{%
	\clearpage
	\begin{landscape}
		\begin{table}
			\centering
			\caption{Viral family name, number of sequence, minimum, maximum and average sequence length, and average $N_A$, $N_C$, $N_G$, $N_T$}
			\begin{tabular}{|l | c c c c c c c c |} \hline
				Family & \# sequence & Min length & Max length & Avg length & Avg $N_A$ & Avg $N_C$ & Avg $N_G$ & Avg $N_T$ \\ \hline
				Blumeviridae & 34 & 3612.00 & 4744.00 & 4247.59 & 1104.85 & 959.65 & 994.06 & 1189.03\\
				Bornaviridae & 19 & 5572.00 & 9006.00 & 8730.74 & 2493.58 & 1864.32 & 1933.11 & 2439.74\\
				Botourmiaviridae & 165 & 958.00 & 5234.00 & 2607.88 & 602.54 & 623.37 & 723.52 & 658.45\\
				Bromoviridae & 125 & 1938.00 & 3644.00 & 2791.70 & 754.28 & 557.18 & 663.09 & 817.16\\
				Caliciviridae & 55 & 6434.00 & 8513.00 & 7633.36 & 1955.00 & 1995.89 & 1882.18 & 1800.29\\
				Casjensviridae & 60 & 54417.00 & 63971.00 & 59167.30 & 12699.53 & 16831.58 & 16886.03 & 12750.15\\
				Caulimoviridae & 111 & 545.00 & 13221.00 & 7557.92 & 2622.24 & 1483.11 & 1622.64 & 1829.93\\
				Chaacviridae & 3 & 10198.00 & 10798.00 & 10590.33 & 3120.33 & 2099.00 & 2131.00 & 3240.00\\
				Chaseviridae & 30 & 50725.00 & 57429.00 & 53436.43 & 14847.23 & 11713.60 & 12712.93 & 14162.67\\
				Chrysoviridae & 124 & 646.00 & 4220.00 & 2993.94 & 797.73 & 654.73 & 868.04 & 673.44\\
				Chuviridae & 52 & 1566.00 & 12996.00 & 8314.12 & 2338.31 & 1877.65 & 1924.63 & 2173.52\\
				Circoviridae & 283 & 648.00 & 4706.00 & 2075.54 & 558.73 & 463.52 & 486.42 & 566.87\\
				Closteroviridae & 80 & 555.00 & 19296.00 & 12470.34 & 3612.10 & 2363.55 & 2832.50 & 3662.19\\
				Coronaviridae & 74 & 25423.00 & 36549.00 & 28649.42 & 7734.05 & 5136.28 & 6183.97 & 9595.11\\
				Corticoviridae & 2 & 10079.00 & 10584.00 & 10331.50 & 3266.00 & 1876.00 & 2466.50 & 2723.00\\
				Cremegaviridae & 2 & 14939.00 & 17738.00 & 16338.50 & 4477.50 & 3937.50 & 3404.00 & 4519.50\\
				Crevaviridae & 4 & 83412.00 & 95815.00 & 91473.00 & 31640.50 & 13642.75 & 14205.25 & 31984.50\\
				Cruliviridae & 9 & 799.00 & 6691.00 & 3675.33 & 1189.33 & 681.00 & 757.44 & 1047.56\\
				Curvulaviridae & 16 & 1634.00 & 2383.00 & 2027.12 & 470.00 & 543.12 & 581.94 & 432.06\\
				Cystoviridae & 21 & 2322.00 & 7051.00 & 4507.38 & 961.29 & 1264.67 & 1248.33 & 1033.10\\
				Deltaflexiviridae & 4 & 6735.00 & 8327.00 & 7871.50 & 1521.00 & 2301.25 & 1804.00 & 2245.25\\
				Demerecviridae & 101 & 14504.00 & 128602.00 & 113646.65 & 34213.77 & 22765.17 & 22759.50 & 33908.22\\
				Dicistroviridae & 28 & 7835.00 & 10436.00 & 9329.18 & 2832.18 & 1725.96 & 1923.82 & 2847.21\\
				Discoviridae & 15 & 1091.00 & 6519.00 & 2913.00 & 822.00 & 632.00 & 605.80 & 853.20\\
				Drexlerviridae & 119 & 37655.00 & 54438.00 & 49347.62 & 13424.13 & 11270.55 & 11725.51 & 12927.43\\
				Druskaviridae & 3 & 102105.00 & 103257.00 & 102560.33 & 21843.67 & 29221.00 & 29859.33 & 21636.33\\
				Duinviridae & 6 & 3543.00 & 4458.00 & 3888.17 & 1056.50 & 856.00 & 857.67 & 1118.00\\
				Duneviridae & 6 & 39290.00 & 46976.00 & 43646.83 & 13859.67 & 7272.33 & 6520.83 & 15994.00\\
				Endornaviridae & 40 & 9636.00 & 23635.00 & 14493.67 & 4893.88 & 2786.15 & 3240.93 & 3572.72\\
				Euroniviridae & 3 & 24648.00 & 29384.00 & 26283.33 & 8126.00 & 5842.00 & 5691.33 & 6624.00\\
				Fiersviridae & 302 & 3211.00 & 5067.00 & 3793.80 & 896.41 & 941.81 & 971.92 & 983.67\\
				Filoviridae & 16 & 13065.00 & 19114.00 & 17376.69 & 5269.31 & 3885.31 & 3696.38 & 4525.69\\
				Fimoviridae & 159 & 980.00 & 7291.00 & 2512.94 & 853.79 & 395.53 & 380.10 & 883.52\\
				Flaviviridae & 169 & 1011.00 & 22780.00 & 9855.40 & 2614.54 & 2242.76 & 2723.79 & 2274.31\\
				Forsetiviridae & 2 & 43978.00 & 47186.00 & 45582.00 & 18538.00 & 5287.00 & 7861.50 & 13895.50\\ \hline
			\end{tabular}
			\label{tab:supporting_family_stat2}
		\end{table}
	\end{landscape}
	\clearpage
}

\afterpage{%
	\clearpage
	\begin{landscape}
		\begin{table}
			\centering
			\caption{Viral family name, number of sequence, minimum, maximum and average sequence length, and average $N_A$, $N_C$, $N_G$, $N_T$}
			\begin{tabular}{|l | c c c c c c c c |} \hline
				Family & \# sequence & Min length & Max length & Avg length & Avg $N_A$ & Avg $N_C$ & Avg $N_G$ & Avg $N_T$ \\ \hline
				Fredfastierviridae & 3 & 43365.00 & 43783.00 & 43546.00 & 11601.67 & 9808.00 & 9856.67 & 12279.67\\
				Fusariviridae & 38 & 3253.00 & 10776.00 & 6847.63 & 1832.16 & 1504.05 & 1641.18 & 1870.24\\
				Fuselloviridae & 10 & 14796.00 & 24186.00 & 16690.30 & 5161.00 & 3134.30 & 3384.70 & 5010.30\\
				Gammaflexiviridae & 4 & 6827.00 & 9525.00 & 8223.50 & 1773.75 & 3008.75 & 1702.25 & 1738.75\\
				Geminiviridae & 816 & 2383.00 & 3763.00 & 2701.93 & 704.61 & 556.02 & 630.00 & 811.30\\
				Genomoviridae & 254 & 1309.00 & 2826.00 & 2190.00 & 550.62 & 594.52 & 532.52 & 512.35\\
				Globuloviridae & 4 & 18212.00 & 28332.00 & 21530.50 & 5177.00 & 4946.75 & 5759.00 & 5647.75\\
				Graaviviridae & 2 & 39784.00 & 42271.00 & 41027.50 & 7176.00 & 13421.00 & 13244.00 & 7186.50\\
				Grimontviridae & 10 & 75106.00 & 112127.00 & 89328.00 & 26681.40 & 18199.10 & 17223.20 & 27224.30\\
				Guelinviridae & 10 & 16747.00 & 19089.00 & 18297.50 & 6461.20 & 2699.30 & 2991.20 & 6145.80\\
				Hadakaviridae & 11 & 859.00 & 2539.00 & 1394.00 & 362.36 & 277.36 & 375.27 & 379.00\\
				Hafunaviridae & 12 & 56593.00 & 77672.00 & 71940.17 & 14879.92 & 20533.17 & 20770.33 & 15756.75\\
				Haloferuviridae & 3 & 35722.00 & 38059.00 & 36794.67 & 8789.00 & 10094.00 & 11266.33 & 6645.33\\
				Hantaviridae & 147 & 324.00 & 6761.00 & 3772.86 & 1213.63 & 638.73 & 807.94 & 1112.56\\
				Hepadnaviridae & 26 & 3018.00 & 3542.00 & 3189.00 & 808.35 & 818.15 & 681.96 & 880.54\\
				Hepeviridae & 13 & 6555.00 & 7312.00 & 6955.77 & 1444.23 & 1958.08 & 1796.38 & 1757.08\\
				Herelleviridae & 137 & 203.00 & 167431.00 & 138558.36 & 46349.77 & 23727.21 & 26266.80 & 42214.58\\
				Hypoviridae & 40 & 7381.00 & 18371.00 & 12286.15 & 3218.18 & 2941.85 & 3035.75 & 3090.38\\
				Hytrosaviridae & 2 & 124279.00 & 190032.00 & 157155.50 & 52546.00 & 26344.00 & 27304.00 & 50961.50\\
				Iflaviridae & 46 & 8832.00 & 10984.00 & 9939.28 & 2954.98 & 1755.46 & 2138.98 & 3089.87\\
				Inoviridae & 74 & 4772.00 & 10638.00 & 7006.72 & 1571.51 & 1721.42 & 1828.30 & 1885.49\\
				Intestiviridae & 18 & 89781.00 & 98001.00 & 93488.00 & 32712.94 & 13910.39 & 14324.56 & 32540.11\\
				Iridoviridae & 29 & 1392.00 & 288858.00 & 156330.21 & 48320.34 & 30163.24 & 30486.34 & 47360.28\\
				Kitaviridae & 29 & 1724.00 & 8893.00 & 5135.69 & 1319.86 & 987.28 & 1149.48 & 1679.07\\
				Kleczkowskaviridae & 2 & 206821.00 & 207623.00 & 207222.00 & 61537.50 & 40014.50 & 45177.00 & 60493.00\\
				Kolmioviridae & 15 & 1547.00 & 1735.00 & 1663.87 & 355.87 & 480.47 & 463.87 & 363.67\\
				Kyanoviridae & 76 & 144311.00 & 252401.00 & 185396.74 & 56901.86 & 35163.88 & 39096.28 & 54234.72\\
				Lavidaviridae & 6 & 17276.00 & 29767.00 & 22079.83 & 7568.17 & 3639.33 & 3706.00 & 7166.33\\
				Leishbuviridae & 3 & 661.00 & 5981.00 & 2607.33 & 834.33 & 485.33 & 525.67 & 762.00\\
				Leisingerviridae & 2 & 26111.00 & 31007.00 & 28559.00 & 8541.00 & 5858.00 & 7279.50 & 6880.50\\
				Lipothrixviridae & 10 & 31324.00 & 41171.00 & 37568.10 & 11729.40 & 7071.30 & 6635.40 & 12132.00\\
				Lispiviridae & 30 & 6450.00 & 15623.00 & 11702.17 & 3870.67 & 2162.03 & 2244.40 & 3425.07\\
				Malacoherpesviridae & 2 & 207439.00 & 211518.00 & 209478.50 & 60403.00 & 44491.00 & 45137.50 & 59447.00\\
				Marnaviridae & 21 & 6360.00 & 9562.00 & 8794.62 & 2442.52 & 1824.29 & 1916.95 & 2610.86\\
				Marseilleviridae & 14 & 3538.00 & 380011.00 & 225563.14 & 63226.64 & 49179.79 & 49347.43 & 63809.29\\ \hline
			\end{tabular}
			\label{tab:supporting_family_stat3}
		\end{table}
	\end{landscape}
	\clearpage
}

\afterpage{%
	\clearpage
	\begin{landscape}
		\begin{table}
			\centering
			\caption{Viral family name, number of sequence, minimum, maximum and average sequence length, and average $N_A$, $N_C$, $N_G$, $N_T$}
			\begin{tabular}{|l | c c c c c c c c |} \hline
				Family & \# sequence & Min length & Max length & Avg length & Avg $N_A$ & Avg $N_C$ & Avg $N_G$ & Avg $N_T$ \\ \hline
				Matonaviridae & 4 & 9621.00 & 9762.00 & 9693.75 & 1516.75 & 3502.75 & 3133.75 & 1540.50\\
				Matshushitaviridae & 2 & 17036.00 & 19604.00 & 18320.00 & 2962.50 & 5869.50 & 6381.50 & 3106.50\\
				Mayoviridae & 10 & 2231.00 & 6228.00 & 4078.60 & 1066.50 & 810.10 & 989.40 & 1212.60\\
				Medioniviridae & 2 & 20268.00 & 25068.00 & 22668.00 & 5965.00 & 6561.00 & 4558.50 & 5583.50\\
				Megabirnaviridae & 6 & 7180.00 & 8985.00 & 8294.83 & 1875.00 & 1939.00 & 2399.00 & 2081.83\\
				Mesoniviridae & 15 & 18939.00 & 20949.00 & 20077.87 & 6765.53 & 4325.13 & 3129.73 & 5857.47\\
				Mesyanzhinovviridae & 18 & 47545.00 & 64096.00 & 59816.94 & 11646.83 & 19296.67 & 18980.17 & 9893.28\\
				Metaviridae & 2 & 7396.00 & 7510.00 & 7453.00 & 2427.50 & 1777.50 & 1549.00 & 1699.00\\
				Microviridae & 62 & 4129.00 & 6478.00 & 5177.18 & 1386.40 & 1052.24 & 1140.34 & 1598.19\\
				Mimiviridae & 16 & 73689.00 & 1572370.00 & 1155047.19 & 425426.31 & 151712.38 & 153476.62 & 424431.88\\
				Mitoviridae & 117 & 2148.00 & 4955.00 & 2846.35 & 877.00 & 545.36 & 578.22 & 845.77\\
				Molycolviridae & 2 & 124169.00 & 124692.00 & 124430.50 & 41647.50 & 20790.00 & 24332.00 & 37661.00\\
				Mymonaviridae & 38 & 3744.00 & 11563.00 & 8184.76 & 2191.97 & 1847.97 & 1875.03 & 2269.79\\
				Mypoviridae & 3 & 2579.00 & 9874.00 & 5325.67 & 1838.67 & 893.00 & 1027.00 & 1567.00\\
				Nairoviridae & 147 & 1443.00 & 14854.00 & 6308.70 & 2036.27 & 1260.44 & 1401.37 & 1610.62\\
				Nanoviridae & 114 & 957.00 & 1111.00 & 1005.55 & 303.89 & 162.66 & 231.63 & 307.37\\
				Narnaviridae & 17 & 1455.00 & 9885.00 & 3557.00 & 928.59 & 794.71 & 848.12 & 985.59\\
				Naryaviridae & 3 & 1788.00 & 2085.00 & 1947.67 & 639.67 & 337.67 & 338.00 & 632.33\\
				Nenyaviridae & 5 & 1661.00 & 2931.00 & 1998.80 & 593.40 & 374.40 & 437.00 & 594.00\\
				Nimaviridae & 2 & 305119.00 & 309286.00 & 307202.50 & 92663.00 & 62846.00 & 62983.50 & 88710.00\\
				Nodaviridae & 30 & 1175.00 & 3628.00 & 2409.70 & 614.90 & 642.13 & 593.03 & 559.63\\
				Nudiviridae & 14 & 96944.00 & 231621.00 & 147162.79 & 48097.00 & 25620.21 & 25554.64 & 47890.93\\
				Nyamiviridae & 23 & 4708.00 & 13295.00 & 9909.87 & 2704.57 & 2384.57 & 2391.61 & 2429.13\\
				Orlajensenviridae & 11 & 17049.00 & 17510.00 & 17397.00 & 2918.36 & 6091.91 & 5849.09 & 2537.64\\
				Orthoherpesviridae & 136 & 165.00 & 241087.00 & 134009.69 & 30292.34 & 37057.08 & 36561.10 & 30099.18\\
				Orthomyxoviridae & 191 & 508.00 & 2427.00 & 1730.04 & 565.30 & 340.38 & 419.36 & 404.99\\
				Pachyviridae & 5 & 71443.00 & 78833.00 & 74320.00 & 27252.40 & 11124.40 & 12844.80 & 23098.40\\
				Papillomaviridae & 208 & 5748.00 & 8809.00 & 7597.20 & 2206.78 & 1550.24 & 1790.64 & 2049.54\\
				Paramyxoviridae & 92 & 14796.00 & 21523.00 & 16270.03 & 5203.63 & 3331.50 & 3421.03 & 4313.87\\
				Partitiviridae & 162 & 1186.00 & 2499.00 & 1890.64 & 507.31 & 518.17 & 346.98 & 518.18\\
				Parvoviridae & 227 & 3411.00 & 6334.00 & 4916.39 & 1541.87 & 1081.06 & 1094.38 & 1199.08\\
				Paulinoviridae & 2 & 5688.00 & 5804.00 & 5746.00 & 804.50 & 1541.00 & 1975.00 & 1425.50\\
				Peduoviridae & 108 & 18281.00 & 40555.00 & 33441.85 & 7988.25 & 8852.69 & 9001.21 & 7599.69\\
				Peribunyaviridae & 451 & 585.00 & 8905.00 & 4090.36 & 1495.58 & 652.55 & 761.91 & 1180.32\\
				Permutotetraviridae & 3 & 2482.00 & 5698.00 & 4011.33 & 997.67 & 929.67 & 1128.33 & 955.67\\ \hline
			\end{tabular}
			\label{tab:supporting_family_stat4}
		\end{table}
	\end{landscape}
	\clearpage
}

\afterpage{%
	\clearpage
	\begin{landscape}
		\begin{table}
			\centering
			\caption{Viral family name, number of sequence, minimum, maximum and average sequence length, and average $N_A$, $N_C$, $N_G$, $N_T$}
			\begin{tabular}{|l | c c c c c c c c |} \hline
				Family & \# sequence & Min length & Max length & Avg length & Avg $N_A$ & Avg $N_C$ & Avg $N_G$ & Avg $N_T$ \\ \hline
				Pervagoviridae & 2 & 72534.00 & 72979.00 & 72756.50 & 22550.00 & 13838.50 & 13868.50 & 22499.50\\
				Phasmaviridae & 97 & 1040.00 & 7740.00 & 3818.56 & 1338.51 & 625.15 & 765.16 & 1089.73\\
				Phenuiviridae & 450 & 461.00 & 9760.00 & 3941.68 & 1214.75 & 766.00 & 869.07 & 1091.85\\
				Phycodnaviridae & 59 & 285.00 & 1473473.00 & 166979.15 & 51369.73 & 32096.90 & 32042.61 & 51469.92\\
				Picobirnaviridae & 15 & 1688.00 & 2666.00 & 2078.93 & 594.67 & 438.00 & 491.00 & 555.27\\
				Picornaviridae & 203 & 2086.00 & 10101.00 & 7799.63 & 2069.96 & 1802.47 & 1761.30 & 2165.90\\
				Plectroviridae & 6 & 4491.00 & 8273.00 & 7184.17 & 2719.50 & 652.83 & 1063.00 & 2748.83\\
				Pleolipoviridae & 16 & 7048.00 & 16992.00 & 11349.50 & 2556.75 & 3051.56 & 3150.88 & 2590.31\\
				Pneumoviridae & 10 & 13350.00 & 15225.00 & 14729.60 & 5300.40 & 2687.80 & 2704.00 & 4037.40\\
				Polycipiviridae & 9 & 10315.00 & 12155.00 & 11498.56 & 3680.00 & 2120.33 & 2220.22 & 3478.00\\
				Polydnaviriformidae & 346 & 263.00 & 140906.00 & 8701.72 & 2796.08 & 1567.32 & 1568.05 & 2770.28\\
				Polymycoviridae & 54 & 890.00 & 2470.00 & 1792.56 & 338.17 & 542.13 & 525.15 & 387.11\\
				Polyomaviridae & 142 & 3962.00 & 14334.00 & 5217.19 & 1502.44 & 1114.36 & 1091.63 & 1508.76\\
				Pootjesviridae & 10 & 143349.00 & 158568.00 & 153196.80 & 39322.80 & 37903.80 & 37574.90 & 38395.30\\
				Portogloboviridae & 2 & 20222.00 & 20424.00 & 20323.00 & 6268.00 & 3835.50 & 3974.50 & 6245.00\\
				Pospiviroidae & 41 & 246.00 & 396.00 & 329.27 & 66.34 & 95.44 & 94.83 & 72.66\\
				Potyviridae & 244 & 1103.00 & 11519.00 & 7962.33 & 2547.37 & 1510.59 & 1848.71 & 2055.66\\
				Poxviridae & 77 & 296.00 & 359853.00 & 148718.92 & 49760.71 & 24699.12 & 24727.68 & 49531.42\\
				Qinviridae & 16 & 1601.00 & 6585.00 & 3862.88 & 1007.06 & 978.75 & 998.25 & 878.81\\
				Quadriviridae & 4 & 3685.00 & 4942.00 & 4269.50 & 1180.25 & 1032.50 & 1222.25 & 834.50\\
				Retroviridae & 93 & 266.00 & 13246.00 & 8259.47 & 2455.45 & 1917.72 & 1837.30 & 2049.00\\
				Rhabdoviridae & 352 & 993.00 & 16133.00 & 11516.48 & 3657.11 & 2240.64 & 2506.55 & 3112.18\\
				Roniviridae & 4 & 26253.00 & 29110.00 & 27197.75 & 7523.75 & 7310.25 & 4804.50 & 7559.25\\
				Rountreeviridae & 39 & 16687.00 & 18899.00 & 17991.59 & 6121.13 & 2832.90 & 2759.10 & 6278.46\\
				Rudiviridae & 19 & 20269.00 & 36493.00 & 31298.68 & 11373.37 & 4428.95 & 4464.95 & 11031.42\\
				Salasmaviridae & 34 & 18379.00 & 28950.00 & 22618.79 & 7468.91 & 3849.68 & 3764.50 & 7535.71\\
				Saparoviridae & 2 & 52643.00 & 54291.00 & 53467.00 & 9617.00 & 17623.50 & 18398.00 & 7828.50\\
				Sarthroviridae & 8 & 502.00 & 872.00 & 766.88 & 226.50 & 174.00 & 158.12 & 208.25\\
				Schitoviridae & 115 & 59080.00 & 103910.00 & 73818.29 & 19903.55 & 17375.44 & 17138.30 & 19400.99\\
				Schizomimiviridae & 2 & 370920.00 & 1421182.00 & 896051.00 & 340731.50 & 107230.00 & 108384.50 & 339705.00\\
				Secoviridae & 197 & 229.00 & 13198.00 & 5928.11 & 1574.44 & 1179.27 & 1425.53 & 1748.87\\
				Sedoreoviridae & 458 & 528.00 & 5792.00 & 1865.13 & 586.25 & 339.33 & 434.15 & 505.40\\
				Simuloviridae & 3 & 16492.00 & 18925.00 & 17535.33 & 3580.67 & 5362.67 & 5671.00 & 2921.00\\
				Sinhaliviridae & 9 & 5877.00 & 5991.00 & 5910.89 & 1152.33 & 1632.56 & 1412.44 & 1713.56\\
				Smacoviridae & 88 & 1881.00 & 3028.00 & 2547.85 & 620.78 & 603.68 & 598.12 & 725.26\\ \hline
			\end{tabular}
			\label{tab:supporting_family_stat5}
		\end{table}
	\end{landscape}
	\clearpage
}

\section{Phylogenetic analysis data}\label{sec:phylogenetic_data}
In this section, we present the description of each dataset used for the phylogenetic analysis.
\afterpage{%
	\clearpage
	\begin{landscape}
		\begin{table}[H]
			\centering
			\caption{Accession, group, sequence length of ebola virus data}
			\begin{tabular}{|c | c c c c c c||c | c c c c c c | } \hline
				Accession & Group & Length & $N_A$ & $N_C$ & $N_G$ & $N_T$ & Accession & Group & Length & $N_A$ & $N_C$ & $N_G$ & $N_T$ \\ \hline
				FJ217161.1 & BDBV & 18940 & 5964 & 4324 & 3632 & 5020 & KC545393.1 & BDBV & 18939 & 5974 & 4293 & 3636 & 5036\\
				KC545395.1 & BDBV & 18939 & 5975 & 4290 & 3635 & 5039 & KC545394.1 & BDBV & 18939 & 5974 & 4292 & 3637 & 5036\\
				KC545396.1 & BDBV & 18939 & 5975 & 4293 & 3635 & 5036 & FJ217162.1 & TAFV & 18935 & 6020 & 4371 & 3630 & 4914\\
				AF522874.1 & RESTV & 18891 & 5937 & 3929 & 3746 & 5279 & AB050936.1 & RESTV & 18890 & 5927 & 3924 & 3762 & 5277\\
				JX477166.1 & RESTV & 18891 & 5935 & 3920 & 3755 & 5281 & FJ621585.1 & RESTV & 18796 & 5900 & 3898 & 3747 & 5251\\
				FJ621583.1 & RESTV & 18887 & 5928 & 3929 & 3767 & 5263 & JX477165.1 & RESTV & 18887 & 5924 & 3929 & 3774 & 5260\\
				FJ968794.1 & SUDV & 18875 & 5905 & 4034 & 3756 & 5180 & KC242783.2 & SUDV & 18875 & 5911 & 4028 & 3750 & 5186\\
				EU338380.1 & SUDV & 18875 & 5914 & 4032 & 3750 & 5179 & AY729654.1 & SUDV & 18875 & 5920 & 4071 & 3732 & 5152\\
				JN638998.1 & SUDV & 18875 & 5931 & 4059 & 3729 & 5156 & KC545389.1 & SUDV & 18874 & 5924 & 4080 & 3731 & 5139\\
				KC545390.1 & SUDV & 18874 & 5925 & 4080 & 3730 & 5139 & KC545391.1 & SUDV & 18874 & 5924 & 4080 & 3731 & 5139\\
				KC545392.1 & SUDV & 18874 & 5923 & 4081 & 3732 & 5138 & KC589025.1 & SUDV & 18875 & 5921 & 4047 & 3734 & 5173\\
				KC242801.1 & EBOV & 18959 & 6061 & 4037 & 3752 & 5109 & NC$\_$002549.1 & EBOV & 18959 & 6061 & 4035 & 3752 & 5111\\
				KC242791.1 & EBOV & 18959 & 6061 & 4037 & 3752 & 5109 & KC242792.1 & EBOV & 18959 & 6047 & 4052 & 3756 & 5104\\
				KC242793.1 & EBOV & 18958 & 6043 & 4052 & 3761 & 5102 & KC242794.1 & EBOV & 18959 & 6039 & 4063 & 3762 & 5095\\
				AY354458.1 & EBOV & 18961 & 6054 & 4051 & 3747 & 5109 & KC242796.1 & EBOV & 18959 & 6055 & 4049 & 3748 & 5107\\
				KC242799.1 & EBOV & 18959 & 6054 & 4050 & 3748 & 5107 & KC242784.1 & EBOV & 18958 & 6061 & 4028 & 3750 & 5119\\
				KC242786.1 & EBOV & 18958 & 6063 & 4025 & 3749 & 5121 & KC242787.1 & EBOV & 18958 & 6062 & 4025 & 3750 & 5121\\
				KC242789.1 & EBOV & 18958 & 6062 & 4023 & 3750 & 5123 & KC242785.1 & EBOV & 18958 & 6060 & 4026 & 3752 & 5120\\
				KC242790.1 & EBOV & 18958 & 6060 & 4025 & 3751 & 5122 & KC242788.1 & EBOV & 18958 & 6063 & 4032 & 3749 & 5114\\
				KC242800.1 & EBOV & 18958 & 6042 & 4052 & 3762 & 5102 & KM034555.1 & EBOV & 18950 & 6049 & 4049 & 3753 & 5099\\
				KM034562.1 & EBOV & 18957 & 6051 & 4050 & 3756 & 5100 & KM233039.1 & EBOV & 18953 & 6052 & 4051 & 3751 & 5099\\
				KM034557.1 & EBOV & 18956 & 6052 & 4052 & 3753 & 5099 & KM034560.1 & EBOV & 18952 & 6050 & 4051 & 3752 & 5099\\
				KM233050.1 & EBOV & 18956 & 6053 & 4050 & 3753 & 5100 & KM233053.1 & EBOV & 18957 & 6053 & 4051 & 3753 & 5100\\
				KM233057.1 & EBOV & 18954 & 6052 & 4052 & 3751 & 5099 & KM233063.1 & EBOV & 18955 & 6053 & 4052 & 3751 & 5099\\
				KM233072.1 & EBOV & 18949 & 6049 & 4049 & 3752 & 5099 & KM233110.1 & EBOV & 18956 & 6053 & 4053 & 3752 & 5098\\
				KM233070.1 & EBOV & 18959 & 6055 & 4052 & 3753 & 5099 & KM233099.1 & EBOV & 18953 & 6052 & 4052 & 3751 & 5098\\
				KM233097.1 & EBOV & 18953 & 6051 & 4052 & 3752 & 5098 & KM233109.1 & EBOV & 18958 & 6053 & 4055 & 3753 & 5097\\
				KM233096.1 & EBOV & 18953 & 6054 & 4049 & 3751 & 5099 & KM233103.1 & EBOV & 18950 & 6051 & 4050 & 3751 & 5098\\
				KJ660346.2 & EBOV & 18959 & 6053 & 4052 & 3755 & 5099 & KJ660347.2 & EBOV & 18959 & 6054 & 4052 & 3754 & 5099\\
				KJ660348.2 & EBOV & 18959 & 6055 & 4051 & 3754 & 5099 &            &      &       &      &      &      &     \\\hline
			\end{tabular}
			\label{tab: ebolavirus}
		\end{table}
	\end{landscape}
	\clearpage
}

\begin{table}[H]
	\centering
	\caption{Accession, group, sequence length of mammalian mitochondria data}
	\begin{tabular}{|c | c c c c c c|} \hline
		Accession & Groups & Length & $N_A$ & $N_C$ & $N_G$ & $N_T$ \\ \hline
		V00662.1 & Primates & 16569 & 5123 & 5176 & 2176 & 4094\\
		D38116.1 & Primates & 16563 & 5189 & 5084 & 2104 & 4186\\
		D38113.1 & Primates & 16554 & 5154 & 5099 & 2133 & 4168\\
		D38114.1 & Primates & 16364 & 5059 & 5022 & 2160 & 4123\\
		X99256.1 & Primates & 16472 & 5039 & 5231 & 2256 & 3946\\
		Y18001.1 & Primates & 16521 & 5195 & 5047 & 2169 & 4110\\
		AY863426.1 & Primates & 16389 & 5243 & 4953 & 2049 & 4137\\
		D38115.1 & Primates & 16389 & 5007 & 5317 & 2168 & 3897\\
		NC$\_$002083.1 & Primates & 16499 & 5031 & 5403 & 2176 & 3889\\
		NC$\_$002764.1 & Primates & 16586 & 5306 & 5027 & 2116 & 4137\\
		U20753.1 & Carnivore & 17009 & 5543 & 4454 & 2406 & 4606\\
		U96639.2 & Carnivore & 16727 & 5290 & 4267 & 2366 & 4804\\
		EU442884.2 & Carnivore & 16774 & 5293 & 4265 & 2398 & 4812\\
		EF551003.1 & Carnivore & 16990 & 5418 & 4513 & 2478 & 4581\\
		EF551002.1 & Carnivore & 16964 & 5397 & 4508 & 2467 & 4592\\
		DQ402478.1 & Carnivore & 16868 & 5270 & 4285 & 2601 & 4712\\
		AF303110.1 & Carnivore & 17020 & 5258 & 4355 & 2676 & 4731\\
		AF303111.1 & Carnivore & 17017 & 5253 & 4346 & 2692 & 4726\\
		EF212882.1 & Carnivore & 16805 & 5338 & 4000 & 2518 & 4949\\
		AJ002189.1 & Artiodactyla & 16680 & 5790 & 4384 & 2210 & 4296\\
		AF010406.1 & Artiodactyla & 16616 & 5594 & 4289 & 2181 & 4552\\
		AF533441.1 & Artiodactyla & 16640 & 5569 & 4313 & 2189 & 4569\\
		V00654.1 & Artiodactyla & 16338 & 5460 & 4237 & 2198 & 4443\\
		AY488491.1 & Artiodactyla & 16355 & 5421 & 4298 & 2261 & 4375\\
		NC$\_$007441.1 & Artiodactyla & 16498 & 5542 & 4358 & 2164 & 4434\\
		NC$\_$008830.1 & Artiodactyla & 16719 & 5786 & 4340 & 2222 & 4371\\
		NC$\_$010640.1 & Artiodactyla & 16524 & 5519 & 4404 & 2205 & 4396\\
		X72204.1 & Cetacea & 16402 & 5374 & 4527 & 2140 & 4361\\
		NC$\_$005268.1 & Cetacea & 16390 & 5354 & 4609 & 2162 & 4265\\
		NC$\_$001321.1 & Cetacea & 16398 & 5359 & 4474 & 2182 & 4383\\
		NC$\_$005270.1 & Cetacea & 16412 & 5374 & 4626 & 2153 & 4259\\
		NC$\_$005275.1 & Cetacea & 16324 & 5377 & 4525 & 2040 & 4382\\
		NC$\_$006931.1 & Cetacea & 16386 & 5357 & 4573 & 2164 & 4292\\
		NC$\_$001788.1 & Perissodactyla & 16670 & 5394 & 4819 & 2198 & 4259\\
		X97336.1 & Perissodactyla & 16829 & 5663 & 4630 & 2131 & 4405\\
		Y07726.1 & Perissodactyla & 16832 & 5623 & 4707 & 2169 & 4333\\
		NC$\_$001640.1 & Perissodactyla & 16660 & 5358 & 4754 & 2236 & 4312\\
		AJ238588.1 & Rodentia & 16507 & 5301 & 4041 & 2071 & 5094\\
		AJ001562.1 & Rodentia & 16602 & 5386 & 3913 & 2096 & 5207\\
		AJ001588.1 & Lagomorpha & 17245 & 5429 & 4584 & 2350 & 4882\\
		X88898.2 & Erinaceomorpha & 17447 & 5937 & 3503 & 2185 & 5822\\ \hline
	\end{tabular}
	\label{tab: mammalianMT}
\end{table}

\begin{table}[H]
	\centering
	\caption{Accession, group, sequence length of rhinovirus data}
	\tiny
	\begin{tabular}{|c | c c c c c c||c | c c c c c c|} \hline
		Accession & Group & Length & $N_A$ & $N_C$ & $N_G$ & $N_T$  & Accession & Group & Length & $N_A$ & $N_C$ & $N_G$ & $N_T$ \\ \hline
		AF499637.1 & HEV & 7458 & 2243 & 1677 & 1662 & 1876 & AF546702.1 & HEV & 7406 & 2209 & 1649 & 1681 & 1867\\
		AY751783.1 & A & 7137 & 2348 & 1362 & 1428 & 1999 & DQ473485.1 & B & 7208 & 2338 & 1436 & 1443 & 1991\\
		DQ473486.1 & B & 7216 & 2381 & 1428 & 1431 & 1976 & DQ473488.1 & B & 7214 & 2382 & 1456 & 1447 & 1929\\
		DQ473489.1 & B & 7223 & 2365 & 1476 & 1463 & 1919 & DQ473490.1 & B & 7212 & 2362 & 1417 & 1420 & 2013\\
		DQ473491.1 & A & 7145 & 2301 & 1383 & 1454 & 2007 & DQ473492.1 & A & 7140 & 2297 & 1371 & 1443 & 2029\\
		DQ473493.1 & A & 7134 & 2370 & 1287 & 1442 & 2035 & DQ473494.1 & A & 7120 & 2389 & 1314 & 1437 & 1980\\
		DQ473496.1 & A & 7106 & 2363 & 1342 & 1389 & 2012 & DQ473497.1 & A & 7025 & 2309 & 1322 & 1396 & 1998\\
		DQ473499.1 & A & 7123 & 2333 & 1301 & 1427 & 2062 & DQ473500.1 & A & 7135 & 2336 & 1344 & 1409 & 2046\\
		DQ473504.1 & A & 7143 & 2253 & 1380 & 1402 & 2108 & DQ473505.1 & A & 7141 & 2247 & 1404 & 1409 & 2081\\
		DQ473506.1 & A & 7149 & 2415 & 1349 & 1412 & 1972 & DQ473507.1 & A & 7143 & 2407 & 1350 & 1426 & 1960\\
		DQ473508.1 & A & 7148 & 2371 & 1389 & 1390 & 1998 & DQ473510.1 & A & 7137 & 2406 & 1313 & 1424 & 1994\\
		DQ473511.1 & A & 7036 & 2367 & 1288 & 1386 & 1995 & EF077279.1 & C & 6944 & 2176 & 1503 & 1509 & 1756\\
		EF077280.1 & C & 7015 & 2153 & 1550 & 1500 & 1812 & EF173414.1 & A & 7125 & 2369 & 1310 & 1403 & 2043\\
		EF173415.1 & A & 7124 & 2299 & 1396 & 1416 & 2013 & EF173420.1 & B & 7219 & 2356 & 1500 & 1477 & 1886\\
		EF173423.1 & B & 7216 & 2384 & 1456 & 1407 & 1969 & EF173425.1 & B & 7215 & 2426 & 1412 & 1432 & 1944\\
		EF186077.2 & C & 7134 & 2296 & 1549 & 1520 & 1769 & EF582385.1 & C & 7099 & 2195 & 1565 & 1473 & 1866\\
		EF582386.1 & C & 7114 & 2304 & 1492 & 1480 & 1838 & EF582387.1 & C & 7086 & 2261 & 1533 & 1513 & 1779\\
		FJ445111.1 & A & 7137 & 2388 & 1284 & 1388 & 2074 & FJ445112.1 & B & 7212 & 2402 & 1391 & 1417 & 2002\\
		FJ445113.1 & A & 7108 & 2376 & 1387 & 1407 & 1938 & FJ445114.1 & A & 7134 & 2375 & 1329 & 1428 & 2002\\
		FJ445115.1 & A & 7133 & 2377 & 1321 & 1426 & 2009 & FJ445116.1 & A & 7140 & 2298 & 1356 & 1437 & 2049\\
		FJ445117.1 & A & 7143 & 2321 & 1352 & 1421 & 2049 & FJ445118.1 & A & 7119 & 2386 & 1340 & 1423 & 1970\\
		FJ445119.1 & A & 7135 & 2354 & 1310 & 1421 & 2048 & FJ445121.1 & A & 7134 & 2376 & 1328 & 1403 & 2026\\
		FJ445122.1 & A & 7129 & 2327 & 1369 & 1445 & 1971 & FJ445123.1 & A & 7126 & 2358 & 1277 & 1399 & 2091\\
		FJ445124.1 & B & 7211 & 2403 & 1397 & 1413 & 1996 & FJ445125.1 & A & 7123 & 2332 & 1299 & 1416 & 2075\\
		FJ445126.1 & A & 7131 & 2354 & 1292 & 1414 & 2069 & FJ445127.1 & A & 7133 & 2375 & 1287 & 1408 & 2062\\
		FJ445128.1 & A & 7133 & 2336 & 1347 & 1429 & 2019 & FJ445129.1 & A & 7138 & 2380 & 1311 & 1391 & 2056\\
		FJ445130.1 & B & 7223 & 2405 & 1423 & 1405 & 1989 & FJ445131.1 & A & 7129 & 2397 & 1275 & 1436 & 2019\\
		FJ445132.1 & A & 7114 & 2352 & 1406 & 1424 & 1932 & FJ445133.1 & A & 7132 & 2342 & 1275 & 1439 & 2074\\
		FJ445134.1 & A & 7109 & 2360 & 1364 & 1404 & 1981 & FJ445135.1 & A & 7118 & 2371 & 1328 & 1395 & 2022\\
		FJ445136.1 & A & 7152 & 2349 & 1350 & 1425 & 2025 & FJ445137.1 & B & 7216 & 2334 & 1517 & 1473 & 1891\\
		FJ445138.1 & A & 7134 & 2353 & 1326 & 1418 & 2036 & FJ445139.1 & A & 7133 & 2364 & 1327 & 1413 & 2029\\
		FJ445140.1 & A & 7136 & 2342 & 1323 & 1381 & 2088 & FJ445141.1 & A & 7134 & 2414 & 1287 & 1401 & 2031\\
		FJ445142.1 & A & 7140 & 2233 & 1363 & 1415 & 2128 & FJ445143.1 & A & 7139 & 2390 & 1315 & 1391 & 2042\\
		FJ445144.1 & A & 7139 & 2340 & 1361 & 1416 & 2022 & FJ445145.1 & A & 7131 & 2371 & 1270 & 1391 & 2095\\
		FJ445146.1 & A & 7141 & 2320 & 1352 & 1411 & 2058 & FJ445147.1 & A & 7162 & 2353 & 1383 & 1429 & 1997\\
		FJ445148.1 & A & 7139 & 2390 & 1339 & 1393 & 2016 & FJ445149.1 & A & 7135 & 2377 & 1325 & 1435 & 1998\\
		FJ445151.1 & B & 7211 & 2316 & 1499 & 1504 & 1890 & FJ445152.1 & A & 7161 & 2375 & 1362 & 1427 & 1997\\
		FJ445153.1 & B & 7216 & 2372 & 1461 & 1452 & 1931 & FJ445154.1 & A & 7136 & 2368 & 1281 & 1406 & 2077\\
		FJ445155.1 & B & 7224 & 2369 & 1421 & 1433 & 2001 & FJ445156.1 & A & 7138 & 2389 & 1341 & 1421 & 1984\\
		FJ445157.1 & A & 7116 & 2357 & 1339 & 1406 & 2011 & FJ445158.1 & A & 7116 & 2336 & 1333 & 1434 & 2013\\
		FJ445159.1 & A & 7116 & 2331 & 1334 & 1437 & 2014 & FJ445160.1 & A & 7123 & 2325 & 1325 & 1454 & 2019\\
		FJ445161.1 & B & 7230 & 2382 & 1376 & 1460 & 2011 & FJ445162.1 & B & 7201 & 2392 & 1420 & 1410 & 1977\\
		FJ445163.1 & A & 7140 & 2355 & 1334 & 1412 & 2039 & FJ445164.1 & B & 7213 & 2388 & 1374 & 1420 & 2028\\
		FJ445165.1 & A & 7150 & 2235 & 1388 & 1416 & 2111 & FJ445166.1 & A & 7152 & 2227 & 1387 & 1425 & 2113\\
		FJ445167.1 & A & 7124 & 2394 & 1295 & 1409 & 2025 & FJ445168.1 & B & 7221 & 2376 & 1515 & 1450 & 1878\\
		FJ445169.1 & B & 7233 & 2338 & 1430 & 1463 & 2002 & FJ445170.1 & A & 7110 & 2371 & 1382 & 1413 & 1944\\
		FJ445171.1 & A & 7134 & 2316 & 1344 & 1446 & 2026 & FJ445172.1 & B & 7207 & 2403 & 1410 & 1418 & 1975\\
		FJ445173.1 & A & 7133 & 2387 & 1299 & 1381 & 2066 & FJ445174.1 & B & 7208 & 2404 & 1400 & 1384 & 2018\\
		FJ445175.1 & A & 7140 & 2319 & 1317 & 1433 & 2071 & FJ445176.1 & A & 7146 & 2284 & 1339 & 1382 & 2140\\
		FJ445177.1 & A & 7132 & 2347 & 1309 & 1449 & 2021 & FJ445178.1 & A & 7137 & 2331 & 1369 & 1420 & 2017\\
		FJ445179.1 & A & 7093 & 2318 & 1336 & 1417 & 2022 & FJ445180.1 & A & 7136 & 2387 & 1314 & 1425 & 2010\\
		FJ445181.1 & A & 7129 & 2333 & 1379 & 1441 & 1972 & FJ445182.1 & A & 7128 & 2334 & 1347 & 1420 & 2027\\
		FJ445183.1 & A & 7145 & 2356 & 1379 & 1429 & 1981 & FJ445184.1 & A & 7152 & 2241 & 1381 & 1406 & 2121\\
		FJ445185.1 & A & 7132 & 2344 & 1384 & 1444 & 1957 & FJ445186.1 & B & 7217 & 2377 & 1401 & 1447 & 1992\\
		FJ445187.1 & B & 7224 & 2372 & 1411 & 1466 & 1971 & FJ445188.1 & B & 7216 & 2322 & 1497 & 1490 & 1907\\
		FJ445189.1 & A & 7119 & 2370 & 1328 & 1408 & 2011 & FJ445190.1 & A & 7132 & 2392 & 1307 & 1389 & 2044\\
		L05355.1 & B & 7212 & 2313 & 1460 & 1475 & 1964 & L24917.1 & A & 7124 & 2383 & 1331 & 1412 & 1998\\
		V01149.1 & HEV & 7440 & 2206 & 1737 & 1711 & 1786 & X02316.1 & A & 7102 & 2324 & 1347 & 1418 & 2013\\ \hline
	\end{tabular}
	\label{tab: rhinovirus}
\end{table}

\begin{table}[H]
	\centering
	\caption{Accession, group, sequence length of coronavirus data}
	\begin{tabular}{|c | c c c c c c|}\hline
		Accession & Group & Length & $N_A$ & $N_C$ & $N_G$ & $N_T$ \\ \hline
		AF304460.1 & Group 1 & 27317 & 7420 & 4549 & 5903 & 9445\\
		AF353511.1 & Group 1 & 28033 & 6937 & 5382 & 6397 & 9317\\
		NC$\_$005831.2 & Group 1 & 27553 & 7253 & 3979 & 5516 & 10805\\
		AY391777.1 & Group 2 & 30738 & 8485 & 4658 & 6655 & 10940\\
		U00735.2 & Group 2 & 31032 & 8490 & 4713 & 6774 & 11055\\
		AF391542.1 & Group 2 & 31028 & 8486 & 4743 & 6772 & 11027\\
		AF220295.1 & Group 2 & 31100 & 8544 & 4711 & 6790 & 11055\\
		NC$\_$003045.1 & Group 2 & 31028 & 8487 & 4752 & 6767 & 11022\\
		AF208067.1 & Group 2 & 31233 & 8087 & 5591 & 7466 & 10089\\
		AF201929.1 & Group 2 & 31276 & 8117 & 5548 & 7422 & 10189\\
		AF208066.1 & Group 2 & 31112 & 8030 & 5534 & 7416 & 10132\\
		NC$\_$001846.1 & Group 2 & 31357 & 8138 & 5614 & 7487 & 10118\\
		NC$\_$001451.1 & Group 3 & 27608 & 7967 & 4479 & 5993 & 9169\\
		EU095850.1 & Group 3 & 27657 & 7969 & 4513 & 6066 & 9108\\
		AY278488.2 & Group 4 & 29725 & 8465 & 5941 & 6185 & 9134\\
		AY278741.1 & Group 4 & 29727 & 8455 & 5940 & 6188 & 9144\\
		AY278491.2 & Group 4 & 29742 & 8475 & 5942 & 6183 & 9142\\
		AY278554.2 & Group 4 & 29736 & 8476 & 5942 & 6185 & 9133\\
		AY282752.2 & Group 4 & 29736 & 8476 & 5939 & 6185 & 9136\\
		AY283794.1 & Group 4 & 29711 & 8453 & 5937 & 6184 & 9137\\
		AY283795.1 & Group 4 & 29705 & 8447 & 5936 & 6187 & 9135\\
		AY283796.1 & Group 4 & 29711 & 8453 & 5936 & 6185 & 9137\\
		AY283797.1 & Group 4 & 29706 & 8451 & 5935 & 6184 & 9135\\
		AY283798.2 & Group 4 & 29711 & 8453 & 5935 & 6185 & 9138\\
		AY291451.1 & Group 4 & 29729 & 8457 & 5940 & 6188 & 9144\\
		NC$\_$004718.3 & Group 4 & 29751 & 8481 & 5940 & 6187 & 9143\\
		AY297028.1 & Group 4 & 29715 & 8458 & 5934 & 6187 & 9135\\
		AY572034.1 & Group 4 & 29540 & 8402 & 5911 & 6154 & 9073\\
		AY572035.1 & Group 4 & 29518 & 8395 & 5907 & 6151 & 9065\\
		NC$\_$006577.2 & Group 5 & 29926 & 8331 & 3895 & 5699 & 12001\\
		NC$\_$001564.2 & Flaviviridae outgroup & 10682 & 2618 & 2531 & 2919 & 2614\\
		NC$\_$004102.1 & Flaviviridae outgroup & 9646 & 1889 & 2893 & 2724 & 2140\\
		NC$\_$001512.1 & Togaviridae outgroup & 11835 & 3676 & 2860 & 2859 & 2440\\
		NC$\_$001544.1 & Togaviridae outgroup & 11657 & 3220 & 2901 & 3065 & 2416\\\hline
	\end{tabular}
	\label{tab: coronavirus}
\end{table}

\begin{table}[H]
	\centering
	\caption{Accession, group, sequence length of influenza A virus data}
	\begin{tabular}{|c | c c c c c c|}\hline
		Accession & Group & Length & $N_A$ & $N_C$ & $N_G$ & $N_T$ \\\hline
		HM370969.1 & H1N1 & 1419 & 453 & 263 & 330 & 373\\
		CY138562.1 & H1N1 & 1422 & 437 & 259 & 343 & 383\\
		CY149630.1 & H1N1 & 1433 & 441 & 261 & 346 & 385\\
		KC608160.1 & H1N1 & 1398 & 409 & 251 & 357 & 381\\
		AM157358.1 & H1N1 & 1413 & 418 & 259 & 355 & 381\\
		AB470663.1 & H1N1 & 1422 & 418 & 252 & 359 & 393\\
		AB546159.1 & H1N1 & 1410 & 421 & 260 & 351 & 378\\
		HQ897966.1 & H1N1 & 1410 & 422 & 246 & 353 & 389\\
		EU026046.2 & H1N1 & 1433 & 439 & 263 & 347 & 384\\
		FJ357114.1 & H1N1 & 1433 & 438 & 253 & 350 & 392\\
		GQ411894.1 & H1N1 & 1413 & 430 & 260 & 346 & 376\\
		CY140047.1 & H1N1 & 1433 & 440 & 261 & 347 & 385\\
		KM244078.1 & H1N1 & 1410 & 447 & 261 & 335 & 367\\
		HQ185381.1 & H5N1 & 1350 & 406 & 240 & 339 & 365\\
		HQ185383.1 & H5N1 & 1350 & 408 & 240 & 336 & 366\\
		EU635875.1 & H5N1 & 1350 & 397 & 248 & 347 & 358\\
		FM177121.1 & H5N1 & 1370 & 407 & 245 & 350 & 368\\
		AM914017.1 & H5N1 & 1350 & 398 & 243 & 344 & 365\\
		KF572435.1 & H5N1 & 1350 & 403 & 247 & 345 & 355\\
		AF509102.2 & H5N1 & 1366 & 401 & 257 & 344 & 364\\
		AB684161.1 & H5N1 & 1350 & 404 & 235 & 348 & 363\\
		EF541464.1 & H5N1 & 1350 & 396 & 246 & 349 & 359\\
		JF699677.1 & H5N1 & 1350 & 404 & 236 & 348 & 362\\
		GU186511.1 & H5N1 & 1370 & 407 & 244 & 345 & 374\\
		EU500854.1 & H7N3 & 1453 & 475 & 284 & 339 & 355\\
		CY129336.1 & H7N3 & 1428 & 470 & 278 & 332 & 348\\
		CY076231.1 & H7N3 & 1420 & 467 & 286 & 327 & 340\\
		CY039321.1 & H7N3 & 1434 & 470 & 288 & 333 & 343\\
		AY646080.1 & H7N3 & 1453 & 485 & 284 & 329 & 355\\
		KF259734.1 & H7N9 & 1398 & 478 & 290 & 321 & 309\\
		KF938945.1 & H7N9 & 1404 & 483 & 287 & 322 & 312\\
		KF259688.1 & H7N9 & 1413 & 490 & 291 & 320 & 312\\
		KC609801.1 & H7N9 & 1426 & 488 & 292 & 332 & 314\\
		CY014788.1 & H7N9 & 1460 & 500 & 306 & 337 & 317\\
		CY186004.1 & H7N9 & 1422 & 494 & 303 & 317 & 308\\
		DQ017487.1 & H2N2 & 1467 & 445 & 281 & 355 & 386\\
		CY005540.1 & H2N2 & 1467 & 455 & 284 & 344 & 384\\
		JX081142.1 & H2N2 & 1457 & 446 & 265 & 349 & 397\\\hline
	\end{tabular}
	\label{tab: influenza}
\end{table}

\begin{table}[H]
	\centering
	\caption{Accession, group, sequence length of bacteria 16S rDNA data}
	\begin{tabular}{|c | c c c c c c|}\hline
		Accession & Family & Length & $N_A$ & $N_C$ & $N_G$ & $N_T$ \\ \hline
		KY486204.1 & Methylobacteriaceae & 1104 & 263 & 265 & 345 & 230\\
		KY486205.1 & Xanthomonadaceae & 761 & 194 & 165 & 256 & 146\\
		KY486206.1 & Xanthomonadaceae & 1452 & 361 & 340 & 464 & 287\\
		KY486207.1 & Intrasporangiaceae & 1253 & 301 & 295 & 395 & 262\\
		KY486218.1 & Microbacteriaceae & 1195 & 296 & 286 & 372 & 241\\
		KY486219.1 & Pseudomonadaceae & 1099 & 282 & 250 & 339 & 228\\
		KY927407.1 & Bacillaceae & 718 & 179 & 179 & 206 & 154\\
		KY486220.1 & Paenibacillaceae & 1335 & 327 & 326 & 419 & 263\\
		KY486221.1 & Enterobacteriaceae & 1339 & 335 & 314 & 430 & 260\\
		KY486222.1 & Xanthomonadaceae & 1337 & 335 & 313 & 422 & 267\\
		KY486223.1 & Microbacteriaceae & 1334 & 317 & 314 & 435 & 268\\
		KY486209.1 & Rhodanobacteraceae & 1366 & 332 & 319 & 447 & 268\\
		KY486210.1 & Enterobacteriaceae & 1356 & 338 & 324 & 431 & 262\\
		KY486232.1 & Enterobacteriaceae & 1350 & 337 & 318 & 432 & 262\\
		KY019246.1 & Enterobacteriaceae & 1346 & 335 & 318 & 431 & 262\\
		KY013009.1 & Enterobacteriaceae & 1351 & 337 & 318 & 434 & 262\\
		KY927404.1 & Microbacteriaceae & 742 & 184 & 177 & 232 & 149\\
		KY486211.1 & Enterobacteriaceae & 1365 & 337 & 325 & 441 & 262\\
		KY013011.1 & Staphylococcaceae & 1035 & 278 & 226 & 299 & 231\\
		KY019245.1 & Enterobacteriaceae & 1344 & 334 & 317 & 430 & 262\\
		KY013010.1 & Bacillaceae & 1343 & 336 & 318 & 420 & 269\\
		KY486208.1 & Enterobacteriaceae & 1269 & 318 & 298 & 400 & 253\\
		KY486212.1 & Microbacteriaceae & 1364 & 341 & 324 & 436 & 263\\
		KY486213.1 & Xanthomonadaceae & 1387 & 345 & 322 & 442 & 278\\
		KY486228.1 & Enterobacteriaceae & 1356 & 337 & 325 & 429 & 265\\
		KY486224.1 & Enterobacteriaceae & 1346 & 335 & 317 & 431 & 262\\
		KY486225.1 & Enterobacteriaceae & 1294 & 329 & 312 & 401 & 251\\
		KY486226.1 & Enterobacteriaceae & 1347 & 338 & 316 & 432 & 261\\
		KY927405.1 & Enterobacteriaceae & 753 & 189 & 169 & 252 & 143\\
		KY486227.1 & Enterobacteriaceae & 1345 & 336 & 317 & 431 & 261\\
		KY927408.1 & Microbacteriaceae & 785 & 193 & 188 & 243 & 161\\
		KY486214.1 & Enterobacteriaceae & 1411 & 352 & 330 & 455 & 274\\
		KY927406.1 & Microbacteriaceae & 796 & 201 & 183 & 266 & 146\\
		KY486215.1 & Enterobacteriaceae & 1319 & 338 & 309 & 417 & 255\\
		KY486216.1 & Enterobacteriaceae & 1345 & 335 & 317 & 430 & 263\\
		KY486217.1 & Enterobacteriaceae & 1341 & 335 & 315 & 431 & 260\\
		KY486229.1 & Enterobacteriaceae & 1345 & 335 & 318 & 432 & 260\\
		KY486230.1 & Pseudomonadaceae & 1346 & 345 & 306 & 416 & 279\\
		KY486231.1 & Enterobacteriaceae & 1322 & 325 & 321 & 419 & 255\\
		KY019244.1 & Enterobacteriaceae & 1337 & 331 & 315 & 428 & 260\\ \hline
	\end{tabular}
	\label{tab: bacteriaRDNA}
\end{table}

\begin{table}[H]
	\centering
	\caption{Accession, group, sequence length of bacteria data}
	\begin{tabular}{|c | c c c c c c|}\hline
		Accession & Family & Length & $N_A$ & $N_C$ & $N_G$ & $N_T$ \\ \hline
		CP001598.1 & Bacillaceae & 5227419 & 1685408 & 930043 & 919269 & 1692699\\ 
		AE016879.1 & Bacillaceae & 5227293 & 1685374 & 930007 & 919244 & 1692668\\
		CP001215.1 & Bacillaceae & 5230115 & 1667671 & 974191 & 876267 & 1711986\\
		AE017225.1 & Bacillaceae & 5228663 & 1685622 & 930391 & 919481 & 1693169\\
		CP000976.1 & Borreliaceae & 931674 & 335148 & 129225 & 127787 & 339514\\
		CP000048.1 & Borreliaceae & 922307 & 321202 & 137556 & 137547 & 326002\\
		CP000993.1 & Borreliaceae & 930981 & 335785 & 129350 & 126839 & 339007\\
		CP000049.1 & Borreliaceae & 917330 & 322493 & 133424 & 133693 & 327720\\
		CP000246.1 & Clostridiaceae & 3256683 & 1148078 & 470943 & 453276 & 1184386\\
		CP000312.1 & Clostridiaceae & 2897393 & 1017083 & 423439 & 395046 & 1061825\\
		BA000016.3 & Clostridiaceae & 3031430 & 1060154 & 446732 & 419228 & 1105316\\
		CP000527.1 & Desulfovibrionaceae & 3462887 & 639427 & 1092219 & 1089813 & 641428\\
		AE017285.1 & Desulfovibrionaceae & 3570858 & 659017 & 1127624 & 1127109 & 657108\\
		CP002297.1 & Desulfovibrionaceae & 3532052 & 652227 & 1113805 & 1116142 & 649878\\
		AM260480.1 & Burkholderiaceae & 2912490 & 486037 & 972789 & 972298 & 481366\\
		CP000091.1 & Burkholderiaceae & 2726152 & 480326 & 883953 & 887595 & 474278\\
		CP000578.1 & Rhodobacteraceae & 1219053 & 192966 & 416292 & 420323 & 189472\\
		CP001151.1 & Rhodobacteraceae & 1297647 & 204455 & 445520 & 446045 & 201627\\
		AM295250.1 & Staphylococcaceae & 2566424 & 833636 & 449856 & 438966 & 843965\\
		AE015929.1 & Staphylococcaceae & 2499279 & 837991 & 405441 & 396707 & 859140\\
		AP006716.1 & Staphylococcaceae & 2685015 & 907537 & 437414 & 443072 & 896992\\
		CP001837.1 & Staphylococcaceae & 2658366 & 878689 & 445972 & 454367 & 879338\\
		AL590842.1 & Yersiniaceae & 4653728 & 1219520 & 1102670 & 1114185 & 1217353\\
		CP001585.1 & Yersiniaceae & 4640720 & 1216182 & 1097565 & 1112625 & 1214348\\
		AE009952.1 & Yersiniaceae & 4600755 & 1200303 & 1090469 & 1101384 & 1208599\\
		CP001593.1 & Yersiniaceae & 4553586 & 1187961 & 1077463 & 1093635 & 1194527\\
		CP001671.1 & Enterobacteriaceae & 5131397 & 1271011 & 1298314 & 1297349 & 1264723\\
		CP000468.1 & Enterobacteriaceae & 5082025 & 1256126 & 1285309 & 1283517 & 1256945\\
		CP001383.1 & Enterobacteriaceae & 4650856 & 1145625 & 1187110 & 1177854 & 1140266\\
		AE005674.2 & Enterobacteriaceae & 4607202 & 1133784 & 1176618 & 1167963 & 1128831\\ \hline
	\end{tabular}
	\label{tab: bacteria}
\end{table}


\begin{thebibliography}{10}
	
	\bibitem{nei1996phylogenetic}
	Masatoshi Nei.
	\newblock Phylogenetic analysis in molecular evolutionary genetics.
	\newblock {\em Annual review of genetics}, 30(1):371--403, 1996.
	
	\bibitem{bellgard1999dynamic}
	Matthew~I Bellgard, Takeshi Itoh, Hidemi Watanabe, Tadashi Imanishi, and
	Takashi Gojobori.
	\newblock Dynamic evolution of genomes and the concept of genome space.
	\newblock {\em Annals of the New York Academy of Sciences}, 870(1):293--300,
	1999.
	
	\bibitem{sievers2011fast}
	Fabian Sievers, Andreas Wilm, David Dineen, Toby~J Gibson, Kevin Karplus,
	Weizhong Li, Rodrigo Lopez, Hamish McWilliam, Michael Remmert, Johannes
	S{\"o}ding, et~al.
	\newblock Fast, scalable generation of high-quality protein multiple sequence
	alignments using clustal omega.
	\newblock {\em Molecular systems biology}, 7(1):539, 2011.
	
	\bibitem{katoh2013mafft}
	Kazutaka Katoh and Daron~M Standley.
	\newblock Mafft multiple sequence alignment software version 7: improvements in
	performance and usability.
	\newblock {\em Molecular biology and evolution}, 30(4):772--780, 2013.
	
	\bibitem{edgar2004muscle}
	Robert~C Edgar.
	\newblock Muscle: multiple sequence alignment with high accuracy and high
	throughput.
	\newblock {\em Nucleic acids research}, 32(5):1792--1797, 2004.
	
	\bibitem{hozumi2021umap}
	Yuta Hozumi, Rui Wang, Changchuan Yin, and Guo-Wei Wei.
	\newblock Umap-assisted k-means clustering of large-scale sars-cov-2 mutation
	datasets.
	\newblock {\em Computers in biology and medicine}, 131:104264, 2021.
	
	\bibitem{chen2022omicron}
	Jiahui Chen and Guo-Wei Wei.
	\newblock Omicron ba. 2 (b. 1.1. 529.2): high potential for becoming the next
	dominant variant.
	\newblock {\em The journal of physical chemistry letters}, 13(17):3840--3849,
	2022.
	
	\bibitem{chen2022persistent}
	Jiahui Chen, Yuchi Qiu, Rui Wang, and Guo-Wei Wei.
	\newblock Persistent laplacian projected omicron ba. 4 and ba. 5 to become new
	dominating variants.
	\newblock {\em Computers in biology and medicine}, 151:106262, 2022.
	
	\bibitem{bleher2021topological}
	Michael Bleher, Lukas Hahn, Maximilian Neumann, Juan~Angel Patino-Galindo,
	Mathieu Carriere, Ulrich Bauer, Raul Rabadan, and Andreas Ott.
	\newblock Topological data analysis identifies emerging adaptive mutations in
	sars-cov-2.
	\newblock {\em arXiv preprint arXiv:2106.07292}, 2021.
	
	\bibitem{patino2021recombination}
	Juan~{\'A}ngel Pati{\~n}o-Galindo, Ioan Filip, Ratul Chowdhury, Costas~D
	Maranas, Peter~K Sorger, Mohammed AlQuraishi, and Raul Rabadan.
	\newblock Recombination and lineage-specific mutations linked to the emergence
	of sars-cov-2.
	\newblock {\em Genome Medicine}, 13:1--14, 2021.
	
	\bibitem{vinga2014alignment}
	Susana Vinga.
	\newblock Alignment-free methods in computational biology, 2014.
	
	\bibitem{zielezinski2017alignment}
	Andrzej Zielezinski, Susana Vinga, Jonas Almeida, and Wojciech~M Karlowski.
	\newblock Alignment-free sequence comparison: benefits, applications, and
	tools.
	\newblock {\em Genome biology}, 18:1--17, 2017.
	
	\bibitem{bonham2014alignment}
	Oliver Bonham-Carter, Joe Steele, and Dhundy Bastola.
	\newblock Alignment-free genetic sequence comparisons: a review of recent
	approaches by word analysis.
	\newblock {\em Briefings in bioinformatics}, 15(6):890--905, 2014.
	
	\bibitem{bernard2016alignment}
	Guillaume Bernard, Cheong~Xin Chan, and Mark~A Ragan.
	\newblock Alignment-free microbial phylogenomics under scenarios of sequence
	divergence, genome rearrangement and lateral genetic transfer.
	\newblock {\em Scientific reports}, 6(1):28970, 2016.
	
	\bibitem{zielezinski2019benchmarking}
	Andrzej Zielezinski, Hani~Z Girgis, Guillaume Bernard, Chris-Andre Leimeister,
	Kujin Tang, Thomas Dencker, Anna~Katharina Lau, Sophie R{\"o}hling, Jae~Jin
	Choi, Michael~S Waterman, et~al.
	\newblock Benchmarking of alignment-free sequence comparison methods.
	\newblock {\em Genome biology}, 20:1--18, 2019.
	
	\bibitem{jun2010whole}
	Se-Ran Jun, Gregory~E Sims, Guohong~A Wu, and Sung-Hou Kim.
	\newblock Whole-proteome phylogeny of prokaryotes by feature frequency
	profiles: An alignment-free method with optimal feature resolution.
	\newblock {\em Proceedings of the National Academy of Sciences},
	107(1):133--138, 2010.
	
	\bibitem{sims2009whole}
	Gregory~E Sims, Se-Ran Jun, Guohong~Albert Wu, and Sung-Hou Kim.
	\newblock Whole-genome phylogeny of mammals: evolutionary information in genic
	and nongenic regions.
	\newblock {\em Proceedings of the National Academy of Sciences},
	106(40):17077--17082, 2009.
	
	\bibitem{blaisdell1986measure}
	B~Edwin Blaisdell.
	\newblock A measure of the similarity of sets of sequences not requiring
	sequence alignment.
	\newblock {\em Proceedings of the National Academy of Sciences},
	83(14):5155--5159, 1986.
	
	\bibitem{wu1997measure}
	Tiee-Jian Wu, John~P Burke, and Daniel~B Davison.
	\newblock A measure of dna sequence dissimilarity based on mahalanobis distance
	between frequencies of words.
	\newblock {\em Biometrics}, pages 1431--1439, 1997.
	
	\bibitem{wu2001statistical}
	Tiee-Jian Wu, Ya-Ching Hsieh, and Lung-An Li.
	\newblock Statistical measures of dna sequence dissimilarity under markov chain
	models of base composition.
	\newblock {\em Biometrics}, 57(2):441--448, 2001.
	
	\bibitem{wu2005optimal}
	Tiee-Jian Wu, Ying-Hsueh Huang, and Lung-An Li.
	\newblock Optimal word sizes for dissimilarity measures and estimation of the
	degree of dissimilarity between dna sequences.
	\newblock {\em Bioinformatics}, 21(22):4125--4132, 2005.
	
	\bibitem{korf2009applying}
	Ian~F Korf and Alan~B Rose.
	\newblock Applying word-based algorithms: the imeter.
	\newblock {\em Plant Systems Biology}, pages 287--301, 2009.
	
	\bibitem{li2008introduction}
	Ming Li, Paul Vit{\'a}nyi, et~al.
	\newblock {\em An introduction to Kolmogorov complexity and its applications},
	volume~3.
	\newblock Springer, 2008.
	
	\bibitem{otu2003new}
	Hasan~H Otu and Khalid Sayood.
	\newblock A new sequence distance measure for phylogenetic tree construction.
	\newblock {\em Bioinformatics}, 19(16):2122--2130, 2003.
	
	\bibitem{tribus1971energy}
	Myron Tribus and Edward~C McIrvine.
	\newblock Energy and information.
	\newblock {\em Scientific American}, 225(3):179--190, 1971.
	
	\bibitem{vinga2003alignment}
	Susana Vinga and Jonas Almeida.
	\newblock Alignment-free sequence comparison—a review.
	\newblock {\em Bioinformatics}, 19(4):513--523, 2003.
	
	\bibitem{yu2013real}
	Chenglong Yu, Troy Hernandez, Hui Zheng, Shek-Chung Yau, Hsin-Hsiung Huang,
	Rong~Lucy He, Jie Yang, and Stephen S-T Yau.
	\newblock Real time classification of viruses in 12 dimensions.
	\newblock {\em PloS one}, 8(5):e64328, 2013.
	
	\bibitem{deng2011novel}
	Mo~Deng, Chenglong Yu, Qian Liang, Rong~L He, and Stephen S-T Yau.
	\newblock A novel method of characterizing genetic sequences: genome space with
	biological distance and applications.
	\newblock {\em PloS one}, 6(3):e17293, 2011.
	
	\bibitem{jeffrey1990chaos}
	H~Joel Jeffrey.
	\newblock Chaos game representation of gene structure.
	\newblock {\em Nucleic acids research}, 18(8):2163--2170, 1990.
	
	\bibitem{randic2013milestones}
	Milan Randi{\'c}, Marjana Novi{\v{c}}, and Dejan Plav{\v{s}}i{\'c}.
	\newblock Milestones in graphical bioinformatics.
	\newblock {\em International Journal of Quantum Chemistry}, 113(22):2413--2446,
	2013.
	
	\bibitem{burma1992genome}
	Pradeep~Kumar Burma, Alok Raj, Jayant~K Deb, and Samir~K Brahmachari.
	\newblock Genome analysis: a new approach for visualization of sequence
	organization in genomes.
	\newblock {\em Journal of biosciences}, 17:395--411, 1992.
	
	\bibitem{almeida2001analysis}
	Jonas~S Almeida, Joao~A Carrico, Antonio Maretzek, Peter~A Noble, and Madilyn
	Fletcher.
	\newblock Analysis of genomic sequences by chaos game representation.
	\newblock {\em Bioinformatics}, 17(5):429--437, 2001.
	
	\bibitem{deschavanne1999genomic}
	Patrick~J Deschavanne, Alain Giron, Joseph Vilain, Guillaume Fagot, and Bernard
	Fertil.
	\newblock Genomic signature: characterization and classification of species
	assessed by chaos game representation of sequences.
	\newblock {\em Molecular biology and evolution}, 16(10):1391--1399, 1999.
	
	\bibitem{hao2000fractals}
	Bai-Lin Hao.
	\newblock Fractals from genomes--exact solutions of a biology-inspired problem.
	\newblock {\em Physica A: Statistical Mechanics and its Applications},
	282(1-2):225--246, 2000.
	
	\bibitem{hoang2015new}
	Tung Hoang, Changchuan Yin, Hui Zheng, Chenglong Yu, Rong~Lucy He, and Stephen
	S-T Yau.
	\newblock A new method to cluster dna sequences using fourier power spectrum.
	\newblock {\em Journal of theoretical biology}, 372:135--145, 2015.
	
	\bibitem{yin2014measure}
	Changchuan Yin, Ying Chen, and Stephen S-T Yau.
	\newblock A measure of dna sequence similarity by fourier transform with
	applications on hierarchical clustering.
	\newblock {\em Journal of theoretical biology}, 359:18--28, 2014.
	
	\bibitem{saw2019alignment}
	Ajay~Kumar Saw, Garima Raj, Manashi Das, Narayan~Chandra Talukdar,
	Binod~Chandra Tripathy, and Soumyadeep Nandi.
	\newblock Alignment-free method for dna sequence clustering using fuzzy
	integral similarity.
	\newblock {\em Scientific reports}, 9(1):3753, 2019.
	
	\bibitem{yu2010novel}
	Chenglong Yu, Qian Liang, Changchuan Yin, Rong~L He, and Stephen S-T Yau.
	\newblock A novel construction of genome space with biological geometry.
	\newblock {\em DNA research}, 17(3):155--168, 2010.
	
	\bibitem{lum2013extracting}
	Pek~Y Lum, Gurjeet Singh, Alan Lehman, Tigran Ishkanov, Mikael
	Vejdemo-Johansson, Muthu Alagappan, John Carlsson, and Gunnar Carlsson.
	\newblock Extracting insights from the shape of complex data using topology.
	\newblock {\em Scientific reports}, 3(1):1236, 2013.
	
	\bibitem{chan2013topology}
	Joseph~Minhow Chan, Gunnar Carlsson, and Raul Rabadan.
	\newblock Topology of viral evolution.
	\newblock {\em Proceedings of the National Academy of Sciences},
	110(46):18566--18571, 2013.
	
	\bibitem{nguyen2022topological}
	Dong Quan~Ngoc Nguyen, Phuong Dong~Tan Le, Lin Xing, and Lizhen Lin.
	\newblock A topological characterization of dna sequences based on chaos
	geometry and persistent homology.
	\newblock In {\em 2022 International Conference on Computational Science and
		Computational Intelligence (CSCI)}, pages 1591--1597. IEEE, 2022.
	
	\bibitem{wang2020persistent}
	Rui Wang, Duc~Duy Nguyen, and Guo-Wei Wei.
	\newblock Persistent spectral graph.
	\newblock {\em International journal for numerical methods in biomedical
		engineering}, 36(9):e3376, 2020.
	
	\bibitem{chen2021evolutionary}
	Jiahui Chen, Rundong Zhao, Yiying Tong, and Guo-Wei Wei.
	\newblock Evolutionary de rham-hodge method.
	\newblock {\em Discrete and continuous dynamical systems. Series B},
	26(7):3785, 2021.
	
	\bibitem{sun2021geometric}
	Nan Sun, Shaojun Pei, Lily He, Changchuan Yin, Rong~Lucy He, and Stephen S-T
	Yau.
	\newblock Geometric construction of viral genome space and its applications.
	\newblock {\em Computational and Structural Biotechnology Journal},
	19:4226--4234, 2021.
	
	\bibitem{sims2009alignment}
	Gregory~E Sims, Se-Ran Jun, Guohong~A Wu, and Sung-Hou Kim.
	\newblock Alignment-free genome comparison with feature frequency profiles
	(ffp) and optimal resolutions.
	\newblock {\em Proceedings of the National Academy of Sciences},
	106(8):2677--2682, 2009.
	
	\bibitem{qi2004whole}
	Ji~Qi, Bin Wang, and Bai-Iin Hao.
	\newblock Whole proteome prokaryote phylogeny without sequence alignment:
	Ak-string composition approach.
	\newblock {\em Journal of molecular evolution}, 58:1--11, 2004.
	
	\bibitem{hu2023rational}
	Ye-Fan Hu, Terrence Tsz-Tai Yuen, Hua-Rui Gong, Bingjie Hu, Jing-Chu Hu,
	Xuan-Sheng Lin, Li~Rong, Coco~Luyao Zhou, Lin-Lei Chen, Xiaolei Wang, et~al.
	\newblock Rational design of a booster vaccine against covid-19 based on
	antigenic distance.
	\newblock {\em Cell Host \& Microbe}, 31(8):1301--1316, 2023.
	
	\bibitem{carlsson2009topology}
	Gunnar Carlsson.
	\newblock Topology and data.
	\newblock {\em Bulletin of the American Mathematical Society}, 46(2):255--308,
	2009.
	
	\bibitem{edelsbrunner2008persistent}
	Herbert Edelsbrunner, John Harer, et~al.
	\newblock Persistent homology-a survey.
	\newblock {\em Contemporary mathematics}, 453(26):257--282, 2008.
	
	\bibitem{cang2017topologynet}
	Zixuan Cang and Guo-Wei Wei.
	\newblock Topologynet: Topology based deep convolutional and multi-task neural
	networks for biomolecular property predictions.
	\newblock {\em PLoS computational biology}, 13(7):e1005690, 2017.
	
	\bibitem{nguyen2019mathematical}
	Duc~Duy Nguyen, Zixuan Cang, Kedi Wu, Menglun Wang, Yin Cao, and Guo-Wei Wei.
	\newblock Mathematical deep learning for pose and binding affinity prediction
	and ranking in d3r grand challenges.
	\newblock {\em Journal of computer-aided molecular design}, 33:71--82, 2019.
	
	\bibitem{nguyen2020mathdl}
	Duc~Duy Nguyen, Kaifu Gao, Menglun Wang, and Guo-Wei Wei.
	\newblock Mathdl: mathematical deep learning for d3r grand challenge 4.
	\newblock {\em Journal of computer-aided molecular design}, 34:131--147, 2020.
	
	\bibitem{memoli2022persistent}
	Facundo M{\'e}moli, Zhengchao Wan, and Yusu Wang.
	\newblock Persistent laplacians: Properties, algorithms and implications.
	\newblock {\em SIAM Journal on Mathematics of Data Science}, 4(2):858--884,
	2022.
	
	\bibitem{liu2023algebraic}
	Jian Liu, Jingyan Li, and Jie Wu.
	\newblock The algebraic stability for persistent laplacians.
	\newblock {\em arXiv preprint arXiv:2302.03902}, 2023.
	
	\bibitem{wang2021hermes}
	Rui Wang, Rundong Zhao, Emily Ribando-Gros, Jiahui Chen, Yiying Tong, and
	Guo-Wei Wei.
	\newblock Hermes: Persistent spectral graph software.
	\newblock {\em Foundations of data science (Springfield, Mo.)}, 3(1):67, 2021.
	
	\bibitem{wei2024persistent}
	Xiaoqi Wei and Guo-Wei Wei.
	\newblock Persistent sheaf laplacians.
	\newblock {\em Foundations of data science (Springfield, Mo.)},
	doi:10.3934/fods.2024033, 2024.
	
	\bibitem{wang2023persistent}
	Rui Wang and Guo-Wei Wei.
	\newblock Persistent path laplacian.
	\newblock {\em Foundations of data science (Springfield, Mo.)}, 5(1):26, 2023.
	
	\bibitem{chen2023persistent}
	Dong Chen, Jian Liu, Jie Wu, and Guo-Wei Wei.
	\newblock Persistent hyperdigraph homology and persistent hyperdigraph
	laplacians.
	\newblock {\em Foundations of Data Science}, 5(4):558--588, 2023.
	
	\bibitem{qiu2023persistent}
	Yuchi Qiu and Guo-Wei Wei.
	\newblock Persistent spectral theory-guided protein engineering.
	\newblock {\em Nature computational science}, 3(2):149--163, 2023.
	
	\bibitem{meng2021persistent}
	Zhenyu Meng and Kelin Xia.
	\newblock Persistent spectral--based machine learning (perspect ml) for
	protein-ligand binding affinity prediction.
	\newblock {\em Science advances}, 7(19):eabc5329, 2021.
	
	\bibitem{hozumi2024analyzing}
	Yuta Hozumi and Guo-Wei Wei.
	\newblock Analyzing single cell rna sequencing with topological nonnegative
	matrix factorization.
	\newblock {\em Journal of Computational and Applied Mathematics}, page 115842,
	2024.
	
	\bibitem{cottrell2023plpca}
	Sean Cottrell, Rui Wang, and Guo-Wei Wei.
	\newblock Plpca: Persistent laplacian-enhanced pca for microarray data
	analysis.
	\newblock {\em Journal of Chemical Information and Modeling}, 2023.
	
	\bibitem{yu2024optimal}
	Hongyu Yu and Stephen S-T Yau.
	\newblock The optimal metric for viral genome space.
	\newblock {\em Computational and Structural Biotechnology Journal},
	23:2083--2096, 2024.
	
	\bibitem{skowronski2017serial}
	Danuta~M Skowronski, Catharine Chambers, Gaston De~Serres, Suzana Sabaiduc,
	Anne-Luise Winter, James~A Dickinson, Jonathan~B Gubbay, Kevin Fonseca,
	Steven~J Drews, Hugues Charest, et~al.
	\newblock Serial vaccination and the antigenic distance hypothesis: effects on
	influenza vaccine effectiveness during a (h3n2) epidemics in canada,
	2010--2011 to 2014--2015.
	\newblock {\em The Journal of infectious diseases}, 215(7):1059--1099, 2017.
	
	\bibitem{starr2020deep}
	Tyler~N Starr, Allison~J Greaney, Sarah~K Hilton, Daniel Ellis, Katharine~HD
	Crawford, Adam~S Dingens, Mary~Jane Navarro, John~E Bowen, M~Alejandra
	Tortorici, Alexandra~C Walls, et~al.
	\newblock Deep mutational scanning of sars-cov-2 receptor binding domain
	reveals constraints on folding and ace2 binding.
	\newblock {\em cell}, 182(5):1295--1310, 2020.
	
	\bibitem{chen2023topological}
	Jiahui Chen, Daniel~R Woldring, Faqing Huang, Xuefei Huang, and Guo-Wei Wei.
	\newblock Topological deep learning based deep mutational scanning.
	\newblock {\em Computers in Biology and Medicine}, 164:107258, 2023.
	
	\bibitem{wee2024rapid}
	JunJie Wee and Guo-Wei Wei.
	\newblock Rapid response to fast viral evolution using alphafold 3-assisted
	topological deep learning.
	\newblock {\em arXiv preprint arXiv:2411.12370}, 2024.
	
	\bibitem{letunic2024interactive}
	Ivica Letunic and Peer Bork.
	\newblock Interactive tree of life (itol) v6: recent updates to the
	phylogenetic tree display and annotation tool.
	\newblock {\em Nucleic Acids Research}, page gkae268, 2024.
	
\end{thebibliography}
\end{document}